\documentclass[journal]{IEEEtran}
\IEEEoverridecommandlockouts
\usepackage{macros}
\allowdisplaybreaks

\begin{document}

\title{Quantum Pufferfish Privacy: A Flexible
Privacy Framework for Quantum Systems}
\author{Theshani~Nuradha,~\IEEEmembership{Graduate Student Member,~IEEE,} Ziv~Goldfeld,~\IEEEmembership{Member,~IEEE,}~and
       Mark~M.~Wilde,~\IEEEmembership{Fellow,~IEEE}%
\thanks{T. Nuradha, Z. Goldfeld, and M.~M. Wilde are with the school of Electrical and Computer Engineering, Cornell University, Ithaca, New York 14850, USA. (e-mail: pt388@cornell.edu, goldfeld@cornell.edu, wilde@cornell.edu)}%
\thanks{T. Nuradha and M.~M. Wilde acknowledge support from the National Science Foundation under Grant No.~1907615 and 2315398.
Z. Goldfeld is partially supported by NSF grants CCF-2046018, DMS-2210368, and CCF-2308446, and the
IBM Academic Award.}
}

\maketitle

 \begin{abstract}
We propose a  versatile privacy framework for quantum systems, termed \textit{quantum pufferfish privacy} (QPP). Inspired by classical pufferfish privacy, our formulation generalizes and addresses limitations of quantum differential privacy by offering flexibility in specifying private information, feasible measurements, and domain knowledge. We show that QPP can be equivalently formulated in terms of the Datta--Leditzky information spectrum divergence, thus providing the first  operational interpretation thereof. We reformulate this divergence as a semi-definite program  and derive several properties of it, which are then used to prove convexity,  composability, and post-processing of QPP mechanisms. Parameters that guarantee QPP of the depolarization mechanism are also derived. We analyze the privacy-utility tradeoff of general QPP mechanisms and, again, study the depolarization mechanism as an explicit instance. The QPP framework is then applied to privacy auditing for identifying privacy violations via a hypothesis testing pipeline that leverages quantum algorithms. Connections to quantum fairness and other quantum divergences are also explored and several variants of QPP are examined.
 \end{abstract}
 
 \begin{IEEEkeywords}
 Auditing privacy, privacy-utility tradeoff, pufferfish privacy, quantum differential privacy, quantum generalized divergences
 \end{IEEEkeywords}

\setcounter{tocdepth}{2}

\tableofcontents

\section{Introduction}

With a surging interest in quantum and hybrid classical--quantum systems, ensuring privacy of both classical and quantum data has become pivotal.
Privacy-preserving data analysis has been widely studied for classical systems by means of statistical privacy frameworks. {Differential privacy (DP) is an important statistical privacy framework} that enables answering aggregate queries about a database while keeping individual records private~\cite{DMNS06, DR14}. However, DP  accounts for one type of private information only (namely, records of individual users), and it does not allow encoding domain knowledge into the framework. To address these limitations, a versatile generalization of~DP, termed Pufferfish Privacy (PP), has been proposed~\cite{KM14}. PP~allows for customizing which information is regarded as private and explicitly integrates distributional assumptions into the definition~\cite{KM14,SWC17,ZOC20}. {PP has found use in several applications, including smart metering~\cite{SmartMeteringPP1,SmartMeteringPP2} and trajectory monitoring with location tracking~\cite{Monitoring1,Monitoring2} (see also Figure~1 of~\cite{nuradha2022pufferfishJ} for an explicit example related to salary releases, where PP is applicable).}
Information-theoretic formulations
of classical DP and PP have been proposed in~\cite{CY16} and~\cite{nuradha2022information}, respectively.  

Quantum DP (QDP) is a generalization of the classical DP notion and has been proposed in~\cite{QDP_computation17}. See also~\cite{aaronson2019gentle} for DP of quantum measurements and~\cite{hirche2023quantum} for an information-theoretic interpretation of QDP. 
Connections to quantum stability through private learning have been studied in~\cite{quek2021private}. Moreover,~\cite{du2021quantum} has explored how quantum classifiers can be made private by using the intrinsic noise of existing quantum systems. See also~\cite{QML_DP17, measurementQLDP22,QDP_LASSO22,QML_DPwatkins23,huang2023certified} for applications of DP in quantum machine learning. Additionally, privacy amplification of quantum and quantum inspired algorithms has been analysed using QDP and classical DP notions in~\cite{privacyAmplificationQDPandDP22}. However, similar to the classical case, the versatility of QDP is limited.

In this paper, we propose a flexible privacy framework for quantum systems, termed quantum PP (QPP), that addresses these limitations. We provide a comprehensive study of QPP, encompassing properties, mechanisms, privacy-utility tradeoffs, as well as the first operational meaning of the Datta--Leditzky information spectrum divergence~\cite{datta2014second} (hereafter abbreviated as the DL~divergence), which arises from our framework.

\subsection{Motivation}

\begin{figure}
\begin{center}
\includegraphics[width=0.36\textwidth]{ 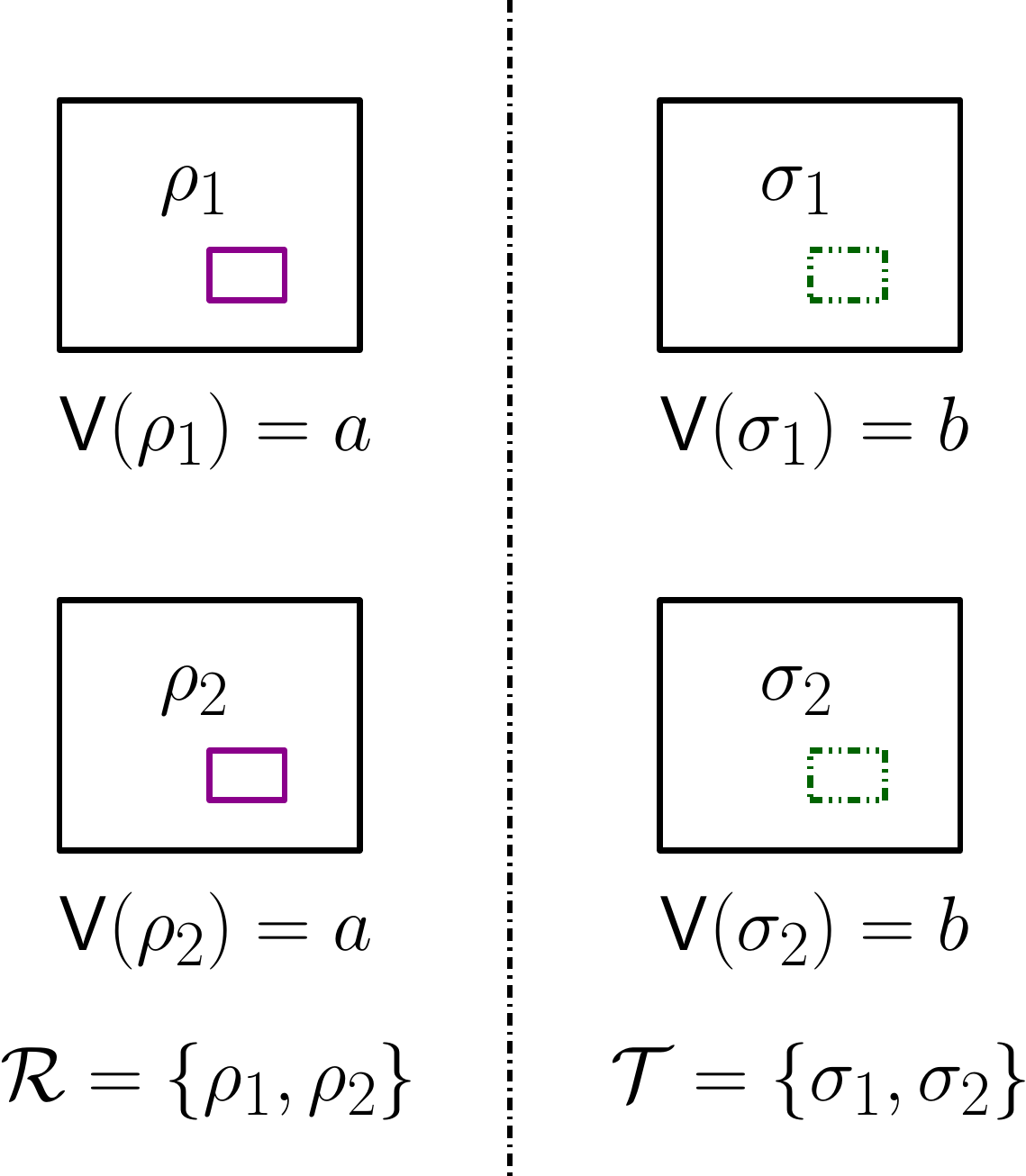} 
  \caption{Depiction of a setup where the goal is to hide whether the amount entanglement $\sV$ present in the bipartite states $\rho_1$, $\rho_2$, $\sigma_1$, and $\sigma_2$ equals $a$ or $b$. In this diagram, large squares represent the entire quantum state, while small rectangles correspond to a specific attribute of that state (i.e., {the amount of entanglement as quantified by the function $\mathsf{V}$}). The specific attribute can take on one of two values, $a$ or $b$, represented by solid or dotted lines, respectively. As the goal is to conceal only the entanglement level, and not necessarily the specific quantum state, we want the sets $\cR=\{\rho_1,\rho_2\}$ and $\cT=\{\sigma_1,\sigma_2\}$ to be indistinguishable.
  }
  \label{fig:example_qpp} 
 \end{center}
\end{figure}

We seek to address key limitations of QDP by exploring more \textit{flexible privacy frameworks} for quantum information processing. As delineated next, flexible secrets, embedding domain knowledge, and relaxing the need for worst-case measurements are considerations central to our approach. 

\medskip
   \noindent\textbf{Flexible secrets.}
QDP guarantees that any pair of states that are classified as neighbors are approximately indistinguishable, i.e., cannot be identified under any possible measurement. 
However, scenarios may arise in which one wants to hide specific properties of the states, as opposed to the state itself (e.g., whether the states possess a certain symmetry or property, have any entanglement with a special subsystem, or have secret correlations with other systems). 
 In such situations, QDP may be an overly pessimistic notion of privacy, which, in turn, hinders utility. As an example, consider hiding the amount of entanglement~$\mathsf{V}$ present in the bipartite states in the set $\{\rho_1,\rho_2,\sigma_1, \sigma_2\}$, for which $\sV(\rho_1)=\sV(\rho_2)=a$ and $\sV(\sigma_1)=\sV(\sigma_2)=b$. As illustrated in \cref{fig:example_qpp}, hiding whether~$\sV$ equals $a$ or $b$ amounts to making the classes $\{\rho_1,\rho_2\}$ and $\{\sigma_1,\sigma_2\}$ indistinguishable. This can be achieved by applying a QDP mechanism to the state space $\{\rho_1,\rho_2,\sigma_1, \sigma_2\}$ {by choosing $(\rho_i, \sigma_j)$ for all $i,j \in \{1,2\}$ as neighbors, with the criterion that $\rho$ and $\sigma$ are neighbors if and only if $|\sV(\rho) - \sV(\sigma) | = |a-b|$}. However, doing so provides a stricter guarantee than required. 
The source of the issue is the inability of QDP to account for secrets concerning collections of states (as opposed to singletons), which is the first issue we aim to address.

\medskip
\noindent\textbf{Domain knowledge.} 
In QDP, a worst-case privacy guarantee is provided for all neighboring states. However, one may possess knowledge about the likelihood of observing different states, e.g., via expert feedback. Referring back to the setting from \cref{fig:example_qpp}, if we have domain knowledge such that observing the states $\rho_1,\rho_2,\sigma_1, \sigma_2$ is prescribed by the probability vector $(p/2, \left(1-p\right)/2, 1/2, 0)$, for $p \in (0,1) $, then the requirement simplifies to the indistinguishability of $\{\rho_1,\rho_2\}$ versus $\{ \sigma_1 \}$. 
Classically, it has been demonstrated that domain knowledge can be leveraged to design privacy mechanisms with increased accuracy and utility~\cite{bassily2013coupled,KM14}. %
This calls for a quantum privacy framework that can also encode domain knowledge. 

\medskip 
\noindent\textbf{Relaxing worst-case measurements.} Another worst-case aspect of QDP is its account of all possible measurements. However, such a requirement might be too stringent in practice, especially in quantum systems. As an example, while a joint measurement can accurately distinguish between entangled but physically separated states, oftentimes only local operations and classical communications (LOCC) are available (e.g., as considered in quantum data-hiding protocols~\cite{TDL01,DLT02,EW02,HLSW04_noUrl,HLS05,LWL16,LPW18_noUrl}). In such cases, one may achieve improved accuracy and utility by relaxing the privacy requirement to account for LOCC measurements only.

\smallskip

In sum, the rapid advancements in quantum technologies requires designing flexible privacy frameworks that can be adjusted to timely needs. Furnishing such a framework is the main objective of our paper.

\subsection{Contributions}

This work proposes a quantum analog of the PP framework that accounts for the three aforementioned aspects. Our formalism enables reasoning about privacy of quantum systems using information-theoretic tools. We provide a comprehensive study of QPP, encompassing properties, mechanisms, and privacy-utility tradeoffs. Our paradigm also gives rise to  the first operational interpretation of the DL~divergence~\cite{datta2014second}. 
The proposed QPP framework comprises four key ingredients:
\begin{enumerate}
\item the set of potential secrets,
\item the set of discriminative pairs that are required to be indistinguishable at the output of the mechanism,
\item the set of data distributions, which encodes domain knowledge on the occurrence of quantum or classical data, and 
\item 
the set of measurements to be accounted for, which is specified based on physical, ethical, or any other constraints.
\end{enumerate}
See \cref{def: qpp} for a formal definition.
QPP guarantees the indistinguishability, under any allowable measurement, of sets of states formed based on the above ingredients.

After defining the operational privacy framework, we observe that when the measurement class contains all possible measurements, QPP can be equivalently posed as a DL~divergence constraint. To the best of our knowledge, this provides the first operational interpretation of the DL~divergence. We then derive an efficiently computable formulation of the DL~divergence as a semi-definite program (SDP), which may be of independent interest. This SDP is utilized to prove properties of the DL~divergence, which are then used in the analysis of QPP mechanisms. These properties include joint quasi-convexity and the data-processing inequality under positive and trace non-increasing maps  (see \cref{Sec:DL-divergence}). Our results also generalize the connection between the hockey-stick divergence and QDP, originally established in~\cite{hirche2023quantum}. Moreover, we show that existing privacy frameworks such as classical DP~\cite{DMNS06,DR14}, classical PP~\cite{KM14}, utility-optimized local DP (not subsumed by classical PP)~\cite{murakami2019utilityOptimizedLDP}, and QDP~\cite{QDP_computation17,hirche2023quantum} are special cases of our QPP~framework.

We then move on to derive properties of QPP mechanisms, encompassing convexity, post-processing, and composability (both parallel and adaptive). As a specific example, we characterize the flip parameter that guarantees QPP of the depolarization mechanism. We also describe how QPP mechanisms implementable on quantum devices can be instantiated to achieve classical PP. We consider the associated privacy-utility tradeoff for QPP mechanisms. Our utility metric captures how invertible the privacy mechanism is, which is formulated as the infimized diamond distance between a post-processing of the mechanism's output and the identity channel. We show that this utility metric can be computed as an SDP, and we analyze the privacy-utility tradeoff of the depolarization mechanism. Lastly, we study optimal privacy-utility tradeoffs of QPP mechanisms and characterize the achievable region in several settings. 

Another application we consider is privacy auditing, which refers to certifying whether a black-box mechanism satisfies a target privacy guarantee. While several auditing methods are available for classical frameworks, there is currently no approach that can handle quantum data. We fill this gap by proposing the first auditing pipeline for quantum privacy mechanisms. In contrast to existing approaches for classical DP and PP, which require first relaxing the privacy notion and only then auditing, our approach audits for QDP directly. Extensions of these ideas to the QPP setting are also considered. 

Finally, we explore connections between QPP, existing quantum privacy frameworks, and figures of merit. First, we examine the connection between quantum privacy and fairness~\cite{fairnessQ_verifying22,Fairness_quantum21}, showing that private quantum mechanisms are fair, and under certain conditions, fair algorithms are private. We also provide bounds on quantum R\'enyi divergences and the trace distance, which stem from QPP. This inspires relaxations of QPP that are defined via these divergences, which, in particular, provide other operational interpretations thereof as privacy metrics. Lastly, we present a variant of QPP that can incorporate entanglement into the framework with the use of reference systems. 

\subsection{Organization}

The rest of our paper is organized as follows. In \cref{Sec:Background}, we introduce notation and preliminaries in quantum information theory and privacy. \cref{Sec:Quantum-Pufferfish-Privacy} presents the QPP framework, its equivalent formulations in terms of the DL~divergence, and special cases of it. In \cref{Sec:DL-divergence}, we focus further on the DL~divergence, reformulate it as an SDP, and use this SDP to prove several properties of it. Properties of general QPP mechanisms and the depolarization mechanism are studied in \cref{Sec:Properties-and-Mechanisms-for-QPP}. We analyze the privacy-utility tradeoff of QPP mechanisms in \cref{Sec:Quantifying-Privacy-Utility-Tradeoff}, while the privacy auditing application is considered in \cref{Sec:Auditing-Privacy-Frameworks}. Connections to existing privacy frameworks and to other quantum divergences are explored in \cref{Sec:Connections-to-the-Tools-of-Interest-to-Quantum-community}. 
Then, we propose several relaxations and variants of QPP in \cref{sec:variants-of-QPP}. 
\cref{Sec:summary} summarizes our main contributions and provides concluding remarks.

\section{Preliminaries and Background} \label{Sec:Background}

\subsection{Notation}

Sets are denoted by calligraphic letters, e.g., $\cX$. For $k,n \in\NN$, we use $\cX^{n \times k}$ to denote the database space of $n\times k$ matrices; columns correspond to different attributes while rows to different individuals. The $(i,j)$th entry of $x\in\cX^{n\times k}$ is denoted as $x(i,j)$. The $i$th row and $j$th column of $x$ are denoted by $x(i,\cdot)$ and $x(\cdot,j)$, respectively. We denote by $(\Omega,\cF,\PP)$ the underlying probability space on which all random variables (RVs) are defined, with $\EE$ designating expectation. RVs are denoted by upper case letters, e.g., $X$, with $P_X$ representing the corresponding probability law. For $X\sim P_X$, we interchangeably use $\supp(X)$ and $\supp(P_X)$ for the support. The joint law of $(X,Y)$ is denoted by $P_{XY}$, while $P_{Y|X}$ designates the (regular) conditional probability of $Y$ given $X$. 
Conventions for $n\times k$-dimensional random variables are the same as for deterministic elements. The space of all Borel probability measures on $\cS\subseteq\RR^d$ is denoted by $\cP(\cS)$. 
The Kullback--Leibler (KL) divergence between $P, Q\in\cP(\cX)$ with $P\ll Q$ is given by $\dkl(P\|Q) \coloneqq \EE_P\!\left[\ln\!\left(\frac{d P}{d Q}\right)\right],$ where $\frac{d P}{d Q}$ is the Radon--Nikodym derivative of $P$ with respect to  $Q$. For $(X,Y)\sim P_{XY}$, the mutual information between $X$ and $Y$ is denoted by $\sI(X;Y)\coloneqq  \mathsf{D}(P_{XY}\|P_X \otimes P_Y)$.

We now review basic concepts from quantum information theory and refer the reader to~\cite{wilde2017quantum,khatri2020principles} for more details. A (classical or quantum) system $R$ is identified with a finite-dimensional Hilbert space~$\cH_R$. We denote the set of linear operators acting on $\cH_R$ by $\cL(\cH_R)$. The support of a linear operator $X \in \cL(\cH_R)$ is defined to be the orthogonal complement of its kernel, and it is denoted by $\supp(X)$.
Let $\T(C)$ 
denote the transpose of $C$. 
The partial transpose of $C \in \cL(\cH_A \otimes \cH_B)$ on the subsystem $A$ is represented as $\T_A(C)$.
Let $\Tr\!\left[C \right]$ denote the trace of $C$, and let $\Tr_A \!\left[C\right]$ denote the partial trace of $C$ over the subsystem $A$. The trace norm  of a matrix $B$ is defined as $\left\|B\right\|_1 \coloneqq \Tr[\sqrt{B^\dagger B} ]$. For operators $A$ and $B$, the notation $A \geq B$ indicates that $A-B$ is a positive semi-definite (PSD) operator, while $A > B$ indicates that $A-B$ is a positive definite operator.

A quantum state $\rho_R\in\cL(\cH_R)$ on $R$ is a PSD, unit-trace operator acting on $\cH_R$. We denote the set of all density operators in 
$\cL(\cH_R)$ by $\cD(\cH_R)$. A state $\rho_R$ of rank one is called pure, and we may choose a normalized vector $| \psi \rangle \in \cH_R$ satisfying $\rho_R= | \psi \rangle\!\langle \psi | $ in this case. Otherwise,
$\rho_R$ is called a mixed state. By the spectral decomposition theorem, every mixed state can be written as a convex combination
of pure, orthogonal states. 
A quantum channel $\cN: \cL(\cH_A ) \to \cL(\cH_B)$ is a linear, 
completely positive and trace-preserving (CPTP) map from $\cL(\cH_A)$ to $\cL(\cH_B)$. We denote the adjoint of $\cN$ by $\cN^\dagger$. A measurement of a quantum system $R$ is described by a
positive operator-valued measure (POVM) $\{\rM_y\}_{y \in \cY}$, which is defined to be a collection of PSD operators  satisfying $\sum_{y \in \cY} \rM_y= \rI_{\cH_R}$, where $\cY$ is a finite alphabet. The Born rule dictates that, after applying the above POVM to $\rho \in \cD(\cH_R)$, the probability of observing the outcome $y$ is given by $\Tr\!\left[\rM_y \rho \right]$.

\subsection{Quantum Divergences}

We define several quantum divergences that will be used throughout this work.
We call a distinguishability measure $\boldsymbol{\sD}(\cdot \Vert \cdot)$ a generalized divergence~\cite{SW12} if it satisfies the data-processing inequality; i.e., for every channel $\cN$, state $\rho$, and PSD operator~$\sigma$, 
\begin{equation}\label{eq:generalized-divergence}
    \boldsymbol{\sD}(\rho \Vert \sigma) \geq \boldsymbol{\sD}\!\left(\cN(\rho) \Vert \cN(\sigma) \right).
\end{equation}

The normalized trace distance between the states $\rho$ and $\sigma$ is defined as
\begin{equation}
\label{eq:normalized-TD}
    \sT(\rho,\sigma)\coloneqq \frac{1}{2} \left\| \rho -\sigma\right\|_1,
\end{equation} 
 while the fidelity between them is defined as~\cite{Uhl76_nourl}
 \begin{equation}
  \sF(\rho,\sigma)\coloneqq  \left\| \sqrt{\rho} \sqrt{\sigma}\right\|^2_1.   
 \end{equation}
The diamond distance between the two channels $\cN,\cM : \cL(\cH_A) \to \cL(\cH_B)$ is defined as~\cite{Kit97_noUrl}
 \begin{equation}
 \left\| \cN-\cM\right\|_\diamond\coloneqq  \sup_{\rho_{RA}} \left\| \cN_{A\to B}(\rho_{RA})- \cM_{A\to B}(\rho_{RA}) \right\|_1,   
 \end{equation}
 where the optimization in the definition is over every reference system $R$ and bipartite density operator~$\rho_{RA}$ (with $R$ allowed to be arbitrarily large). It is well known, however, that it suffices to perform the optimization over pure bipartite states such that the dimension of the reference system $R$ is equal to the dimension of the channel input system $A$.

The Petz--R\'enyi quantum relative entropy of order $\alpha \in (0,1) \cup (1, \infty)$ of a state $\rho$ with respect to  a PSD operator~$\sigma$ is given by~\cite{P85,P86}
\begin{equation}\label{eq:petz renyi}
  \sD_\alpha(\rho\Vert \sigma) \coloneqq  \frac{1}{\alpha-1} \ln\Tr[ \rho^\alpha \sigma^{1- \alpha}]
\end{equation}
if $\alpha \in (0,1) \lor
(\alpha > 1 \land \supp(\rho)\subseteq\supp(\sigma))$ 
and $\infty$ otherwise.
{It is a generalized divergence for $\alpha \in [0,1) \cup (1,2]$~\cite{P86}.} 
The special case of $\alpha\to 1$ is called the quantum relative entropy and amounts to
\begin{equation} 
\sD(\rho \| \sigma) \equiv \sD_1(\rho \| \sigma)\coloneqq  \lim_{\alpha \to 1} \sD_\alpha(\rho \Vert \sigma)=\Tr\!\left [\rho (\ln \rho - \ln \sigma) \right] \label{eq: limit alpha 1}
\end{equation}
when $\supp(\rho)\subseteq\supp(\sigma)$ and it is equal to $+\infty$ otherwise. 
The quantum entropy of a state $\rho$ is defined as
\begin{equation}
\sS(\rho)\coloneqq  -\Tr\!\left[\rho \ln \rho \right] .   
\end{equation}
Equivalently, $\sS(\rho)=-\sD_1(\rho\| \rI)$, where $\rI$ is the identity operator.  

Fix $\alpha \in (0,1) \cup (1, \infty)$. The sandwiched R\'enyi relative entropy of a state $\rho$ and a PSD operator~$\sigma$ is defined as~\cite{muller2013quantum, wilde2014strong}
\begin{equation} \label{eq:sandwiched-renyi-def}
\widetilde{\sD}_\alpha(\rho\Vert \sigma)  \coloneqq \frac{1}{\alpha-1} \ln \Tr\!\left[ \left( \sigma^{\frac{1-\alpha}{2\alpha}}\rho \sigma^{\frac{1- \alpha}{2\alpha}} \right)^\alpha \right]
\end{equation}
if $\alpha \in (0,1) \, \lor  (\alpha \in (1, \infty)\land  \ \supp(\rho)\subseteq\supp(\sigma))$ 
and $\infty$ otherwise.
It is a generalized divergence for $\alpha \in [1/2,1)\cup(1,\infty)$~\cite{FL13} (see also~\cite{W18opt}).

Fix $\delta \in [0,1]$, a state $\rho$, and a PSD operator~$\sigma$. The Datta--Leditzky information spectrum divergences  are defined as follows~\cite{datta2014second}:
\begin{subequations}
\begin{align}
\underline{\sD}^{\delta}(\rho\Vert\sigma) &  \coloneqq  \sup\left\{  \gamma
\in\mathbb{R}:\operatorname{Tr}[\left(  \rho-e^{\gamma}\sigma\right)
_{+}]\geq1-\delta\right\} ,  
\label{eq:1st-DL-def} \\
\overline{\sD}^{\delta}(\rho\Vert\sigma) &  \coloneqq  \inf\left\{  \gamma
\in\mathbb{R}:\operatorname{Tr}[\left(  \rho-e^{\gamma}\sigma\right)
_{+}]\leq\delta\right\}  ,
\label{eq:2nd-DL-def}
\end{align}\label{eq:DL_div}%
\end{subequations}
where
\begin{equation}
\left(  A\right)  _{+}\coloneqq \sum_{i:a_{i}\geq0}a_{i}|i\rangle\!\langle i|    
\end{equation}
for a Hermitian operator $A~=~\sum_{i}a_{i}|i\rangle\!\langle i|$. Hereafter we abbreviate these divergences as DL~divergences. 
Proposition~4.3 of~\cite{datta2014second} shows that 
\begin{equation}
\underline{\sD}^{\delta}(\rho\Vert\sigma)=\overline{\sD}%
^{1-\delta}(\rho\Vert\sigma), \label{eq:inf-spec-equality}
\end{equation}
and so we can speak of a single DL~divergence, which we set hereafter to be $\overline{\sD}^{\delta}$ from~\eqref{eq:2nd-DL-def}. Slightly rewriting~\eqref{eq:DL_div}, we have the equivalent representations:
\begin{subequations}
\begin{align}
\underline{\sD}^{\delta}(\rho\Vert\sigma) &  =\ln \sup\left\{
\lambda\geq0:\operatorname{Tr}[\left(  \rho-\lambda\sigma\right)  _{+}%
]\geq1-\delta\right\}  
\label{eq:DL-def-rewrite-1}\\
\overline{\sD}^{\delta}(\rho\Vert\sigma) &  =  \ln \inf\left\{
\lambda\geq0:\operatorname{Tr}[\left(  \rho-\lambda\sigma\right)  _{+}%
]\leq\delta\right\}   \label{eq: overline DL}.
\end{align}\label{eq:DL_rewrite}%
\end{subequations}

The max-relative entropy of a state $\rho$ and a PSD
operator~$\sigma$ is defined as~\cite{datta2009min}
\begin{align}
\sD_{\max}(\rho\Vert\sigma) & \coloneqq\ln \inf\left\{  \lambda:\rho\leq\lambda
\sigma\right\} \label{eq:D-max-def} \\
& = \ln \sup_{0\psd \rM \psd \rI} \frac{\Tr\!\left[\rM \rho\right]}{\Tr\!\left[\rM \sigma\right]}, \label{eq:D-max-def_ALT}
\end{align}
and the smooth max-relative entropy is defined for $\delta \in [0,1]$ as 
\begin{equation}
\label{eq:smooth-max-rel}
\sD_{\max}^{\delta}(\rho\Vert\sigma)\coloneqq\inf_{\widetilde{\rho}\,:\frac{1}%
{2}\left\Vert \widetilde{\rho}-\rho\right\Vert _{1}\leq\delta}\sD_{\max
}(\widetilde{\rho}\Vert\sigma),
\end{equation}
with the optimization taken over every state $\widetilde{\rho}$. These quantities have been given an operational meaning in~\cite{wang2019resource}.

The Thompson metric~\cite{thompson1963certain} is defined in terms of the max-relative entropy as 
\begin{equation}
    \sD_T(\rho \Vert \sigma) \coloneqq \max \{\sD_{\max}(\rho \Vert \sigma), \sD_{\max}(\sigma \Vert \rho)\},
    \label{eq:Thompson-metric-def}
\end{equation}
and it has been given an operational meaning in~\cite{salzmann2021symmetric,regula2022postselected}.

\subsection{Classical and Quantum Privacy} 

In this section, we provide background on the existing definitions of privacy for both classical and quantum systems, starting from classical DP and proceeding to quantum DP thereafter. 

\medskip
\subsubsection{Classical Differential and Pufferfish Privacy}

DP allows for answering queries about aggregate quantities while protecting the individual entries in a database~\cite{DMNS06}. To this end, the output of a differential privacy mechanism should be indistinguishable for neighboring databases, defined as those that differ only in a single record (row). Formally, we say~that $x,x'\in \cX^{n\times k}$ are neighbors, denoted $x\sim x'$,  if $x(i,\cdot)\neq x'(i,\cdot)$ for some $i\in\{1,\ldots,n\}$, and they agree on all other rows. We also note that a randomized privacy mechanism~$A$, as mentioned below, is described by a (regular) conditional probability distribution $P_{A|X}$ for its output given the data.

\begin{definition}[Classical differential privacy] \label{def: DP}
Fix $\varepsilon \geq 0$ and $\delta \in [0,1]$. A~randomized mechanism ${A}: \cX^{n \times k} \to \cY$ is $(\varepsilon,\delta)$-differentially private if 
\begin{equation}
\PP\big(A(x) \in \cB \big) \leq e^{\varepsilon} \hspace{1mm} \PP\big(A(x') \in \cB\big) + \delta,\label{eq:dp_def}
\end{equation}
for all $x \sim x'$ with $x,x' \in \cX^{n \times k}$ and $\cB \subseteq \cY $ measurable.
\end{definition}

As is evident from the above definition, DP aims to conceal whether any particular individual (row) is in fact part of the database or not. While being a powerful and widely applicable privacy framework, it is often appropriate to consider even broader frameworks. Pufferfish privacy~\cite{KM14} is a versatile generalization of DP that not only allows flexibility in the definition of secrets but also enables the integration of domain knowledge of the database space $\cX^{n \times k}$. The PP framework consists of three components: 
\begin{enumerate}
    \item A set of secrets $\mathcal{S}\subseteq\cX^{n \times k}$ of measurable subsets;
    \item  A set of secret pairs $\mathcal{Q} \subseteq \mathcal{S} \times \mathcal{S}$ that need to be indistinguishable in the $(\varepsilon,\delta)$ sense (cf.,~\eqref{eq:pp_def} below),
    \item A class of data distributions $\Theta \subseteq \cP(\cX^{n \times k})$ that captures prior beliefs or domain knowledge.
\end{enumerate}

As formulated next, PP aims to guarantee that all secret pairs in $\cQ$ are indistinguishable with respect to the prior beliefs $P_X \in \Theta$.  
\begin{definition}[Classical pufferfish privacy]\label{def:pp_def}
Fix $\varepsilon \geq 0$ and $\delta \in [0,1]$. A randomized mechanism ${A}: \cX^{n \times k} \to \cY$ is $(\varepsilon,\delta)$-private in the pufferfish framework $(\mathcal{S},\mathcal{Q},\Theta)$ if 
\begin{equation}
\PP\big(A(X) \in \cB \big| \cR\big) \leq e^{\varepsilon} \hspace{1mm} \PP\big(A(X) \in \cB \big | \cT\big) + \delta,\label{eq:pp_def}
\end{equation}
for all $P_X \in \Theta$, $(\cR,\cT) \in \mathcal{Q}$ with $P_X(\cR),P_X(\cT)> 0$, and $\cB \subseteq \cY $ measurable.
\end{definition}

DP from \cref{def: DP} is a special case of PP when $\cS=\cX^{n\times k}$, the set $\cQ$ contains all neighboring pairs of databases, and $\Theta=\cP(\cX^{n \times k})$ (i.e., there are no distributional assumptions, and privacy is guaranteed in the worst case). Other important examples that are subsumed by PP include (i) generic DP~\cite{SoK_DP22}, which allows for arbitrary neighboring relationships, and (ii) attribute privacy~\cite{ZOC20}, which privatizes global properties of a database (e.g., a column that corresponds to some sensitive information, such as salary).

\medskip
\subsubsection{Quantum Differential Privacy}

QDP lifts the notion of DP to the space of quantum states, with the neighboring relation typically defined either in terms of 
closeness in trace distance~\cite{QDP_computation17}, reachability by a single local operation~\cite{aaronson2019gentle},\footnote{Given two quantum states $\rho$ and $\sigma$ of $n$ registers each, call them neighbors if it is possible to reach either $\sigma$ from $\rho$ or $\rho$ from $\sigma$ by performing a general quantum channel 
on a single register only.} or by quantum Wasserstein distance of order 1~\cite{quantumWasserstein1_21}. We denote two states being neighbors by $\rho \sim \sigma$.

\begin{definition}[Quantum differential privacy~\cite{QDP_computation17,hirche2023quantum}] \label{def: QDP}
Fix $\varepsilon \geq 0$ and $\delta \in [0,1]$. Let $\cD$ be a set of quantum states, and let $\cA$ be a quantum algorithm (viz., a quantum channel). The algorithm~$\cA$ is $(\varepsilon,\delta)$-differentially private if 
\begin{equation} \Tr\!\left[\rM \cA(\rho)\right] \leq e^\varepsilon \Tr\!\left[\rM \cA(\sigma)\right]+ \delta.
\end{equation}
for every measurement operator $\rM$ (i.e., satisfying  $0 \psd \rM \psd \rI$) and all $\rho, \sigma \in \cD$ such that $\rho \sim \sigma$.
\end{definition}
This definition reduces to classical DP for discrete-output mechanisms with an appropriate choice of the measurement set. See \cref{rem:classical-PP} below for further details.

\section{Quantum Pufferfish Privacy (QPP)}\label{Sec:Quantum-Pufferfish-Privacy}

Inspired by the versatility of the classical PP framework, we propose a quantum variant thereof. Termed QPP, our framework allows for customizing the notion of private states, tailoring the feasible set of measurements to the application of interest, and incorporating domain knowledge of the state distribution into the model. As such, the QPP framework can generate a rich class of privacy definitions for both classical and quantum systems, and for hybrid classical--quantum systems as well.

\subsection{Framework}

The QPP framework requires a domain expert to specify four components: a set $\cS$ of potential secrets, a set $\cQ \subseteq \cS \times \cS$ of discriminative pairs, a set $\Theta$ of data distributions, and a set~$\cM$ of measurements. We expand on and explicitly define each component next. 

\textbf{Set $\cS$ of potential secrets:} Secrets are modeled as subsets of density operators that share a certain property (these subsets are merely singletons in the QDP case). The set $\cS$ is a collection of such secret subsets. For example, if one aims to privatize the resource value $\sV$ of a state, then the corresponding set of secrets is $\cS=\bigcup_{i=1}^n\cT_i$, where
\begin{equation}
    \cT_i = \big\{\rho \in \cD(\cH): \sV(\rho)=a_i\big\} 
\end{equation}
and $\{a_i\}_{i=1}^n$ are the possible values that $\sV$ can take (recall that, in \cref{fig:example_qpp}, we considered a setup relevant to hiding the resource value $\sV$ being $a$ or $b$).

\textbf{Set $\cQ$ of discriminative pairs:} This is a subset of $\cS \times \cS$ that specifies which pairs of elements from $\cS$ should be indistinguishable.  
Namely, if $(\cT_1,\cT_2) \in \cQ$, then the goal of the privacy mechanism is to conceal whether the input belongs to $\cT_1$ or $\cT_2$. Note that $\rho \in \cT_1 \Rightarrow \rho \notin \cT_2$. We require that $\cQ$ is symmetric, i.e., that $(\cT_i,\cT_j) \in \cQ$ if and only if $(\cT_j,\cT_i) \in \cQ $.
Proceeding with the same example, we can set 
\begin{equation}
\cQ=\bigcup_{i\neq j} \{(\cT_i,\cT_j)\}.
\end{equation}

\textbf{Set $\Theta$ of data distributions:} A collection of probability distributions $P_X \in \cP(\cX)$ over a finite space~$\cX$ that indexes an ensemble of density operators $\{\rho^x\}_{x\in\cX}$. Taking $X\sim P_X\in\Theta$, the matrix-valued random  variable $\rho^X$ models a density operator that is randomly chosen according to $P_X$. Proceeding with the same example,  $\{\rho^x\}_{x\in\cX}= \{\sigma: \sigma \in \cT_i, \ \cT_i \in \cS \} \subset \cD(\cH)$.
The set $\Theta$ can be 
understood as capturing beliefs that the adversary has regarding the state of the system. 

 {In the above example, we have considered a subset of density operators (i.e., $\{\rho^x\}_{x\in\cX} \subset \cD(\cH)$). There could be applications where we have to consider all density operators. To incorporate this, we choose the following: Fix $k\in\NN$ and let $\mathfrak{F}_k\subset 2^{\cD(\cH)}$ be the collection of all finite subsets of $\cD(\cH)$ with $k$ elements. For each $\cF\in\mathfrak{F}_k$, we write $\cP(\cF)$ for the class of all distributions supported on $\cF$, and define 
\begin{equation}\label{eq:finite_all_distributions}
\cP_k\big(\cD(\cH)\big)\coloneqq \bigcup_{\cF\in\mathfrak{F}_k}\cP(\cF).  
\end{equation}
Every distribution $P\in\cP_k\big(\cD(\cH)\big)$ is supported on exactly $k$ density operators. Note that all density operators outside of the underlying  finite set comprise of the null set. We associate a random variable $X\sim P=P_X$ with each such distribution and write $\cX=\supp(P_X)$ for its support. Note the slight abuse in notation, as the support of $P_X$ changes with the distribution, which is not reflected in the generic indexing set $\cX$. The set of data distributions in the QPP framework is now taken as $\Theta\subseteq\cP_k\big(\cD(\cH)\big)$ for some $k\in \mathbb{N}$.}

\textbf{Set $\cM$ of measurements:} This set is a subset of all possible measurements, i.e., $\cM \subseteq \{ \rM:  0 \psd \rM  \psd \rI \}$. The choice of $\cM$ gives the flexibility to consider only measurements that are possible under physical, legal, or ethical constraints. 

\begin{remark}[Designing QPP frameworks]
QPP allows system designers to explicitly encode their assumptions into the privacy framework. 
Setting the framework $(\cS,\cQ,\Theta,\cM)$ to accurately reflect the application of interest is crucial for obtaining meaningful privacy guarantees and to optimize utility.  
Explicit assumptions can also help account for ethical or fairness concerns associated with quantum systems; cf. \cref{rem:Privacy-implying-fairness} for a concrete example concerning quantum fairness and how it is incorporated within QPP. 
\end{remark}

Now, we are ready to present a formal definition of the quantum analog of PP, which we call QPP. 

\begin{definition}[Quantum pufferfish privacy] \label{def: qpp}
Fix $\varepsilon \geq 0$ and $\delta \in [0,1]$. 
A quantum algorithm $\cA$ is $(\varepsilon,\delta)$-private in the quantum pufferfish framework $(\mathcal{S},\mathcal{Q}, \Theta, \cM )$ if for all $ P_X \in \Theta$, $(\cR,\cT) \in \mathcal{Q}$ with $P_X(\cR),P_X(\cT)> 0$,
 and all $\rM \in \cM$, the following inequality holds:
\begin{equation}
\Tr\!\left[\rM \cA(\rho^\cR)\right] \leq e^\varepsilon \Tr\!\left[\rM \cA(\rho^\cT)\right] + \delta, \label{eq:qpp_def}
\end{equation}
where
\begin{align}
    \rho^\cR  & \coloneqq \sum_{\{x:\rho^x \in \cR\}} q_\cR(x) \rho^x, \\ 
    q_\cR(x)  & \coloneqq \frac{P_X(x)}{P_X(\cR)},\\
    P_X(\cR) & \coloneqq \sum_{\{x:\rho^x \in \cR\}} P_X(x),
\end{align}
and $\rho^\cT$ is defined similarly but with $\cT$ instead of $\cR$.
We say that an algorithm ${\cA}$ satisfies $\varepsilon$-QPP if it satisfies $(\varepsilon,0)$-QPP. 
\end{definition}

Evidently, discriminative secret pairs in $\cQ$ are indistinguishable at the output of a QPP mechanism $\cA$ in the $(\varepsilon, \delta)$-sense, under every measurement from the class $\cM$. 

\begin{remark}[Semantics of the QPP framework]
    Informally, the QPP framework provides the following privacy guarantee for fixed $(\cR,\cT) \in \cQ$ and $P_X \in \Theta$: For a state $\rho^X$ chosen according to $X \sim P_X$ and input to the quantum channel~$\cA$, an adversary applying a measurement $\rM \in \cM$ on the channel output~$\cA( \rho^X)$ draws the same conclusions regardless of whether $\rho^X$ belongs to $\cR$ or $ \cT$.
\end{remark}

\begin{remark}[Incorporating entanglement] \label{rem: Incorporating entanglement through reference systems}
We can incorporate entanglement in the QPP framework by introducing a reference system. Specifically, we can modify the QPP framework from  $(\cS,\cQ,\Theta,\cM)$ to $(\cS,\cG,\Theta,\cM')$, where
\begin{equation}
\label{eq: set L for reference systems}
\cG\coloneqq \left \{
\begin{array}[c]{c}
(\omega_{RA}^\cR, \omega_{RA}^\cT)  :  \omega_{RA}^\cR,\omega_{RA}^\cT \in \cD(\cH_R\otimes \cH_A), \\
\Tr_R\!\left[\omega_{RA}^\cR\right]= \rho^{\cR},  \Tr_R\!\left[\omega_{RA}^\cT \right]= \rho^{\cT}, \\
(\cR,\cT) \in \cQ
\end{array}
\right\} \end{equation}
is a set of pairs of bipartite states with $\rho^\cR$ and $\rho^\cT$ defined similar to \cref{def: qpp}. We then say that $\cA$ is $(\varepsilon,\delta)$-QPP in that framework if for all $P_X\in\Theta$, $\rM' \in \cM'$, and $(\omega_{RA}^\cR, \omega_{RA}^\cT) \in \cG$, we have
\begin{equation}
\Tr\!\left[\rM' (\cI \otimes \cA)(\omega_{RA}^\cR)\right] \leq e^\varepsilon \Tr\!\left[\rM' (\cI \otimes \cA)(\omega_{RA}^\cT) )\right] + \delta. \label{eq:qpp_def_reference}
\end{equation} 
However, it is unclear whether such a stronger privacy notion would be useful in practical applications. For example, consider $\sigma_1 \coloneqq |0\rangle\!\langle 0| \otimes \rho^\cR  $ and $\sigma_2 \coloneqq |1\rangle\!\langle 1| \otimes \rho^\cT  $ with $(\cR,\cT) \in \cQ$. If a measurement on the reference system can be applied, then a computational-basis measurement distinguishes $\sigma_1$ and $ \sigma_2$ perfectly. Thus, it is important to choose $\cA$ appropriately with a practically applicable $\cM'$, such that the required indistinguishability is achieved. 

{We shall revisit a variant of this framework with quantum divergences in \cref{SubS: Incorporating Entanglement}. The strength of the privacy framework is determined by the underlying quantum divergence. However, note that the problems discussed previously are not completely solved by the variant proposed therein.}
    
\end{remark}

\subsection{Equivalent Formulation of QPP with DL Divergence}

We present an equivalent formulation for $(\varepsilon,\delta)$-QPP by means of the DL~divergence from~\eqref{eq:2nd-DL-def}. To the best of our knowledge, this provides the first operational interpretation of the DL~divergence.

\begin{proposition}[Equivalent formulation of $(\varepsilon,\delta)$-QPP] \label{prop: Equivalent formulation with DS}
    Fix the framework $(\cS,\cQ,\Theta,\bar{\cM})$, with $\bar{\cM}$ corresponding to the set of all possible measurements. 
    Then algorithm~$\cA$ satisfies $(\varepsilon,\delta)$-QPP with respect to the framework $(\cS,\cQ,\cM,\Theta)$ if and only if for all  $P_X \in \Theta$ and $(\cR,\cT) \in \cQ$, we have
    \begin{align}
        &
          \overline{\sD}^{\delta}\!\left( \cA(\rho^\cR) \middle \Vert\cA(\rho^\cT) \right)  \leq \varepsilon.
        \label{eq:QPP-eq-DL-div-expr}
    \end{align}
    
\end{proposition}

\begin{IEEEproof}
 We first show that $(\varepsilon,\delta)$-QPP implies~\eqref{eq:QPP-eq-DL-div-expr}.
For fixed $P_X \in \Theta$ and $(\cR,\cT) \in \cQ$, observe that $(\varepsilon,\delta)$-QPP corresponds to 
\begin{equation}
\sup_{\rM \in \bar{\cM}} \Tr\!\left[ \rM \left(\cA(\rho^\cR) -e^\varepsilon \cA(\rho^\cT) \right) \right] \leq \delta.
\label{eq:qpp-basic-cond}
\end{equation}
Since
\begin{multline}
    \sup_{\rM \in \bar{\cM}} \mspace{-3mu}\Tr\mspace{-3mu}\left[ \rM \left(\cA(\rho^\cR)\mspace{-3mu} -\mspace{-3mu}e^\varepsilon\mspace{-2mu} \cA(\rho^\cT) \right) \right]\mspace{-3mu} \\ =\mspace{-3mu} \Tr\mspace{-3mu}\left[ \left(\cA(\rho^\cR) \mspace{-3mu}-\mspace{-3mu}e^\varepsilon \mspace{-2mu}\cA(\rho^\cT) \right)_{\mspace{-2mu}+}\mspace{-2mu} \right]\mspace{-3mu}, 
\end{multline}
{as a consequence of, e.g.,  \cite[Lemma~II.1]{hirche2023quantum},}
the inequality in~\eqref{eq:qpp-basic-cond} is equivalent to
\begin{equation}
    \Tr\!\left[ \left(\cA(\rho^\cR) -e^\varepsilon \cA(\rho^\cT) \right)_{+} \right] \leq \delta.
\end{equation} 
By the definition in~\eqref{eq:2nd-DL-def}, this leads to $\varepsilon$ being a possible candidate for the optimization in $\overline{\sD}^{\delta}\!\left(\cA(\rho^\cR) \middle \Vert \cA(\rho^\cT) \right)$, and thus implies
\begin{equation}
    \overline{\sD}^{\delta}\!\left(\cA(\rho^\cR) \Vert \cA(\rho^\cT) \right) \leq \varepsilon.
\end{equation}
As this holds for every $P_X \in \Theta$ and $(\cR,\cT) \in \cQ$, we obtain the desired implication $(\varepsilon,\delta)$-QPP $\Rightarrow$~\eqref{eq:QPP-eq-DL-div-expr}. 

\medskip
Next, we show that~\eqref{eq:QPP-eq-DL-div-expr} implies $(\varepsilon,\delta)$-QPP.
Suppose that for all $P_X \in \Theta$ and $(\cR,\cT) \in \cQ$, we have
$\overline{\sD}^{\delta}\!\left( \cA(\rho^\cR) \Vert\cA(\rho^\cT) \right) \leq \varepsilon$. 
Then, for fixed $P_X \in \Theta$ and $(\cR,\cT) \in \cQ$, 
let 
\begin{equation}
    \overline{\sD}^{\delta}\!\left( \cA(\rho^\cR) \middle \Vert\cA(\rho^\cT) \right)= \nu, 
\end{equation}
which implies that
$\Tr\!\left[ \left(\cA(\rho^\cR) -e^\nu \cA(\rho^\cT) \right)_{+} \right] \leq \delta$.
Recalling that $\nu \leq \varepsilon$ and noting that $\lambda\mapsto \Tr\!\left[ \left(\cA(\rho^\cR) -e^\lambda \cA(\rho^\cT) \right)_{+} \right]$ is a monotonically decreasing function (cf. \cite[Lemma~4.2]{datta2014second}), we have
\begin{align}
\sup_{\rM \in \bar{\cM}} &\Tr\!\left[ \rM \left(\cA(\rho^\cR) -e^\varepsilon \cA(\rho^\cT) \right) \right] \notag \\
&\qquad = \Tr\!\left[ \left(\cA(\rho^\cR) -e^\varepsilon \cA(\rho^\cT) \right)_{+} \right]  \\
&\qquad \leq \Tr\!\left[ \left(\cA(\rho^\cR) -e^\nu \cA(\rho^\cT) \right)_{+} \right] \\
&\qquad \leq  \delta.
\end{align}
As $P_X \in \Theta$ and $  (\cR,\cT) \in \cQ$ are arbitrary, $(\varepsilon,\delta)$-QPP follows.
\end{IEEEproof}

\medskip
In the following remark, we highlight how the DL divergence also provides a novel characterization for classical PP. 
\begin{remark}[Classical PP through DL divergence]
For discrete probability distributions $p,q \in \cP(\cY)$, the DL divergence in \cref{eq: overline DL} reduces to 
\begin{multline}\label{eq:classical_DL}
    \overline{\sD}^{\delta}_c(p \Vert q) \coloneqq \\ \ln \inf\left\{
\lambda\geq0: \sum_{y \in \cY} \max \!\left\{ p(y) -\lambda \, q(y),0 \right\} \leq\delta\right\}. 
\end{multline}

A randomized mechanism ${A}: \cX^{n \times k} \to \cY$ is $(\varepsilon,\delta)$-classical PP in the framework $(\mathcal{S}_c,\mathcal{Q}_c,\Theta_c)$ if for all $P_X \in \Theta_c$, $(\cR,\cT) \in \mathcal{Q}_c$ with $P_X(\cR),P_X(\cT)> 0$,
\begin{equation}
\overline{\sD}^{\delta}_c\!\left( P_{A(X)|\cR} \Vert P_{A(X)|\cT} \right) \leq \varepsilon,
\end{equation}
where $P_{A(X)|\cR},  P_{A(X)|\cT}$ are the output distributions conditioned on the secret events $\cR$ and $\cT$, respectively. {See also \cref{rem:approximate_max} for further connections to information-theoretic quantities characterizing classical privacy frameworks.}

We further note that \cref{lem: SDP formulation ISD} below provides a semi-definite programming characterization of the DL divergence, which reduces to a linear program in the classical case.
\end{remark}

\begin{remark}[Operational interpretation of DL~divergence]
\label{rem:op-int-DL-div}
For fixed $P_X \in \Theta$ and $ (\cR,\cT) \in \cQ$, the DL~divergence $\overline{\sD}^{\delta}\!\left( \cA(\rho^\cR) \Vert\cA(\rho^\cT) \right)$ is equal to the minimal $\varepsilon$ that can be achieved for fixed $\delta$ via the indistinguishability condition of the QPP framework $(\cS,\cQ,\Theta,\bar{\cM})$ stated in~\eqref{eq:qpp_def}.
\end{remark}

\begin{remark}[Equivalent formulation with hockey-stick divergence]
\label{rem:Equivalent formulation with hockey-stick divergence}
Another equivalent formulation of QPP arises as a generalization of the information-theoretic equivalence for QDP~\cite{hirche2023quantum}. Specifically, 
 $\cA$ is $(\varepsilon,\delta)$-QPP with respect to the framework $(\cS,\cQ,\Theta,\bar{\cM})$, where $\bar{\cM}= \{ \rM: 0 \psd \rM \psd \rI\}$, if 
\begin{equation}  
\sE_{e^{\varepsilon}}\!\left( \cA(\rho^\cR) \middle \Vert \cA(\rho^\cT) \right) \leq \delta,\end{equation}
for all  $ P_X \in \Theta$ and $(\cR,\cT) \in \cQ$, where $\sE_{\nu}(\rho\| \sigma) \coloneqq  \Tr\!\left[(\rho-\nu \sigma)_{+}\right]$  is the hockey-stick divergence for $\nu \geq 1$~\cite{SW12}. Fixing $P_X$ and $(\cR,\cT)$, the quantity $\sE_{e^{\varepsilon}}\!\left( \cA(\rho^\cR) \| \cA(\rho^\cT) \right)$ is the minimal $\delta$ that can be achieved for fixed $\varepsilon$ under the indistinguishability condition from~\eqref{eq:qpp_def}. 
\end{remark}

\subsection{Reduction to Existing Privacy Frameworks }

The proposed QPP framework  subsumes other important privacy frameworks as special cases. These reductions are presented next.

\medskip
\subsubsection{Quantum DP}
\label{rem: Quantum DP set}
In QDP (\cref{def: QDP}), secrets are singletons, discriminative pairs comprise states satisfying a neighboring relation, while the measurement class $\cM$ includes all possible measurements. QPP recovers the QDP setting by making the following choices {while recalling~\eqref{eq:finite_all_distributions} }\footnote{{For each pair of states $(\rho,\sigma)$, there exists at least one probability distribution that assigns positive probability for these two states, which recovers the definitions of QDP.
}}: 
\begin{equation}
    \begin{aligned}
\cS&=\cD , \\
    \cQ&=\{(\rho,\sigma) : \rho, \sigma \in \cD, \rho \sim \sigma \} ,\\
    \Theta&={ \cP_2\big(\cD(\cH)\big)}, \\ 
    \cM&=\{\rM: 0 \psd \rM  \psd \rI \}.
\end{aligned}
\end{equation}
More generally, one may add flexibility to the QDP formulation by considering other subsets  $\Theta$ {(i.e., $\Theta \subset  \cP_2\big(\cD(\cH)\big)$)}. This can be used, for instance, to treat situations in which only certain neighboring pairs are of interest, namely, by choosing the distributions that assign positive probabilities only to those selected density operators. This can be interpreted as adding domain knowledge to the original QDP framework.

\medskip
\subsubsection{Quantum Local DP}

\label{rem:QLDP}

In quantum local DP (QLDP)~\cite{hirche2023quantum} \footnote{{QLDP is also known as Local differential privacy (under the `extreme setting', as compared to standard QDP) in Section~V.A of~\cite{hirche2023quantum}.}} , we choose secret pairs to be pairs of arbitrary distinct states, while the measurement class includes all possible measurements. Thus, QLDP realizes the same $(\cS,\cQ,\Theta,\cM)$ framework as QDP, except that $\cQ= \{(\rho,\sigma): \rho,\sigma \in \cD  \}$ for QLDP.

\medskip
\subsubsection{Classical PP}\label{rem:classical-PP}
Consider a classical PP framework $(\cS_c,\cQ_c,\Theta_c)$, as specified in \cref{def:pp_def}.
{Assume that $p_X \in \Theta_c$ are discrete probability distributions over the probability space $\cP(\cX^{n \times k})$.}
Let the encoding of the database $x \in \cX^{n \times k}$ be $\rho^x\coloneqq |x \rangle\!\langle x|$, and denote a projective measurement operator corresponding to outcome $y$ as $|y \rangle\!\langle y|$. Here note that $\{ |x \rangle\}_{x \in \cX}$ and  $\{ |y \rangle\}_{y \in \cY}$ are respective orthonormal bases formed related to the input and output alphabets of the classical PP mechanism $\cA_c$.  Then classical PP is obtained from QPP by setting 
\begin{equation}
\begin{aligned}
    \cS&= \big \{ \{ \rho^x: x \in \cR_c\}: \cR_c \in \cS_c \big \} , \\
    \cQ&= \big\{\big(\{ \rho^x: x \in \cR_c\} , \{\rho^x: x \in \cT_c \}\big) : (\cR_c,\cT_c) \in \cQ_c \} \big\} , \\
    \Theta&=\Theta_c  , \\
    \cM&=\left\{ \sum_{y \in \cB} |y \rangle\!\langle y|: \cB \subseteq \cY \right \}.
\end{aligned}
\end{equation}
In this scenario, assuming the output of the algorithm is discrete, we have that
\begin{equation}
\cA(\rho^x)=\sum_{y\in\mathcal{Y},x'\in\mathcal{X}} p(y|x) |y \rangle\!\langle x'|\rho^x |x' \rangle\!\langle y|    
\end{equation}
where $p(y|x)= \PP\big(\cA_c(x)=y \big)$.

\begin{remark}[Utility-optimized privacy models]
    As is evident from above, the measurement set corresponding to classical PP entails  every subset $\cB \subseteq \cY $. However, when some of the outcomes are not sensitive, we may want to relax this requirement to gain utility (cf., e.g.,~\cite{murakami2019utilityOptimizedLDP}). While classical PP does not allow for that, QPP gives extra flexibility in choosing~$\cM$ and adapting it to the application of interest.   
    Indeed, if we only need to privatize outcomes within the set $\cY'\subsetneq\cY$, the smaller measurement set
    $\cM=\left\{ \sum_{y \in \cY'} |y \rangle\!\langle y|: \cB \subseteq \cY' \right \}$ is sufficient. 
\end{remark}

\section{Datta--Leditzky Information Spectrum Divergence} \label{Sec:DL-divergence}

We now focus on the DL~divergence~\cite{datta2014second}, whose operational interpretation in terms of QPP was provided in the previous section (see \cref{rem:op-int-DL-div}), and we study structural properties thereof, which will be useful when analyzing the QPP framework. 
We first formulate a primal and dual SDP to compute the DL~divergence and then use that to prove joint-quasi convexity, the data-processing inequality under positive, trace-preserving maps, and connections to the smooth max-relative entropy.

\subsection{SDP Formulations}

We now present several SDPs for computing the DL~divergence in~\eqref{eq:2nd-DL-def}, which may be of independent interest. (Recall that the other DL divergence in~\eqref{eq:1st-DL-def} is easily obtained by applying the equality in~\eqref{eq:inf-spec-equality}.)

\begin{lemma}[SDP formulation of the DL~divergence]
\label{lem: SDP formulation ISD}
For $\delta \in(0,1)$, a state $\rho$, and a PSD operator~$\sigma$, the following equalities hold%
\begin{subequations}
\begin{align}
\overline{\sD}^{\delta}(\rho\Vert\sigma)  &  =\ln \inf
_{\lambda,Z\geq0}\left\{  \lambda:\operatorname{Tr}[Z]\leq \delta
,\ Z\geq\rho-\lambda\sigma\right\} \label{eq:SDP-info-spec-rel-ent-inf}\\
&  =\ln \sup_{\mu,W\geq0}\left\{
\begin{array}[c]{c}
\operatorname{Tr}[W\rho]-\mu  \delta :\\
\operatorname{Tr}[W\sigma]\leq1,\ 
W\leq\mu I
\end{array}
\right\} \label{eq:SDP-info-spec-rel-ent-sup}.
\end{align}
\end{subequations}
\end{lemma}

\begin{IEEEproof}
Considering~\eqref{eq: overline DL}, fix $\lambda>0$ and first observe that
\begin{equation}
\operatorname{Tr}[\left(  \rho-\lambda\sigma\right)  _{+}]=\sup_{\Lambda
:\,0\leq\Lambda\leq I}\operatorname{Tr}[\Lambda\left(  \rho-\lambda
\sigma\right)  ].
\label{eq:SDP-positive-part}
\end{equation}
Indeed, this follows because, for every $0\leq\Lambda\leq I$, we have that
\begin{align}
\operatorname{Tr}[\Lambda\left(  \rho-\lambda\sigma\right)  ]  &
=\operatorname{Tr}\!\left[\Lambda\left(  \left(  \rho-\lambda\sigma\right)
_{+}-\left(  \rho-\lambda\sigma\right)  _{-}\right)\right  ]\notag \\
&  \leq\operatorname{Tr}\!\left[\Lambda\left(  \rho-\lambda\sigma\right)  _{+}\right]\notag \\
&  \leq\operatorname{Tr}\!\left[\left(  \rho-\lambda\sigma\right)  _{+}\right],
\end{align}
and the inequalities above are all attained by setting $\Lambda$ to be the projection onto the support
of $\left(  \rho-\lambda\sigma\right) _{+}$. The SDP dual of this quantity is given by
\begin{equation}
\operatorname{Tr}\!\left[\left(  \rho-\lambda\sigma\right)  _{+}\right]=\inf_{Z\geq
0}\left\{  \operatorname{Tr}[Z]:Z\geq\rho-\lambda\sigma\right\}  .
\label{eq:dual-positive-part}
\end{equation}
Intuitively, $Z=\left(  \rho-\lambda\sigma\right)  _{+}$ is the smallest choice of a PSD operator that satisfies the constraint $Z\geq\rho-\lambda\sigma$.

We then find from~\eqref{eq:2nd-DL-def},~\eqref{eq: overline DL}, and~\eqref{eq:dual-positive-part} that
\begin{align*}
\overline{\sD}%
_{s}^{\delta}(\rho\Vert\sigma)
&  =\ln \inf\left\{  \lambda\geq0:\operatorname{Tr}[\left(  \rho
-\lambda\sigma\right)  _{+}]\leq \delta\right\} \\
&  =\ln \inf_{\lambda,Z\geq0}\left\{  \lambda:\operatorname{Tr}%
[Z]\leq \delta,\ Z\geq\rho-\lambda\sigma\right\}  ,\numberthis
\label{eq:DL-SDP-dual-form}
\end{align*}
which completes the proof of~\eqref{eq:SDP-info-spec-rel-ent-inf}.

\medskip

The dual forms of these optimization problems are derived from the canonical primal and dual formulations of SDPs, which are respectively given by (cf.~\cite[Definition~2.20]{khatri2020principles})
\begin{equation}
\begin{split}
&\inf_{Y\geq0}\left\{  \operatorname{Tr}[BY]:\Phi^{\dag}(Y)\geq A\right\} , \\
&\sup_{X\geq0}\left\{  \operatorname{Tr}[AX]:\Phi(X)\leq B\right\} ,
\end{split}
\end{equation}
where $A$ and $B$ are Hermitian matrices and $\Phi$ is a Hermiticity-preserving superoperator. Comparing the former to~\eqref{eq:DL-SDP-dual-form}, 
we make the following choices so that the general optimization problem recovers~\eqref{eq:DL-SDP-dual-form} (inside the logarithm):
\begin{align}
Y  &  =%
\begin{bmatrix}
\lambda & 0\\
0 & Z
\end{bmatrix}
,\qquad B=%
\begin{bmatrix}
1 & 0\\
0 & 0
\end{bmatrix}
,\\
\Phi^{\dag}(Y)  &  =%
\begin{bmatrix}
-\operatorname{Tr}[Z] & 0\\
0 & Z+\lambda\sigma
\end{bmatrix}
,\\
A& =%
\begin{bmatrix}
-\delta  & 0\\
0 & \rho
\end{bmatrix}
.
\end{align}
Then, setting%
\begin{equation}
X=%
\begin{bmatrix}
\mu & 0\\
0 & W
\end{bmatrix}
,
\end{equation}
we solve for the map $\Phi(X)$ to find that
\begin{align}
& \operatorname{Tr}[X\Phi^{\dag}(Y)]  \notag  \\& =\operatorname{Tr}\!\left[
\begin{bmatrix}
\mu & 0\\
0 & W
\end{bmatrix}%
\begin{bmatrix}
-\operatorname{Tr}[Z] & 0\\
0 & Z+\lambda\sigma
\end{bmatrix}
\right] \\
&  =-\mu\operatorname{Tr}[Z]+\operatorname{Tr}\!\left[  W\left(  Z+\lambda
\sigma\right)  \right] \\
&  =\operatorname{Tr}[\left(  W-\mu I\right)  Z]+\lambda\operatorname{Tr}%
[W\sigma]\\
&  =\operatorname{Tr}\!\left[
\begin{bmatrix}
\lambda & 0\\
0 & Z
\end{bmatrix}%
\begin{bmatrix}
\operatorname{Tr}[W\sigma] & 0\\
0 & W-\mu I
\end{bmatrix}
\right] \\
&  =\operatorname{Tr}[Y\Phi(X)],
\end{align}
so that%
\begin{equation}
\Phi(X)=%
\begin{bmatrix}
\operatorname{Tr}[W\sigma] & 0\\
0 & W-\mu I
\end{bmatrix}
.
\end{equation}
Plugging into the dual form, we obtain%
\begin{multline}
\sup_{X\geq0}\left\{  \operatorname{Tr}[AX]:\Phi(X)\leq B\right\}  \\ 
=
\sup_{\mu,W\geq0}\left\{  \operatorname{Tr}[W\rho]-\mu\delta  :\operatorname{Tr}[W\sigma]\leq1,W\leq\mu I\right\}. \end{multline}

Choose $\mu=\mu_1  \in (0,1)$ and $ W=\mu_2 I $ such that $\mu_1 \delta < \mu_2 < \mu_1 $, as a strictly feasible solution to the above. For the other SDP formulation from~\eqref{eq:DL-SDP-dual-form}, set $\lambda$ such that $\Tr\!\left[(\rho-\lambda \sigma)_+ \right] \leq \delta$, and $Z~=~(\rho-\lambda \sigma)_{+} \geq 0 $ as a feasible solution. By Slater's condition, we conclude that strong duality holds, and the primal and dual optimal values coincide. 
\end{IEEEproof}

\begin{corollary}[Another formulation of the DL divergence] \label{Cor:another_form_DL}
    DL divergence has the following equivalent formulation: 
    \begin{equation}
        \overline{\sD}^{\delta}(\rho\Vert\sigma) = \ln \sup_{0 \leq W \leq I,  \Tr[W\rho] \geq \delta} \frac{\Tr[W \rho] - \delta }{\Tr[W\sigma]}.
    \end{equation}
\end{corollary}
\begin{IEEEproof}
    Consider the SDP formulation in~\eqref{eq:SDP-info-spec-rel-ent-sup} and set $W' =\frac{W}{\mu}$ therein to arrive at 
    \begin{align}
        \overline{\sD}^{\delta}(\rho\Vert\sigma) &  =\ln \sup_{\mu,W'\geq0}\left\{
\begin{array}[c]{c}
\mu \operatorname{Tr}[W'\rho]-\mu  \delta :\\
\mu \operatorname{Tr}[W'\sigma]\leq1,\ 
W'\leq I
\end{array}
\right\} \\ 
&=\ln \sup_{0 \leq W' \leq I, \Tr[W'\rho] \geq \delta } \frac{\Tr[W' \rho] - \delta }{\Tr[W'\sigma]},
    \end{align}
where the last equality follows from identifying that $\mu=1/\operatorname{Tr}[W'\sigma]$ is the $\mu$ that maximizes the former, given that $\Tr[W'\rho] \geq \delta $. When $ \Tr[W'\rho]< \delta $, the optimum $\mu=0$ and the objective within the supremum becomes zero. 
Replacing $W'$ by $W$, concludes the proof. 
\end{IEEEproof}

\begin{remark}[Approximate-max divergence]\label{rem:approximate_max}
    In [2], the $\delta$-approximate-max divergence is defined as 
    \begin{equation}
        \sD_\infty^\delta(p_Y \Vert p_Z) \coloneqq  \ln \max_{S \in \mathrm{Supp}(Y), \Pr[Y \in S] \geq \delta} \frac{\Pr[Y \in S] -\delta}{\Pr[Z \in S]},
    \end{equation}
    where $Y$ and $Z$ are random variables distributed according to $Y \sim p_Y$ and $Z \sim p_Z$.
    By substituting classical distributions into \cref{Cor:another_form_DL}, we observe that the DL divergence reduces to the approximate-max divergence. Note that the approximate-max divergence has been used to characterize $(\varepsilon, \delta)$-(classical) DP in {\cite[ Remark~3.1]{DR14}}. Thus, this showcases that the equivalence we established for QPP with the DL divergence herein reduces to the existing equivalence for (classical) DP. 
\end{remark}

\subsection{Properties}

\medskip

We derive several properties of the DL~divergence from~\eqref{eq: overline DL}, which are subsequently used in the analysis of the QPP framework. Basic properties of the DL~divergence, including the data-processing inequality, have been proven in \cite[Proposition~4.3]{datta2014second}. Here, we generalize the data-processing inequality to hold for arbitrary positive, trace non-increasing maps (beyond the set of quantum channels) and also establish joint-quasi convexity of the DL~divergence, along with its connection to the smooth max-relative entropy (recall the definition in~\eqref{eq:smooth-max-rel}). The proofs of these properties rely on the SDP formulation from \cref{lem: SDP formulation ISD}.

\begin{proposition}[Properties of the DL~divergence] \label{prop: Properties of DL divergence} 
Fix $\delta \in (0,1)$, and let $\rho,\rho_1,\ldots,\rho_k$ and $\sigma,\sigma_1,\ldots,\sigma_k$ be two collections of states and PSD operators, respectively. The DL~divergence in~\eqref{eq: overline DL} satisfies the following properties:
   \begin{enumerate}
    \item Data-processing inequality:  For every positive, trace non-increasing map $\cN$, we have 
    \begin{equation}
    \overline{\sD}^{\delta}(\rho\Vert\sigma) \geq \overline{\sD}^{\delta}\big(\cN(\rho)\Vert\cN(\sigma) \big)  .
    \end{equation}
    \item Joint-quasi convexity: Let $ p_i \in [0,1]$, for $i\in \{1,\ldots,k\}$, with $\sum_{i=1}^k p_i =1$. Then 
\begin{equation}
\overline{\sD}^{\delta}\!\left(  \sum_{i=1}^k p_i \rho_i \middle \Vert   \sum_{i=1}^k p_i \sigma_i \right) \leq \max_i \overline{\sD}^{\delta}\!\left(   \rho_i \Vert   \sigma_i \right),
\end{equation}
    and, more generally,
    \begin{equation}
    \overline{\sD}^{\delta'}\!\left(  \sum_{i=1}^k p_i \rho_i \middle \Vert   \sum_{i=1}^k p_i \sigma_i \right) \leq \max_i \overline{\sD}^{\delta_i}\!\left(   \rho_i \Vert   \sigma_i \right),
    \end{equation}
     where $\delta'\coloneqq \sum_{i=1}^k p_i \delta_i$ with $\delta_1,\ldots,\delta_k\in(0,1)$.

\item Relation to smooth max-relative entropy: 
\begin{equation}
\overline{\sD}^{\delta}(\rho\Vert\sigma)\leq \sD_{\max}^{\delta
}(\rho\Vert\sigma)\leq\overline{\sD}^{\delta'}\!(\rho
\Vert\sigma)-\ln\!\left(1-\delta'\right),
\label{eq:dmax-inf-spec-ineqs}%
\end{equation}
where $\delta'\coloneqq 1-\sqrt{1-\delta^{2}} \in (0,1)$, and the second inequality above can be  equivalently written as
\begin{equation}
    \sD_{\max}^{\sqrt{\delta(2-\delta)}
}(\rho\Vert\sigma)\leq\overline{\sD}^{\delta}(\rho
\Vert\sigma)-\ln\!\left(1-\delta\right).
\end{equation}

\item Quasi subadditivity: Let $\delta_1, \delta_2 \in (0,1)$ satisfy $\delta_1' + \delta_2' < 1$, with $\delta'_i\coloneqq \sqrt{\delta_i(2-\delta_i)} \in (0,1)$ for $i\in\{1,2\}$. Then
\begin{multline}
 \overline{\sD}^{\delta_1' + \delta_2'}(\rho_1 \otimes \rho_2 \Vert\sigma_1 \otimes \sigma_2) \\  \leq    \overline{\sD}^{\delta_1}(\rho_1
\Vert\sigma_1) + \overline{\sD}^{\delta_2}(\rho_2
\Vert\sigma_2) - \ln\big((1-\delta_1)(1-\delta_2)\big).
\end{multline}
Furthermore, 
\begin{enumerate}
    \item if $\delta_1=\delta_2=0$, then 
    \begin{equation}
        \overline{\sD}^{0}(\rho_1 \otimes \rho_2 \Vert\sigma_1 \otimes \sigma_2)  \leq    \overline{\sD}^{0}(\rho_1
\Vert\sigma_1) + \overline{\sD}^{0}(\rho_2
\Vert\sigma_2).
    \end{equation}
    \item if $\sigma_1, \sigma_2$ are states, then
\begin{equation}
    \overline{\sD}^{\delta}(\rho_1 \otimes \rho_2 \Vert\sigma_1 \otimes \sigma_2)  \leq    \overline{\sD}^{\delta_1}(\rho_1
\Vert\sigma_1) + \overline{\sD}^{\delta_2}(\rho_2
\Vert\sigma_2),
\end{equation}
where 
\begin{equation}
  \delta \coloneqq \min\! \left\{ \delta_1 + e^{\overline{\sD}^{\delta_1}(\rho_1
\Vert\sigma_1)} \delta_2, \  \delta_2+ e^{\overline{\sD}^{\delta_2}(\rho_2
\Vert\sigma_2)} \delta_1 \right \}.  
\end{equation}
\end{enumerate}

\end{enumerate} 
\end{proposition}

\begin{IEEEproof}
    \underline{Property~1:} 
    The statement was proven in \cite[Proposition~4.3]{datta2014second} for a quantum channel $\mathcal{N}$ by using the inequality 
    \begin{equation}
       \Tr\!\left[ (\rho -e^\gamma \sigma)_+\right] \geq \Tr\!\left[ \left( \cN(\rho) -e^\gamma \cN(\sigma) \right)_+\right] ,   
    \end{equation}
    which holds for all $\gamma \in \mathbb{R}$.
    Here, we prove the data-processing inequality, but we generalize it to hold for a positive, trace non-increasing map $\mathcal{N}$. Our derivation relies on the SDP formulation of the DL~divergence from~\eqref{eq:SDP-info-spec-rel-ent-inf}. 
    
    Let $\lambda^\star$ and $Z^\star$ be optimal choices\footnote{When the DL~divergence is finite, the infimum is achieved by a standard continuity plus compactness argument. The stated relations trivially hold when the DL~divergence is infinite.} 
    in the optimization for $\overline{\sD}^{\delta}\!\left(   \rho \Vert   \sigma\right)$, so that $\overline{\sD}^{\delta}\!\left(   \rho \Vert   \sigma\right)= \ln \lambda^\star$, $Z^\star\geq \rho- \lambda^\star \sigma$ with $\Tr\!\left[Z^\star\right] \leq \delta$, and $Z^\star \geq 0$ (indeed, note that the infimum is achieved with $\Tr[(\rho- \lambda^\star \sigma)_+] =\delta$).
    Since $Z^\star -(\rho- \lambda^\star \sigma) \geq 0$, it follows that 
    $ \mathcal{N} \!\big( Z^\star -(\rho- \lambda^\star \sigma) \big)  \geq 0$ from the assumption that $\mathcal{N}$ is a positive map. Consequently, we obtain
    \begin{equation}
        Z' \coloneqq \mathcal{N}( Z^\star )  \geq \cN(\rho) - \lambda^\star \cN(\sigma). 
    \end{equation}
    Furthermore,
     $Z' \geq 0$ since $Z^\star \geq 0$ and $\mathcal{N}$ is a positive map. Additionally, since $\mathcal{N}$ is trace non-increasing, it follows that 
     \begin{equation}
       \operatorname{Tr}[Z']\leq \operatorname{Tr}[Z^\star]\leq \delta.     
     \end{equation}
     Thus, $\lambda^\star$ is a feasible point for $\overline{\sD}^{\delta}\!\big(\cN(\rho)\Vert\cN(\sigma) \big)$. We conclude the proof by noting that the quantity $\overline{\sD}^{\delta}\big(\cN(\rho)\Vert\cN(\sigma) \big)$ involves a minimization over all such feasible points, implying the desired inequality:
     \begin{equation}
      \overline{\sD}^{\delta}\big(\cN(\rho)\Vert\cN(\sigma) \big) \leq \ln(\lambda^\star)= \overline{\sD}^{\delta}\!\left(   \rho \Vert   \sigma\right).   
     \end{equation}
   {Note that this property can also be derived using the proof of   \cite[Lemma~4]{SW12}.}

    \underline{Property~2:} We again consider the SDP from~\eqref{eq:SDP-info-spec-rel-ent-inf}. Let $\lambda^\star_i$ and $Z^\star_i$ be optimal for $\overline{\sD}^{\delta}\!\left(   \rho_i \Vert   \sigma_i \right)$, so that $\overline{\sD}^{\delta}\!\left(   \rho_i \Vert   \sigma_i \right)= \ln(\lambda^\star_i)$, $Z^\star_i \geq \rho_i- \lambda^\star_i \sigma_i$ with $\operatorname{Tr}\!\left[Z^\star_i\right] \leq \delta$, and $Z^\star_i \geq 0$. 
    Define
    \begin{equation}
         Z \coloneqq \sum_{i=1}^k p_i Z^\star_i \geq \sum_{i=1}^k p_i \rho_i - \sum_{i=1}^k \lambda^\star_i p_i \sigma_i.
    \end{equation}
    and observe that $\Tr\!\left[ Z\right] \leq \delta$ and $Z \geq 0$.
    This implies that $Z \geq \sum_{i=1}^k p_i \rho_i -  \max_i  \lambda^\star_i \sum_{j=1}^k p_j \sigma_j $, which suggests that $\max_i  \lambda^\star_i$ and $Z$ are (candidate) infimizers in the SDP formulation of $\overline{\sD}^{\delta}\!\left(  \sum_{i=1}^k p_i \rho_i \middle \Vert   \sum_{i=1}^k p_i \sigma_i \right)$. 
    Consequently, we obtain    
    \begin{equation}
        \overline{\sD}^{\delta}\!\left(  \sum_{i=1}^k p_i \rho_i \middle \Vert   \sum_{i=1}^k p_i \sigma_i \right)  \leq \ln \!\left(\max_i  \lambda^\star_i \right) = \max_i \overline{\sD}^{\delta}\!\left(   \rho_i \Vert   \sigma_i \right).
    \end{equation}
    The proof of the general case follows along the same lines by observing that $\Tr[Z] \leq \sum_{i=1}^k p_i \delta_i$.

{We also present an alternative proof for this using the joint-convexity of hockey-stick divergence in Appendix~\ref{APP:Alternative_proof}.}
    \medskip 

\underline{Property~3}:
From  \cite[Appendix~B]{wang2019resource}, we have that
\begin{align}
    {\sD_{\max}^{\delta}(\rho\Vert\sigma)}  &  =\ln \inf_{\lambda,\widetilde
{\rho},Y\geq0}\left\{
\begin{array}
[c]{c}%
\lambda:\widetilde{\rho}\leq\lambda\sigma,\ \ \operatorname{Tr}[Y]\leq
\delta,\\
\operatorname{Tr}[\widetilde{\rho}]=1,\ \ Y\geq\rho-\widetilde{\rho}%
\end{array}
\right\}  .
\end{align}
Let $\lambda$, $Y$, and $\widetilde{\rho}$ be arbitrary operators satisfying
the constraints for ${\sD_{\max}^{\delta}(\rho\Vert\sigma)}$. Then by
combining the inequalities $\widetilde{\rho}\leq\lambda\sigma$ and $Y\geq
\rho-\widetilde{\rho}$, we get
\begin{equation}
  Y\geq\rho-\lambda\sigma.  
\end{equation}
We see that $\lambda$ and $Y$ satisfy the constraints
needed for $\lambda$ and $Z$, respectively, in the SDP for $\overline
{\sD}^{\delta}(\rho\Vert\sigma)$, whereby
\begin{equation}
   {\overline{\sD}^{\delta}(\rho\Vert\sigma)}\leq\lambda. 
\end{equation}
Since the argument holds for all $\lambda$, $Y$, and $\widetilde{\rho}$
satisfying the constraints in the definition of ${\sD_{\max}^{\delta}(\rho\Vert\sigma)}$,
we further obtain
\begin{equation}
  {\overline{\sD}^{\delta}(\rho\Vert\sigma)}\leq {\sD_{\max
}^{\delta}(\rho\Vert\sigma)}.  
\end{equation}
The proof is concluded by invoking the following lemma (proven in Appendix~\ref{Sec: proof lemma supporting connecting DL to smooth max}).

\begin{lemma}
\label{lem:info-spec-to-sm-dmax-helper} Fix $\lambda>0$, let $\rho$ be a state and $\sigma$ a
positive semi-definite operator, and define $\delta \coloneqq \operatorname{Tr}[\left(  \rho-\lambda\sigma\right)  _{+}]$.
Then%
\begin{equation}
    \sD_{\max}^{\sqrt{\delta\left(  2-\delta\right)  }}(\rho\Vert
\sigma)\leq\ln \lambda-\ln(1-\delta)  .
\end{equation}
\end{lemma}

For fixed $\delta\in(0,1)$, by definition, we have
$\overline{\sD}^\delta(\rho\Vert\sigma)=\ln(\lambda^\star)$ with $\delta=\operatorname{Tr}[\left(  \rho-\lambda^\star\sigma\right)_{+}]$. 
With that, \cref{lem:info-spec-to-sm-dmax-helper} with the reparametrization
 $ \delta \to 1-\sqrt{1-\delta^{2}}$, 
 yields
 \begin{equation}
      \sD_{\max}^{\delta
}(\rho\Vert\sigma)\leq\overline{\sD}^{1-\sqrt{1-\delta^{2}}}(\rho
\Vert\sigma)+\ln\!\left(  \frac{1}{\sqrt{1-\delta^{2}}}\right) .
 \end{equation}
This completes the proof. 
\medskip 

\underline{Property~4}: This follows by invoking Property~3 and using the fact that the smooth max-relative entropy satisfies subadditivity (Appendix~\ref{app:proof-sub-additivity-lemma}) with 
\begin{equation}\label{eq:Dmax-delta}
  \sD_{\max}^{\delta_1 + \delta_2
}(\rho_1 \otimes \rho_2\Vert\sigma_1 \otimes \sigma_2) \leq \sD_{\max}^{\delta_1
}(\rho_1\Vert\sigma_1)+ \sD_{\max}^{\delta_2
}(\rho_2\Vert\sigma_2).  
\end{equation}

Part (a) now follows by taking the limits $\delta_1 \to 0$ and $\delta_2 \to 0$ in~\eqref{eq:Dmax-delta}, and applying Property~3.

To prove Part (b), we use the SDP formulation in~\eqref{eq:SDP-info-spec-rel-ent-inf}. Let $\overline{\sD}^{\delta_i}\!\left(   \rho_i \Vert   \sigma_i \right) =\ln(\lambda^\star_i)$ for $i \in\{1,2\}$. It follows that $Z_i \geq \rho_i- \lambda^\star_i \sigma_i$ with $\Tr\!\left[Z_i\right] \leq \delta$ and $Z_i \geq 0$. Consider that
    \begin{align}
       & (\rho_1 \otimes \rho_2) - \lambda^\star_1 \lambda^\star_2 (\sigma_1 \otimes \sigma_2) \notag \\
        &= (\rho_1 \otimes \rho_2) - \lambda^\star_1 \sigma_1 \otimes \rho_2  + \lambda^\star_1 \sigma_1 \otimes \rho_2 - \lambda^\star_1 \lambda^\star_2 (\sigma_1 \otimes \sigma_2) \notag  \\ 
        &= (\rho_1- \lambda^\star_1 \sigma_1 ) \otimes \rho_2 +  \lambda^\star_1 \sigma_1  \otimes (\rho_2- \lambda^\star_2 \sigma_2 ) \notag \\
        & \leq Z_1 \otimes \rho_2 + \lambda^\star_1 \sigma_1 \otimes Z_2 \eqqcolon Z .
    \end{align}
Observe that $Z \geq 0$ and $\Tr\!\left[Z\right]= \Tr\!\left[Z_1\right] +\lambda^\star_1 \Tr\!\left[Z_2\right]$, since $\Tr\!\left[\rho_1\right]=\Tr\!\left[\sigma_1\right]=1$. Consequently, we have $\Tr\!\left[Z\right] \leq \delta_1 + \lambda^\star_1  \delta_2 $, and $\lambda^\star_1 \lambda^\star_2$ is a candidate infimizer. 
For $\delta'=\delta_1 + \lambda^\star_1  \delta_2$, we now arrive at 
\begin{align}
\overline{\sD}^{\delta'}\!\left(   \rho_1 \otimes \rho_2 \Vert   \sigma_1 \otimes \sigma_2 \right) & \leq  \ln(\lambda^\star_1 \lambda^\star_2)\\
& = \ln(\lambda^\star_1) + \ln(\lambda^\star_2) \\
& = \overline{\sD}^{\delta_1}\!\left(   \rho_1 \Vert   \sigma_1  \right) + \overline{\sD}^{\delta_2}\!\left(  \rho_2 \Vert  \sigma_2 \right).
\end{align}

The above holds for $\delta'=\delta_2 + \lambda^\star_2 \delta_1$ as well, by adding and subtracting $\lambda_2^\star \rho_1 \otimes \sigma_1$ instead of $\lambda_1^\star \sigma_1 \otimes \rho_2$, and then following the same argument.
\end{IEEEproof}

\section{Properties and Mechanisms for QPP} \label{Sec:Properties-and-Mechanisms-for-QPP}

\subsection{Properties of QPP Mechanisms}

Modern guidelines for privacy frameworks~\cite{KL12} render properties such as convexity and post-processing (also known as transformation invariance) as basic requirements for privacy frameworks. Composability is another important property, which implies that a combination of privacy mechanisms is itself private. These properties are known to hold for the classical mutual information PP framework, and all of them, except for composability, hold for the classical PP framework; cf.~\cite[Theorem~2]{nuradha2022information} and \cite[Theorem~5.1]{KM14}, respectively.

\begin{figure}
\centering
\begin{subfigure}{0.4\textwidth}
    \centering
    \includegraphics[width=\linewidth]{ 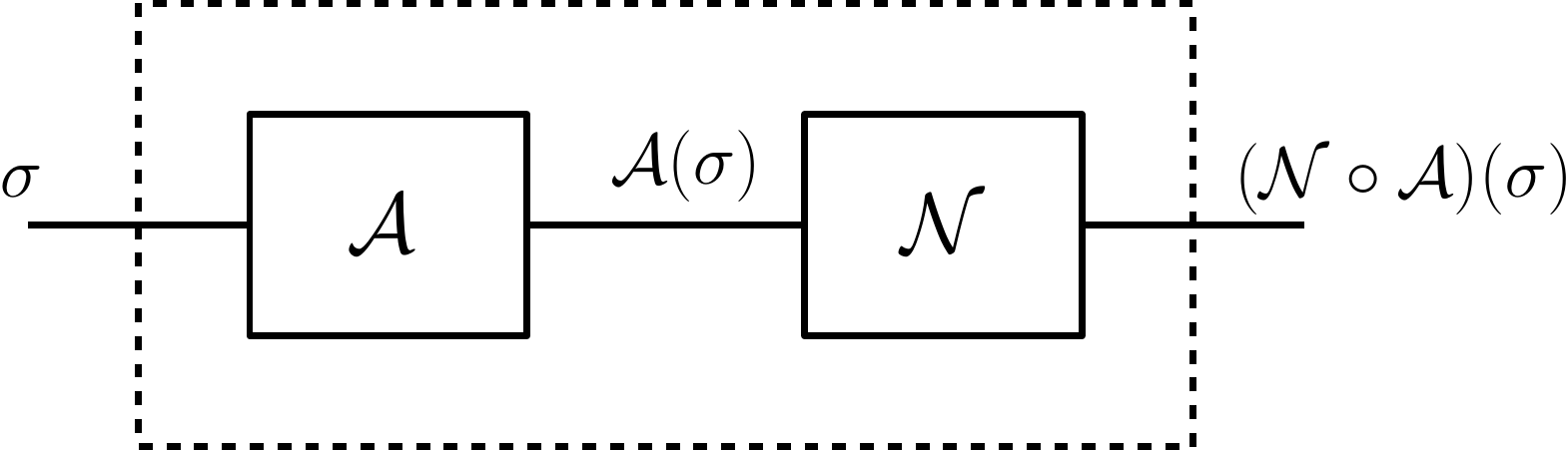}
    \caption{}
    \vspace{3mm}
    \label{fig:post processing}
\end{subfigure}
\begin{subfigure}{0.5\textwidth}
\vspace{-4mm}
    \centering
\hspace{14mm}\includegraphics[width=0.55\linewidth]{ 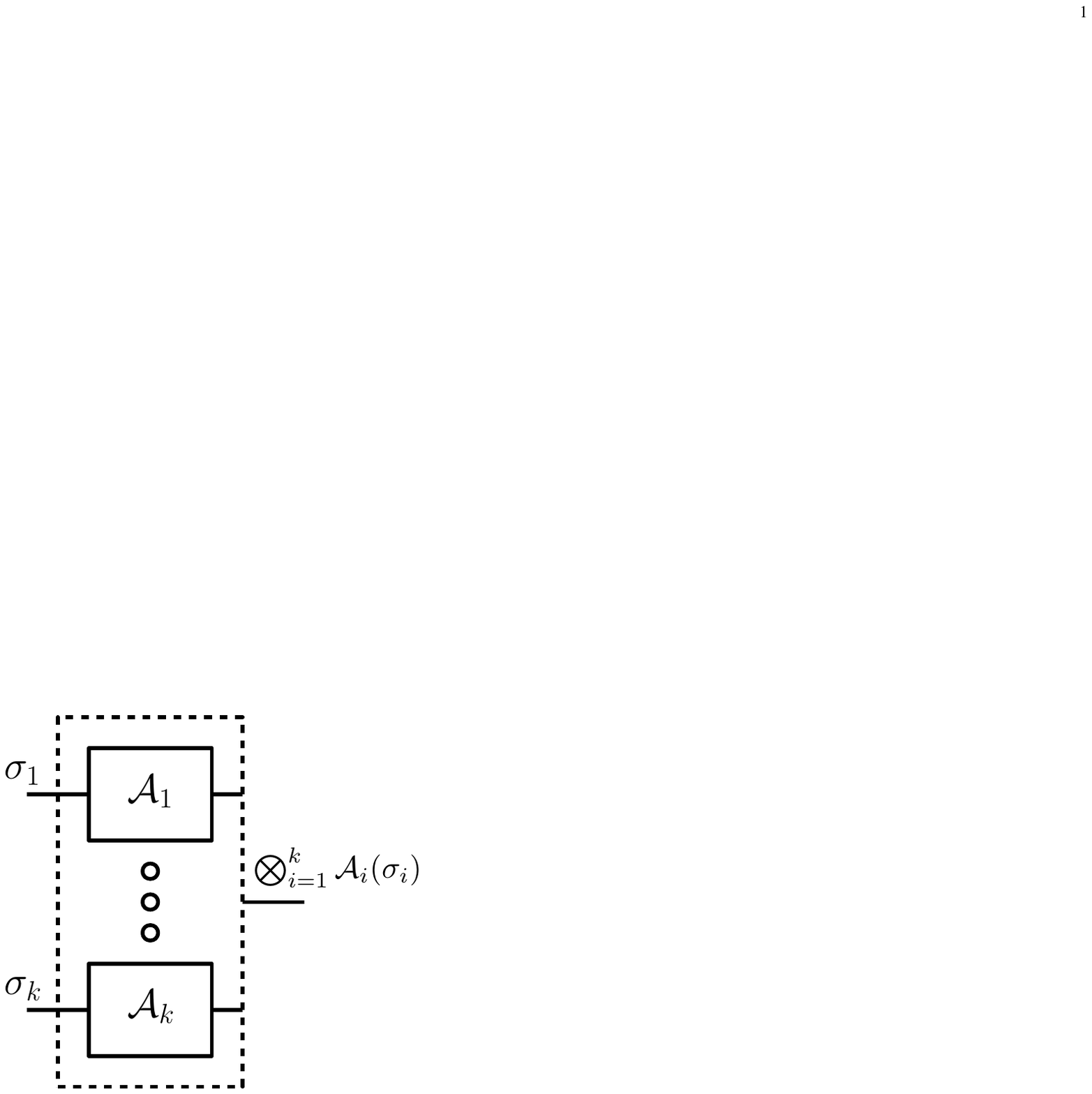}
  \caption{}
     \label{fig:parallel composition}
\end{subfigure}

\caption{Properties of QPP mechanisms: (a) refers to post-processing of QPP algorithm $\cA$; If $\cA$ satisfies QPP, then $\cN \circ \cA$ also satisfies QPP. 
 (b) refers to  parallel composition of $k$ QPP mechanisms; composition of $k$ mechanisms independently in a parallel fashion satisfies QPP if each $\cA_i$ satisfies QPP.
}
\label{fig:properties}
\end{figure}

Before proving these properties for the QPP framework, we discuss their operational interpretation. Convexity means that applying a QPP mechanism that is randomly chosen from a given set of such mechanisms still satisfies  QPP. Post-processing ensures that passing the output of a QPP mechanism $\cA$ through a channel $\cN$ preserves QPP; see \cref{fig:post processing}.  
Parallel composability is illustrated in \cref{fig:parallel composition} and guarantees that QPP holds after applying
\begin{equation}
\cA^{(k)} \coloneqq \bigotimes_{i=1}^k \cA_i=\cA_1\otimes \cA_2 \otimes \cdots \otimes \cA_k   
\end{equation}
to the input $\rho^{X_1} \otimes \rho^{X_2} \otimes \cdots \otimes \rho^{X_k} $, with $X_i\sim P_{X} \in \Theta$, where each $X_i$ is independently chosen.
Informally, the semantic meaning of this property is that after applying $\cA^{(k)}$, the same conclusions can be drawn about the input  $\rho^{X_1} \otimes \rho^{X_2} \otimes \cdots \otimes \rho^{X_k} $ regardless  of whether each $\rho^{X_i}$ belongs to $\cR_i$ or $\cT_i$, where $(\cR_i,\cT_i) \in \cQ$ for all $i \in\{1, \ldots,k\}$.
In this setting, the set of discriminative pairs is taken as
\begin{align}
    &\cQ^{(k)}\coloneqq \notag \\ 
    &\left \{ (\cR^{(k)},\cT^{(k)} ): \begin{aligned}
        &\cR^{(k)} \coloneqq  (\cR_1, \ldots,\cR_k ),  \\
        & \cT^{(k)} \coloneqq (\cT_1 , \ldots, \cT_k )\\ 
               & \forall i \in \{1,\ldots,k\} \
                (\cR_i,\cT_i) \in \cQ
    \end{aligned}  \right \}.
\end{align}
Furthermore, the class of product measurements is $\bigotimes_{i=1}^k \cM_i$ (i.e., the output of  algorithm $\cA_i$  is followed by a  measurement from $\cM_i$, for all $i\in \{1, \ldots, k\}$), while the set of all possible measurements on the $k$ systems, including joint measurements, is denoted by $\bar{\cM}^k$. We note here that one could consider other classes of limited measurements, such as local operations and classical communication (LOCC) measurements and positive-partial-transpose (PPT) measurements~\cite{MWW09}.

The formal statement of these properties is as follows.
\begin{theorem}[Properties of QPP mechanisms] \label{thm:QPP_properties}
The following properties hold: 
\begin{enumerate}[wide, labelindent=10pt]
     \item \underline{Convexity}:  Let 
     $\cA_1,\ldots,\cA_k$ be $(\varepsilon,\delta)$-QPP mechanisms in the framework $(\cS,\cQ,\Theta,\cM)$. Take~$I$ to be a $k$-ary categorical random variable with probability distribution $(p_1,\ldots,p_k)$. Then the mechanism $\cA\coloneqq \cA_I$ (i.e., $\cA=\cA_i$ with probability $p_i$, for $i\in \{1,\ldots,k\}$) also satisfies $(\varepsilon,\delta)$-QPP in the same framework $(\cS,\cQ,\Theta,\cM)$.
     \item \underline{Post-processing}:  If a mechanism $\cA$ satisfies $(\varepsilon,\delta)$-QPP in the framework $(\cS,\cQ,\Theta,\cM)$, then, for a quantum channel $\cN$, the processed mechanism $\cN \circ \cA$ also satisfies $(\varepsilon,\delta)$-QPP in the framework  $(\cS,\cQ,\Theta,\cM')$, where 
     $\cM' \subseteq \left\{ \rM': \cN^\dagger(\rM') \in \cM\right\}$. 
    
     \item \underline{Parallel composability (non-adaptive)}:
     Let $\cA_1,\ldots,\cA_k$ be mechanisms such that $\cA_i$ is $(\varepsilon_i,\delta_i)$-QPP in the framework  $(\cS,\cQ,\Theta,\cM_i)$, for each $i \in \{1,\ldots,k\}$. Then the composed mechanism 
     \begin{equation}
     \cA^{(k)}: \bigotimes_{i=1}^k \sigma_i \mapsto \cA_1(\sigma_1) \otimes \cdots \otimes \cA_k(\sigma_k)    
     \end{equation}
    satisfies $\left(\sum_{i=1}^k \varepsilon_i, \sum_{i=1}^k \delta_i\right)$-QPP in the framework $\left(\cS, \cQ^{(k)},\Theta, \bigotimes_{i=1}^k \cM_i\right)$. 
\end{enumerate}
\end{theorem}
\begin{IEEEproof}
    See Appendix~\ref{App:Proof-of-thm-of-QPP-Properties}.
\end{IEEEproof}

\medskip

More broadly, parallel composition (i.e., Property~3 of \cref{thm:QPP_properties}) holds under particular separable measurements that are defined as follows:
\begin{equation}
    \left\{  \sum_j \rM_1^{(j)} \otimes \cdots \otimes \rM_k^{(j)}: \forall i  \sum_j \rM_i^{(j)}  \in \cM_i, \ \forall i,j \ \rM_i^{(j)}\geq 0 \right\}
\end{equation}
where product measurements considered in \cref{thm:QPP_properties} are a special case.

\medskip
The latter two properties of \cref{thm:QPP_properties} change if one considers a measurement class that comprises all possible measurements $\bar{\cM}^k$,
as opposed to only product measurements. This is one of the main distinctions between the semi-classical and quantum cases, where, for the latter, joint measurements can infer more information and thus privacy degrades. The following theorem accounts for this latter scenario. 

\begin{theorem}[Properties of QPP with $\cM= \bar{\cM}$] \label{Cor: Properties of QPP all measurements}
The following properties hold for the case in which the measurement class is $\bar{\cM}$: 
    \begin{enumerate}[wide, labelindent=10pt]
     \item \underline{Convexity}:  Let 
     $\cA_1,\ldots,\cA_k$ be $(\varepsilon,\delta)$-QPP mechanisms in the framework $(\cS,\cQ,\Theta,\bar{\cM})$. 
     Take $I$ to be a $k$-ary categorical random variable with parameters $(p_1,\ldots,p_k)$. 
     Then the mechanism $\cA\coloneqq \cA_I$ (i.e., $\cA=\cA_i$ with probability $p_i$, for $i\in  \{1,\ldots,k\}$) also satisfies $(\varepsilon,\delta)$-QPP in the framework $(\cS,\cQ,\Theta,\bar{\cM})$.
     
     \item \underline{Post-processing}:  If a mechanism $\cA$ satisfies $(\varepsilon,\delta)$-QPP in the framework $(\cS,\cQ,\Theta,\bar{\cM})$, then, for a quantum channel~$\cN$, the mechanism $\cN \circ \cA$ also satisfies $(\varepsilon,\delta)$-QPP in the framework  $(\cS,\cQ,\Theta,\bar{\cM})$. 
    
     \item \underline{Parallel composability}:
    If $\cA_i$ 
     satisfies $(\varepsilon_i,\delta_i)$-QPP in $(\cS,\cQ,\Theta,\bar{\cM})$ for $i\in\{1,2\}$, then the composed mechanism 
    $ \cA_1 \otimes \cA_2 $ satisfies $(\varepsilon',\delta')$-QPP in the framework $\left(\cS, \cQ^{(2)},\Theta, \bar{\cM}^2 \right)$  where 
    \begin{align}
         \varepsilon'& \coloneqq \varepsilon_1 + \varepsilon_2 + \ln\!\left( \frac{1}{(1-\delta_1)(1-\delta_2)}\right), \\
         \delta'& \coloneqq \sqrt{\delta_1(2-\delta_1)} + \sqrt{\delta_2(2-\delta_2)}.   \label{eq:our-parallel-composition}  
    \end{align}
and $ \cA_1 \otimes \cA_2 $ also satisfies $(\varepsilon_1 + \varepsilon_2, \delta)$-QPP where  \begin{equation} \label{eq:parallelcomposition-similar-QDP}
     \delta \coloneqq \min\{ \delta_1 +e^{\varepsilon_1} \delta_2, \delta_2 + e^{\varepsilon_2} \delta_1 \}.
 \end{equation}
 Observe that, if $\delta_i=0$ for $i \in \{1,2\}$, then $ \cA_1 \otimes \cA_2 $ satisfies  $(\varepsilon_1+ \varepsilon_2)$-QPP for the parallel composed framework.
\end{enumerate}
\end{theorem}

\begin{IEEEproof}
    The proof of \cref{Cor: Properties of QPP all measurements} relies on properties of the DL~divergence established in \cref{prop: Properties of DL divergence}. Items 1), 2), and 3) follow from joint quasi-convexity (Property~2), data processing (Property~1),  and quasi subadditivity (Property~4), respectively. 
\end{IEEEproof}

\begin{remark}[Comparison to existing results]
      In \cite[Corollary~III.3]{hirche2023quantum}, the parallel composition of two mechanisms that satisfy $(\varepsilon_i,\delta_i)$-QDP for $i\in \{1,2\}$ is shown to be $(\varepsilon_1+\varepsilon_2,\delta)$-QDP, where $\delta$ is given in~\eqref{eq:parallelcomposition-similar-QDP}. The proof technique is, however, different from ours. Property~3 of \cref{thm:QPP_properties} also reveals that if one considers a restricted class of measurements (e.g., product measurements), then it is possible to achieve tighter privacy guarantees (namely, $(\varepsilon_1+\varepsilon_2, \delta_1+\delta_2)$-QDP) than those obtained when allowing all joint measurements on the two systems. Also note that $\delta'$ in~\eqref{eq:our-parallel-composition} is independent of $\varepsilon_1$ and $ \varepsilon_2$, whereas $\delta$ in~\eqref{eq:parallelcomposition-similar-QDP} depends on them. Depending on the particular values that the parameters $\varepsilon_i$ and $\delta_i$ take, for $i\in\{1,2\}$, one of these aforementioned results provides sharper privacy guarantees.
\end{remark}

\begin{example}
Here we provide an example to illustrate the distinction between the case of
joint measurements and product measurements, and more generally PPT
measurements (which contain the set of LOCC measurements, as well as product
measurements). Let $\alpha_{d}$ be the maximally mixed state on the
antisymmetric subspace of two $d$-dimensional systems, and let $\sigma_{d}$ be
the maximally mixed state on the symmetric subspace, i.e.,%
\begin{align}
\alpha_{d}  & \coloneqq \frac{I-F}{d\left(  d-1\right)  },\\
\sigma_{d}  & \coloneqq \frac{I+F}{d\left(  d+1\right)  },
\end{align}
where $F\coloneqq \sum_{i,j}|i\rangle\!\langle j|\otimes|j\rangle\!\langle i|$ is the
unitary swap operator. These states are orthogonal and thus perfectly
distinguishable by a joint measurement. Indeed, this measurement is given by
$\left\{  \Pi^{\alpha_{d}},\Pi^{\sigma_{d}}\right\}  $, where $\Pi^{\alpha
_{d}}\coloneqq \left(  I-F\right)  /2$ and $\Pi^{\sigma_{d}}\coloneqq \left(  I+F\right)  /2$.
By setting $\rM=\Pi^{\alpha_{d}}$, we find that $\operatorname{Tr}[\rM\alpha
_{d}]=1$ and $\operatorname{Tr}[\rM\sigma_{d}]=0$.

We can consider a QPP\ framework with $\mathcal{Q}=\left\{  (\alpha_{d}%
,\sigma_{d})\right\}  $ and the set of measurements to be $\mathcal{\bar{M}}$.
In this case, we only have QPP (i.e., the inequality $\operatorname{Tr}%
[\rM\alpha_{d}]\leq e^{\varepsilon}\operatorname{Tr}[\rM\sigma_{d}]+\delta$ is
satisfied) by setting $\varepsilon\geq0$ arbitrary and $\delta\geq1$, which is
a weak privacy guarantee (or really no privacy at all).

However, we can alternatively restrict the measurement operators to PPT
measurement operators, i.e., those $\rM$ which satisfy $0\leq \rM\leq I$ and
$0\leq \rM^{\Gamma}\leq I$, where the $\Gamma$ superscript denotes the partial
transpose. In this case, we find for every such PPT measurement operator $M$
that%
\begin{align}
\operatorname{Tr}[\rM\alpha_{d}]  & =\operatorname{Tr}[\rM^{\Gamma}\alpha
_{d}^{\Gamma}]\\
& =\operatorname{Tr}[\rM^{\Gamma}\left(  I-d\,\Phi^{d}\right)  /\left(  d\left(
d-1\right)  \right)  ]\\
& \leq\operatorname{Tr}[\rM^{\Gamma}\left(  I+d\,\Phi^{d}\right)  /\left(
d\left(  d-1\right)  \right)  ]\\
& =\operatorname{Tr}[\rM\left(  I+F\right)  /\left(  d\left(  d-1\right)
\right)  ]\\
& =\frac{d+1}{d-1}\operatorname{Tr}[\rM\sigma_{d}].
\end{align}
The first equality follows because the partial transpose is its own adjoint,
and the second equality follows by introducing the maximally entangled state
$\Phi^{d}\coloneqq \frac{1}{d}\sum_{i,j}|i\rangle\!\langle j|\otimes|i\rangle\!\langle
j|$. The inequality follows because $0\leq \rM^{\Gamma}$. By applying the above
inequality, we conclude that $\left(  \varepsilon,\delta\right)  $-QPP\ holds
with $\varepsilon=\ln\!\left(  \frac{d+1}{d-1}\right)  $ and $\delta=0$, so that
privacy improves as dimension increases.
\end{example}

\subsubsection{Adaptive Composability} \label{App:adaptive composition}

Adaptive composition refers to the case when each subsequently composed mechanism is chosen based on the outputs of the preceding ones. The goal is to quantify the overall privacy leakage at the output of the adaptively composed mechanism. This idea has been studied in detail for classical privacy settings~\cite{DR14}, and here we explore it for QPP. 

\begin{figure}
    \centering
    \includegraphics[width=\linewidth]{ 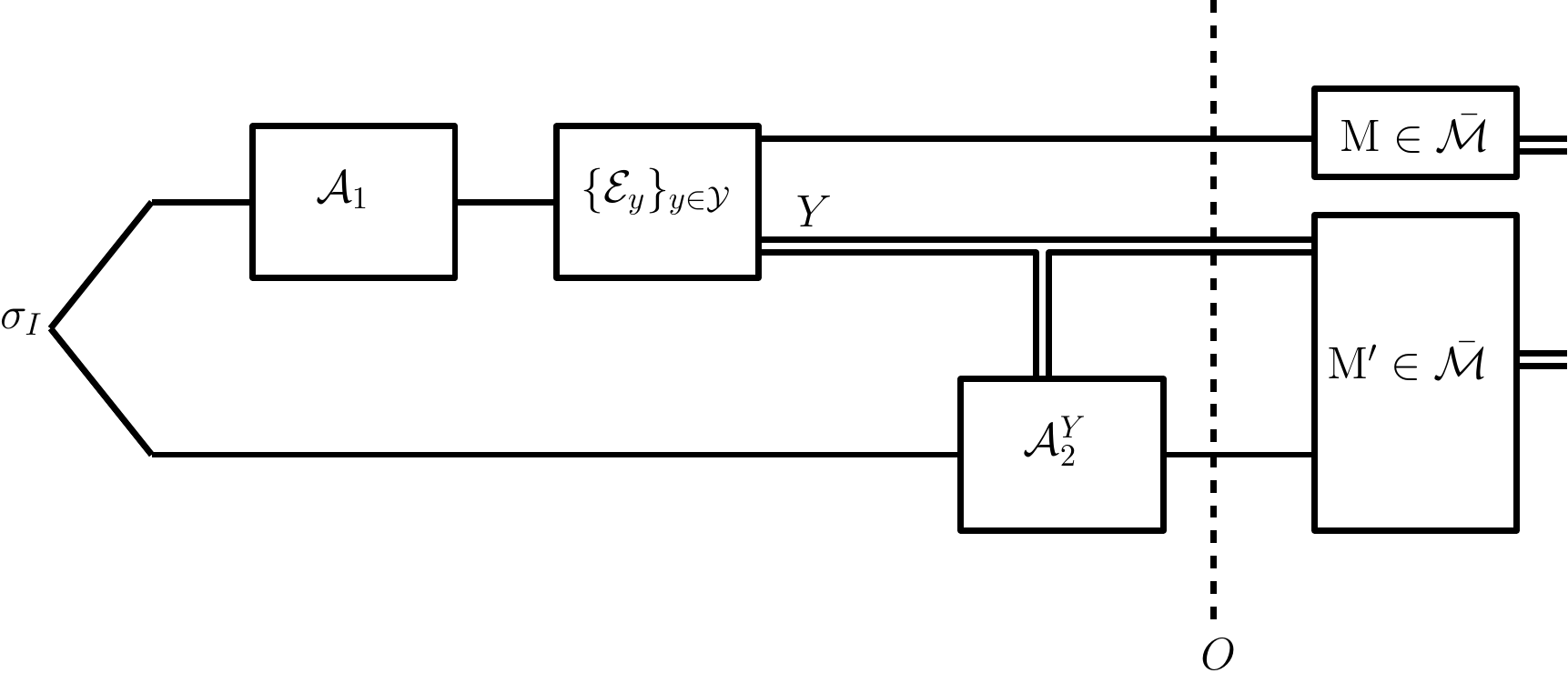}
    \caption{Setup for adaptive composition: On the top system, the channel $\cA_1$ is followed by the quantum instrument $\{ \cE_y\}_{y \in \cY}$, and then the random classical outcome $Y$ is used to choose the channel $\cA_2^Y$. In this setting, we analyse how well an adversary can learn properties of the input state $\sigma_I$ by applying measurements on the output state. 
    }
    \label{fig:adaptive composition}
\end{figure}
We first focus on the setting depicted in \cref{fig:adaptive composition}. 
Fix $X_i \sim P_X \in \Theta$ for $i \in \{1,2\}$, which are independently chosen,
and let the input state be
\begin{equation}\label{eq:initial-input-adaptive}
    \sigma_I \coloneqq \rho^{X_1} \otimes \rho^{X_2}.
\end{equation}

On the top subsystem in \cref{fig:adaptive composition}, the channel $\cA_1$ is followed by the quantum instrument $  \{\cE_y\}_{y \in \cY}$, which is a collection of completely positive maps such that the sum map
\begin{equation}
\overline{\cE} \coloneqq \sum_{y \in \cY} \cE_y    
\end{equation}
is trace preserving~\cite{davies1970operational,D76,Ozawa1984}. Depending on the measurement outcome $y$, the channel $\cA_2^y$ is chosen and applied to the bottom subsystem. The combined output state at stage $O$, as marked in the figure, is
\begin{align}
    \sigma_O & \coloneqq  \sum_{y \in \cY} 
    \cE^y\!\left(\cA_1(\rho^{X_1}) \right)  \otimes  |y \rangle\!\langle y| \otimes   \cA^y_2(\rho^{X_2}) .
\end{align}

We focus on adaptive composition of two quantum mechanisms in the above described setting. Suppose that $\cA_1$ is an $(\varepsilon_1,\delta_1)$-QPP mechanism in the framework $(\cS,\cQ,\Theta,\cM_1)$. 
Suppose furthermore that, for each  outcome $y \in \cY$, the mechanism $\cA_2^{y}$  satisfies $(\varepsilon_2,\delta_2)$-QPP  in the framework $(\cS,\cQ,\Theta,\cM_2)$ in the following sense: for all $(\cR,\cT) \in \cQ$ and $\rM \in \cM_2$, 
\begin{equation} \label{eq:adaptivelyComposable-distinguishability}
     \Tr\!\left[\rM \cA_2^{y}(\rho^\cR) \right] \leq e^{\varepsilon_2}  \Tr\!\left[\rM \cA_2^{y}(\rho^\cT) \right] + \delta_2. 
\end{equation}
Under adaptive composition, we want to guarantee indistinguishability of pairs of states 
\begin{equation}
    \sigma_I^\cR \coloneqq  \rho^{\cR_1} \otimes \rho^{\cR_2}  \ \textnormal{ and } \
     \sigma_I^\cT  \coloneqq \rho^{\cT_1} \otimes \rho^{\cT_2},
\end{equation}
where $(\cR_i,\cT_i)\in \cQ$ for $i \in \{1,2\}$. This means, informally, that the adversary would draw the same conclusions regardless of whether $\rho^{X_i}$ belongs to $\cR_i$ or $\cT_i$, for $i\in\{1,2\}$, when the initial input $\sigma_I$ to the system in \cref{fig:adaptive composition} is given by~\eqref{eq:initial-input-adaptive}. The following proposition provides parameters under which QPP of the adaptively composed mechanism is guaranteed.

\begin{proposition}[Adaptive composition of QPP]\label{Prop:Adaptive-composition-of-QPP}
    Fix the framework $(\cS,\cQ,\Theta,\bar{\cM})$.
   Suppose that $\cA_1$ satisfies $(\varepsilon_1,\delta_1)$-QPP and 
   $\cA_2^y$ satisfies $(\varepsilon_2,\delta_2)$-QPP for every measurement outcome~$y$, as in~\eqref{eq:adaptivelyComposable-distinguishability}. Then the mechanism in \cref{fig:adaptive composition} satisfies $(\varepsilon_1+ \varepsilon_2, \delta_2 + \delta_1 | \cY|)$-QPP in the framework $(\cS,\cQ \times \cQ,\Theta,\bar{\cM} \otimes \bar{\cM} )$ 
   where $|\cY|$ denotes the cardinality of the set $\cY$.
\end{proposition}

\begin{IEEEproof}
    See Appendix~\ref{App:Proof-of-adaptive-com}.
\end{IEEEproof}

\medskip
Note that, when $\delta_i =0$ for $i \in \{1,2\}$, the privacy parameters are additive. However, when $ \delta_i \neq 0$, the privacy parameter $\delta_2 + \delta_1 | \cY|$ degrades linearly with an increasing number of  measurement outcomes.

\begin{remark}[Composability with correlated states]
    In Property~3 of \cref{thm:QPP_properties} and \cref{Prop:Adaptive-composition-of-QPP}, we considered the case in which two mechanisms composed in parallel, receive independent inputs (i.e., the input being $\rho^{X_1} \otimes \rho^{X_2}$ where $X_i \sim P_X \in \Theta$ for $i=\{1,2\}$, which are chosen independently). In Appendix~\ref{App:Composability_with_correlated_states}
    we study  the setting in which the inputs are correlated. There, we observe that QPP shares similar properties related to composability of classical PP frameworks, where the class of $\Theta$ plays a key role in composability to hold in general.
\end{remark}

{In \cref{Prop:Adaptive-composition-of-QPP}, we assume a local structure of measurements conducted in the process, as shown in \cref{fig:adaptive composition}. This assumption is mainly motivated by technical considerations, as we can treat the resulting setting using our existing set of tools. Exploration of advanced adaptive composition techniques, which holds for more general classes of measurements is an interesting avenue for future work. In \cref{SubS: Incorporating Entanglement}, we present a variant of QPP where adaptive composition holds for general measurements and strategies (refer {to} \cref{fig:adaptive-reference-DR} and \cref{rem:properties_referenceSystemVariant}). }

\subsection{Mechanisms for QPP}
We propose mechanisms to achieve $\varepsilon$-QPP and $(\varepsilon,\delta)$-QPP using the depolarization channel. In addition, we provide a general procedure to generate $(\varepsilon,\delta)$-(classical) PP mechanisms using a quantum mechanism satisfying $(\varepsilon,\delta)$-QPP. 

\subsubsection{Depolarization Mechanism}

Let 
\begin{equation}
 \cA^p_{\mathrm{Dep}}(\rho) \coloneqq (1-p) \rho + \frac{p}{d} I,   
 \label{eq:depol-ch-def}
\end{equation}
 where $p \in[0,1]$ and $d$ is the dimension of the Hilbert space on which $\rho$ acts.
\begin{figure}
    \centering
    \includegraphics[width=\linewidth]{ 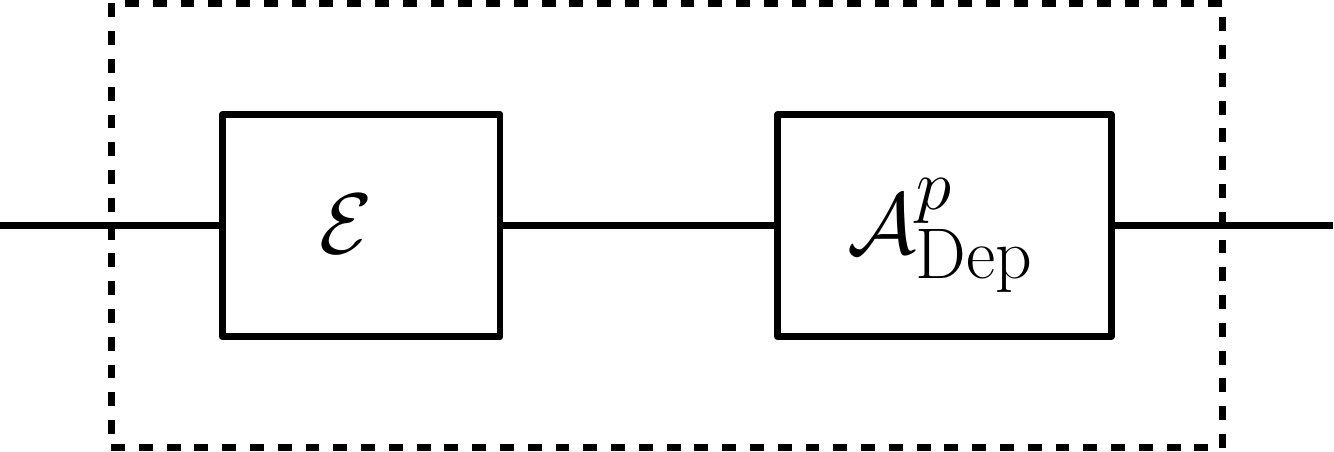}
    \caption{Depolarization mechanism to achieve QPP: This corresponds to a channel $\cE$ followed by a depolarizing channel. Note that we can choose $\cE=\cI$ to be the identity channel as well.}
    \label{fig: depolarizedChannel }
\end{figure}

\begin{theorem}[$\varepsilon$-QPP depolarization mechanism] \label{thm:epsilonQPPDeploarization}
Fix $p \in[0,1]$ and a privacy framework $(\cS,\cQ,\Theta,\cM)$.
Let $\cE$ be a quantum channel. Then $\cA^p_{\mathrm{Dep}}\!\left(\cE(\cdot)\right)$ (in \cref{fig: depolarizedChannel }) is $\varepsilon$-QPP if 
\begin{equation}
p \geq \frac{dK}{dK+e^\varepsilon -1},
\end{equation}
where 
\begin{equation}
K\coloneqq  \sup_{\rM \in \cM} \frac{\left\|\rM \right\|_\infty}{\Tr\!\left[\rM\right]} \times \sup_{\Theta, (\cR,\cT) \in \cQ} \frac{\left \| \cE(\rho^\cR)- \cE(\rho^\cT) \right \|_1}{2}. \end{equation}
This further implies that the depolarization channel with parameter~$p$ achieves $\varepsilon$-QPP whenever
 \begin{equation}\label{eq:privacy-dep-2}
 \varepsilon \geq    \ln\!\left( 1+ \frac{(1-p)dK}{p} \right).\end{equation}
\end{theorem}
\begin{IEEEproof}
Fix $P_X \in \Theta$, $(\cR,\cT) \in \cQ$, and $\rM \in \cM$, and consider that
\begin{align}
    & \frac{\Tr\!\left[\rM \cA^p_{\mathrm{Dep}}\!\left(\cE(\rho^\cR)\right)  \right]}{\Tr\!\left[\rM \cA^p_{\mathrm{Dep}}\!\left(\cE(\rho^\cT)\right)  \right]} -1  \notag \\
    &=  \frac{ (1-p) \Tr\!\left[\rM \cE(\rho^\cR) \right] + \frac{p}{d}\Tr\!\left[\rM\right]}{(1-p) \Tr\!\left[\rM \cE(\rho^\cT) \right] + \frac{p}{d}\Tr\!\left[\rM\right]} -1 \\ 
    &= \frac{ (1-p) \Tr\!\left[\rM \!\left( \cE(\rho^\cR) - \cE(\rho^\cT)\right) \right]}{(1-p) \Tr\!\left[\rM \cE(\rho^\cT) \right] + \frac{p}{d}\Tr\!\left[\rM\right]} \\
    & \leq \frac{ (1-p) {\left |\Tr\!\left[\rM \!\left( \cE(\rho^\cR) - \cE(\rho^\cT)\right) \right] \right |}}{ \frac{p}{d}\Tr\!\left[\rM\right]} \numberthis \label{eq:proof mechanism}
\end{align}
Given the above, if
\begin{equation}
\varepsilon \geq \ln\!\left( 1+ \frac{d(1-p)  {\left |\Tr\!\left[\rM \!\left( \cE(\rho^\cR) - \cE(\rho^\cT)\right) \right] \right |}}{p \Tr\!\left[\rM\right]} \right),
\end{equation}
then 
\begin{equation}
  \frac{\Tr\!\left[\rM \cA^p_{\mathrm{Dep}}\!\left(\cE(\rho^\cR)\right)  \right]}{\Tr\!\left[\rM \cA^p_{\mathrm{Dep}}\!\left(\cE(\rho^\cT)\right)  \right]}  \leq e^\varepsilon.
\end{equation}
Recalling that $(\cR,\cT) \in \cQ$ if and only if  $(\cT,\cR) \in \cQ$, the roles of $\rho^\cR$ and $\rho^\cT$ can be interchanged, and we conclude that
\begin{equation}
e^{-\varepsilon} \leq  \frac{\Tr\!\left[\rM \cA^p_{\mathrm{Dep}}\!\left(\cE(\rho^\cR)\right)  \right]}{\Tr\!\left[\rM \cA^p_{\mathrm{Dep}}\!\left(\cE(\rho^\cT)\right)  \right]}  .
\end{equation}

Consider that
\begin{equation}
\Tr\!\left[\rM \!\left( \cE(\rho^\cR) - \cE(\rho^\cT)\right) \right] \leq \left\| \rM\right\|_\infty \frac{\left\| \cE(\rho^\cR)- \cE(\rho^\cT) \right\|_1}{2}.  
\end{equation}
Indeed, consider the following Jordan--Hahn decomposition $\cE(\rho^\cR) - \cE(\rho^\cT)=P-Q$, where $P$ and $Q$ are the positive and negative parts of $\cE(\rho^\cR) - \cE(\rho^\cT)$, respectively, satisfying $P,Q\geq 0$ and $PQ=0$. Then
\begin{align}
   & \Tr\!\left[\rM \!\left( \cE(\rho^\cR) - \cE(\rho^\cT)\right) \right] \cr
   & =\Tr\!\left[\rM \!\left( P -Q\right) \right] \\
   & \leq \Tr\!\left[\rM P \right] \\ 
   & \leq \left\| \rM\right\|_\infty  \frac{\left\| \cE(\rho^\cR)- \cE(\rho^\cT) \right\|_1}{2},
\end{align}
where the last inequality follows from H\"older's inequality and because $ \left\| \cE(\rho^\cR)- \cE(\rho^\cT) \right\|_1= \Tr[P] + \Tr[Q] =2\Tr[P]$ since  $\Tr\!\left[\cE(\rho^\cR) - \cE(\rho^\cT)\right] = 0=\Tr[P-Q]$.

Collecting all terms and supremizing over $\cM$ and~$\Theta$ and all secret pairs of $\cQ$ yields the desired result. 
\end{IEEEproof}

\medskip
Note that the parameter $K$ derived from \cref{thm:epsilonQPPDeploarization} represents the domain knowledge accessible and incorporated into the privacy model of the $(\mathcal{S},\mathcal{Q},\Theta,\mathcal{M})$ QPP framework. 

\begin{corollary}[$\varepsilon$-QDP with domain knowledge]
Fix $p \in[0,1]$, and a privacy framework $(\cS,\cQ,\Theta,\cM)$ for QDP that encodes domain knowledge. Let $\cE$ be a quantum channel. Then $\cA^p_{\mathrm{Dep}}\!\left(\cE(\cdot)\right)$ is $\varepsilon$-QPP with 
\begin{equation} \varepsilon \geq    \ln\!\left( 1+ \frac{(1-p)d}{2p} k'  \sup_{\rM \in \cM} \frac{\|\rM\|_\infty}{\Tr\!\left[\rM\right]}   \right) ,\end{equation}
where
\begin{align}
k' & \coloneqq \sup_{(\rho^{x_1},\rho^{x_2}) \in \cW_\Theta}  \left \| \cE(\rho^{x_1})- \cE(\rho^{x_2}) \right \|_1  ,  \\
\cW_\Theta & \coloneqq \{(\rho^{x_1},\rho^{x_2}) \in \cQ \ | \  \exists P_X \in \Theta \ \! \ P_X(x_1), P_X(x_2) >0 \}.
\end{align}
\end{corollary}
Note that the domain knowledge encoded into the QDP framework may guide towards an improved accuracy/utility, as opposed to considering all neighboring states as secret pairs and all possible measurements. 
For the QDP framework without domain knowledge, \cite[Theorem~3]{QDP_computation17} shows that  
\begin{equation}
  \varepsilon \geq    \ln\!\left( 1+ \frac{(1-p)d}{2p} \sup_{\rho \sim \sigma} \left \| \rho - \sigma \right \|_1  \right)
\end{equation}
is a sufficient condition to
ensure $(\varepsilon,0)$-QDP. Compared with that
due to the condition
\begin{multline}
\sup_{\rM \in \cM} \frac{\left\|\rM \right\|_\infty}{\Tr\!\left[\rM\right]} \times \sup_{(\rho^{x_1},\rho^{x_2}) \in \cW_\Theta}  \left \| \cE(\rho^{x_1})- \cE(\rho^{x_2}) \right \|_1 \\
\leq \sup_{\rho \sim \sigma} \left \| \rho - \sigma \right \|_1, \end{multline}
a QDP framework that has the capability to incorporate domain knowledge may cause less perturbation to the useful channel output of $\cE$ in some cases. The rightmost inequality holds because $\cW_\Theta$ includes only the neighboring pairs of states such that their occurrence has a positive probability, while $\rho \sim \sigma$ denotes all possible neighboring pairs. Furthermore, we always have that $\frac{\left\|\rM \right\|_\infty}{\Tr\!\left[\rM\right]} \leq 1$ for every measurement operator~$\rM$.

\begin{remark}[Local DP]
For the setup in \cref{rem:QLDP}, \cref{thm:epsilonQPPDeploarization} reduces to 
\begin{equation}
    p \geq \frac{d}{d+e^\varepsilon -1},
\end{equation}
with the choice of the identity channel instead of $\cE$ in \cref{fig: depolarizedChannel }. This occurs because $\left \| \rho -\sigma \right\|_1 \leq 2$, with equality for pairs of orthogonal states, and $\left\|\rM \right\|_\infty \leq \Tr\!\left[\rM\right]$, with equality holding whenever $\rM$ is a rank-one measurement operator. 
This is analogous to a version of the randomized response technique used to achieve classical local DP~\cite{erlingsson2014rappor, LDP1, LDP2}. For a finite alphabet $\cX$ with cardinality $|\cX |$, the randomized response mechanism outputs the true value with probability $1-q$, and it outputs a randomly chosen realization with probability $q/|\cX |$. Then, if
\begin{equation}
q\geq \frac{|\cX|}{|\cX|+e^\varepsilon -1},    
\end{equation}
$\varepsilon$-local differential privacy is achieved. This analogy further suggests that the depolarization mechanism can be considered as a quantum version of the randomized response mechanism that achieves classical privacy guarantees. 
\end{remark}

Considering the scenario in which we want to provide a privacy guarantee for all possible measurements (i.e., $\cM=\bar{\cM}$), next we derive the parameter $p$ to achieve $(\varepsilon,\delta)$-QPP. 

\begin{proposition} [$(\varepsilon,\delta)$-QPP depolarization mechanism]
\label{prop:(eps,delt)-QPP depolarized}
Fix $p \in[0,1]$ and the privacy framework $(\cS,\cQ,\Theta,\cM)$ with $\cM=\{\rM: 0 \psd \rM \psd \rI\}$.
Let $\cE$ be a quantum channel. Then $\cA^p_{\mathrm{Dep}}\!\left(\cE(\cdot)\right)$ is $\varepsilon$-QPP if 
\begin{equation}
p \geq \max\! \left\{0,\frac{d(K'-\delta)}{dK'+e^\varepsilon -1}\right \},\end{equation}
where 
\begin{equation} K'\coloneqq  \sup_{\Theta, (\cR,\cT) \in \cQ} \frac{ \left\| \cE(\rho^\cR)- \cE(\rho^\cT) \right\|_1}{2}.
\end{equation}     
\end{proposition}

\begin{IEEEproof}
The proof follows from the use of the equivalent formulation through the hockey-stick divergence and the properties of this divergence.    
By \cite[Lemma~IV.I]{hirche2023quantum}, we have 
\begin{multline}
\sE_{e^\varepsilon}\!\left( \cA_\mathrm{Dep}\!\left(\cE(\rho^\cR)\right) \Vert \cA_\mathrm{Dep}\!\left(\cE(\rho^\cT)\right)\right) \\
\leq (1-e^\varepsilon)\frac{p}{d}+(1-p) \sE_{e^\varepsilon}\!\left( \cE(\rho^\cR)\Vert \cE(\rho^\cT)\right).
\end{multline}
We also have the property \cite[Lemma~II.4]{hirche2023quantum}
\begin{equation}
 \sE_{e^\varepsilon}\!\left( \cE(\rho^\cR)\Vert \cE(\rho^\cT)\right) \leq \frac{ \left \|\cE(\rho^\cR)- \cE(\rho^\cT) \right \|_1} {2 }.  
\end{equation}
Combining these relations, and supremizing over $\Theta$, and secret pairs $(\cR,\cT) \in \cQ$, we can choose 
\begin{equation}
\delta \geq (1-e^\varepsilon)\frac{p}{d}
+(1-p) K'.
\end{equation}
Then, rearranging the terms we arrive at 
\begin{equation} p \geq \frac{d(K'-\delta)}{dK'+e^\varepsilon -1}. \end{equation}
Since $p \geq 0$, when $K'-\delta \leq 0$, we set $p=0$.
\end{IEEEproof}
\medskip

\subsubsection{Classical PP Mechanisms from QPP Mechanisms}

The QPP formalism provides a direct methodology to design classical PP mechanisms with the assistance of QPP mechanisms. \footnote{In the context of classical PP mechanisms, the main attempt has been the introduction of Wasserstein mechanisms, based on the infinity order Wasserstein distance and its modifications. In particular,~\cite{SWC17} and~\cite{wassersteinSmoothMechanism} introduced these mechanisms to achieve $(\varepsilon,0)$-PP and $(\varepsilon,\delta)$-PP, respectively. However, it is important to note that these approaches may encounter computational intractability challenges.} In this case, we use quantum encoding to convert classical data to quantum data. We denote the quantum encoding of classical data $x \in \cX^{n \times k}$ as $\rho^x \coloneqq |x \rangle\!\langle x|$ {(recall that $p_X \in \Theta_c$ are discrete probability distributions over the probability space $\cP(\cX^{n \times k})$ and this leads to a finite collection of $\{\rho^x\}_x$ quantum encodings)}. Then, we ensure the privacy for the quantum data (quantum encoding) such that the privacy is ensured for the underlying classical data. 

\begin{figure}
    \centering
    \includegraphics[width=\linewidth]{ 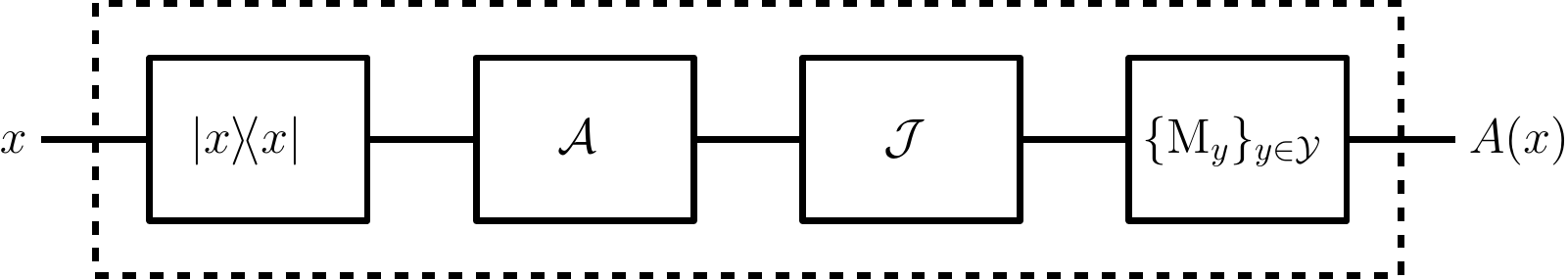}
    \caption{Generation of classical PP mechanisms from QPP mechanism $\cA$: First the classical data is encoded using quantum encoding techniques, then the QPP mechanism $\cA$, and if needed any other channel $\cJ$, and finally the measurement channel.}
    \label{fig:classicalPP}
\end{figure}

\begin{proposition}[$(\varepsilon,\delta)$- (classical) PP mechanism] \label{prop: classical PP mechanism}
Given an ($\varepsilon, \delta$)-QPP mechanism $\cA$ within the framework $(\cS,\cQ,\Theta,\cM)$ when $\rho^x\coloneqq |x\rangle\!\langle x|$ with 
\begin{equation}
 \begin{aligned}
     \cS&= \left \{ \{ \rho^x: x \in \cR_c\}: \cR_c \in \cS_c \right \} , \\
    \cQ&= \left\{\left(\{ \rho^x: x \in \cR_c\} , \{\rho^x: x \in \cT_c \}\right) : (\cR_c,\cT_c) \in \cQ_c \} \right\} , \\
    \Theta&=\{ \{ p_X(x), \rho^x \}_x:  p_X \in \Theta_c \} , \\
    \cM&=\left\{ \rM: 0 \psd \rM \psd \rI \right \},
\end{aligned}   
\end{equation}
any post-processing of $\cA$ by  a quantum channel $\cJ$ followed by applying a POVM $\{\rM_y \}_{y \in \cY} $  denoted as $A:~\cX^{n \times k} \to \cY$ as shown in \cref{fig:classicalPP} is $(\varepsilon,\delta)$-PP in the framework $(\cS_c,\cQ_c,\Theta_c)$.

Furthermore, for a selected post-processing $\cJ$ and POVM $\{\rM_y \}_{y \in \cY} $, it is sufficient for $\cM^\cJ_c \subseteq \cM$ for $A$ to be $(\varepsilon,\delta)$-PP, where 
\begin{equation}\cM^\cJ_c \coloneqq  \left \{ \cJ^\dagger\left(\sum_{y \in \cB} \rM_y \right): \cB \in \cY \right\}. \end{equation}

\end{proposition}

\begin{IEEEproof}
Fix $\cB \subseteq \cY$.
Consider  that
\begin{align}
    & \PP\!\left(A(X) \in \cB \middle | \cR_c\right) \notag
    \\
    &= \frac{ \PP\!\left(\{A(X) \in \cB\} \cap \cR_c\right)}{ \PP\!\left(\cR_c\right)} \\ 
   &  \stackrel{(a)}= \frac{ \sum_{x \in \cR_c} p(x) \PP\!\left(A(x) \in \cB\right)}{ P_X(\cR)} \\
    & \stackrel{(b)}=  \frac{ \sum_{x \in \cR_c} p(x) \sum_{y \in \cB} \PP\!\left(A(x) =y\right)  }{ P_X(\cR)} \\
    & \stackrel{(c)}= \frac{ \sum_{\rho^x \in \cR} p(x) \sum_{y \in \cB} \Tr\!\left[\rM_y \cJ \circ \cA(\rho^x)\right]}{ P_X(\cR)} \\
    &\stackrel{(d)}= \Tr\! \left[ \sum_{y \in \cB} \rM_y \cJ \circ \cA \left( \sum_{\rho^x \in \cR}  \frac{p(x)}{ P_X(\cR)} \rho^x \right) \right] \\ 
     & \stackrel{(e)}= \Tr \!\left[ \cJ^\dagger\left(\sum_{y \in \cB} \rM_y \right) \cA \left( \rho^\cR\right) \right] \\
     & \stackrel{(f)}=\Tr \!\left[ \rM \cA(\rho^\cR) \right],
\end{align}
where: (a) from $\cR\coloneqq \{ \rho^x: x \in \cR_c\}$; (b) from $\cB$ being a collection of $y \in \cY$; (c) from $\rM_y$ being the measurement applied to obtain the outcome $y$; (d) from the linearity of trace operator and quantum channels $\cA, \cJ$; (e) from the definition of $\rho^\cR$, and $\cJ^\dagger$ being the adjoint of $\cJ$; and finally (f) from $\rM~\coloneqq~\cJ^\dagger\left(\sum_{y \in \cB} \rM_y \right) $. 

Similarly, $\PP\!\left(A(X) \in \cB \middle | \cT_c\right)=\Tr\! \left[\rM \cA(\rho^\cT) \right]$.  Then with the assumption that $\cA$ is $(\varepsilon,\delta)$-QPP for $(\cS,\cQ,\Theta, \cM)$ mentioned in the proposition statement, we have 
\begin{equation}\PP\!\left(A(X) \in \cB \middle | \cR\right) \leq e^{\varepsilon} \hspace{1mm} \PP\!\left(A(X) \in \cB \middle | \cT\right) + \delta .
\end{equation}
concluding the proof. 
\end{IEEEproof}

\medskip
Depolarization is a common kind of noise considered in quantum information processing. Thus, this offers a method for designing classical PP mechanisms by combining the results presented in \cref{prop: classical PP mechanism} and \cref{thm:epsilonQPPDeploarization}. However, it is essential to recognize that quantum encoding of classical data would require additional computational resources, shifting the complexity of the mechanism design phase to the encoding phase.

\section{Quantifying Privacy-Utility Tradeoff} \label{Sec:Quantifying-Privacy-Utility-Tradeoff}

In this section, we aim to assess the utility achievable through the implementation of a privatization mechanism while adhering to privacy constraints and characterize the inherent tradeoffs involved in this process. To achieve this, we define a utility metric grounded in an operational approach and demonstrate its representation via an SDP. Subsequently, we leverage this metric to conduct an in-depth analysis of privacy-utility tradeoffs, with a specific emphasis on the depolarization mechanism.

\subsection{Utility Metric}

Let $\cA$ denote a privacy mechanism. 
We focus on assessing the potential of reversing the effects of $\mathcal{A}$ by applying a post-processing mechanism $\mathcal{B}$ to recover the initial input state to $\mathcal{A}$ up to an error $1-\gamma$. 
We define $\gamma$-utility  in terms of how distinguishable $\cB \circ \cA$ is from the identity channel, employing the normalized diamond distance as the distinguishability measure.
\begin{definition}[$\gamma$-utility] \label{def:gamma_utility}
Let $\cA: \cL(\cH_A) \to \cL(\cH_C)$ be a privacy mechanism, and fix $\gamma \in [0,1]$. We say that $\cA$ satisfies $\gamma$-utility if
    \begin{equation} \sU(\cA) \coloneqq  1- \inf_{\cB} \frac{1}{2} \left\| \cI- \cB \circ \cA \right\|_\diamond \geq \gamma, \end{equation}
    where the infimum is taken over every quantum channel $\cB: \cL(\cH_C) \to \cL(\cH_D)$.
\end{definition}

The defined utility metric can be reformulated as an SDP by using the dual SDP form of the diamond distance \cite[Section~4]{Wat09} and rewriting the quantum channel $\cB$ in terms of its Choi matrix $\Gamma_{CD}^\cB$, as well as translating the conditions for $\cB$ to be a channel to conditions on its Choi matrix, namely $\Gamma_{CD}^\cB \geq 0$ and $\Tr_D\!\left[\Gamma_{CD}^\cB \right]= I_c$.

\begin{proposition}[SDP formulation of $\gamma$-utility] \label{prop:SDP formulation of gamma-utility}
The $\gamma$-utility of a privacy mechanism $\cA$ can be formulated as the following SDP: 
\begin{equation}
\sU(\cA)=1- \inf_{\substack{\mu \geq 0 \\ Z_{AD} \geq 0 \\ \Gamma_{CD}^\cB \geq 0} } \left \{
\begin{array}[c]{c}
\mu : \\
Z_{AD} \geq \Gamma_{AD} -\Gamma_{AD}^{\cB \circ \cA}, \\
\mu  I_A \geq \Tr_D\!\left[Z_{AD}\right], \\ \Tr_D\!\left[\Gamma_{CD}^\cB \right]= I_C 
\end{array}
\right \} \geq \gamma, 
\end{equation}
where 
\begin{equation} \label{eq:composedABChoi}
\Gamma_{AD}^{\cB \circ \cA}\coloneqq  \Tr_C\!\left[\left( I_A \otimes \Gamma_{CD}^{\cB} \right)\left(\T_C(\Gamma_{AC}^\cA) \otimes I_D \right) \right],    
\end{equation}
and $\Gamma$ represents the Choi matrix with the subscripts showing the input and the output system of the channel while the superscript indicating the channel considered, with no superscript for the identity channel.\footnote{The Choi matrix of the composed channel $\cB \circ \cA$ is denoted by $\Gamma_{AD}^{\cB \circ \cA}$ and~\eqref{eq:composedABChoi} follows from \cite[Eq.~(3.2.22)]{khatri2020principles}.}
\end{proposition}

{Note that a similar SDP formulation for approximate degradability where the identity channel in \cref{def:gamma_utility} is replaced by the complementary channel is presented in \cite[Proposition~9]{sutter2017approximate}.}

\begin{remark}[Characterizing optimal privacy-utility tradeoffs]
    The optimal utility attained by an $(\varepsilon,\delta)$-QPP mechanism can be characterized as an SDP. 
    To achieve this, we combine the equivalent formulation of QPP via the DL~divergence from \cref{prop: Equivalent formulation with DS} with the SDP formulation of DL divergence in \cref{lem: SDP formulation ISD}. Combining this with the SDP formulated from \cref{prop:SDP formulation of gamma-utility} enables computing the privacy requirements and quantifying utility together. Additionally, we determine the optimal privacy parameters for fixed utility requirements. For a comprehensive discussion of this point, please refer to Appendix~\ref{app:privacyUtilityTradeoff}.
    The utilization of the SDP derived for the DL divergence in this operational task highlights an advantage of the equivalent formulation for QPP using the DL divergence. 
\end{remark}

\subsection{Analysis of Depolarization Mechanism}

We now instantiate $\mathcal{A}$ as the depolarizing channel with parameter $p$, denoted as $\mathcal{A}^p_{\mathrm{Dep}}$ (as defined in~\eqref{eq:depol-ch-def}), and proceed to analyze $\mathcal{U}(\mathcal{A}^p_{\mathrm{Dep}})$.

\begin{proposition}[Utility from depolarization mechanism] \label{prop:Utility from Depolarization mechanism}
Fix $p \in [0,1]$. The depolarization mechanism satisfies $\gamma$-utility if and only if
\begin{equation}   \sU(\cA_{\mathrm{Dep}}^p)
= 1-\frac{p(d^2-1)}{d^2} \geq \gamma.\end{equation}

\end{proposition}
\begin{IEEEproof}
The proof below relies on observing that the optimization term (in the utility metric) is minimized by setting $\cB=\cI$, and then evaluating $\left\| \cI- \cA_{\mathrm{Dep}}^p \right\|_\diamond $ using the Choi states of the channels $\cA_{\mathrm{Dep}}^p$ and $\cI$, due to the joint covariance of the two channels under unitaries (i.e., $\cA_{\mathrm{Dep}}^p \circ \cU = \cU \circ \cA_{\mathrm{Dep}}^p$ and $\cI \circ \cU = \cU \circ \cI$ for every unitary channel $\cU$). 

    Consider that
    \begin{align}
        & \left \| \cI- \cB \circ \cA_{\mathrm{Dep}}^p \right\|_\diamond \notag \\
        & \stackrel{(a)}= \left\| \cU \circ (\cI- \cB \circ \cA_{\mathrm{Dep}}^p) \circ \cU^\dagger \right\|_\diamond \\
        &\stackrel{(b)}=\left \| \cI- \cU \circ \cB \circ \cU^\dagger \circ \cA_{\mathrm{Dep}}^p) \right \|_\diamond  \\
        &\stackrel{(c)}=\int \hspace{-2mm} \dd \cU \, \left \| \cI- \cU \circ \cB \circ \cU^\dagger \circ \cA_{\mathrm{Dep}}^p) \right \|_\diamond  \\
        &\stackrel{(d)}\geq  \left \| \cI- \left( \int \hspace{-2mm} \dd \cU \  \cU \circ \cB \circ \cU^\dagger \right) \circ \cA_{\mathrm{Dep}}^p) \right \|_\diamond,
    \end{align}
    where: (a) follows from the unitary invariance of the diamond norm with $\cU$ representing a unitary channel; (b) from the commutative property of $\cA_{\mathrm{Dep}}^p$ with every unitary channel; (c) with $\dd \cU$ denoting the Haar measure over the unitary group and from the left-hand side being independent of $\cU$; and (d) from the convexity of the diamond norm. 

    Next, observe that $\cB^\star\coloneqq   \int \hspace{-1mm} \dd \cU \  \cU \circ \cB \circ \cU^\dagger  $ is  a quantum channel, and it is in fact equal to a depolarization channel~\cite{HHH99}. 
Then, $\cB^\star=\cA_{\mathrm{Dep}}^q$ for some $q \in [0,1]$.

The composition of two depolarization channels 
\begin{equation}
   \cB^\star=\cA_{\mathrm{Dep}}^q \circ \cA_{\mathrm{Dep}}^p 
\end{equation}
 is also a depolarization channel with parameter
$p^\star\coloneqq  1-(1-p)(1-q)$. The minimum value is attained by the choice $q=0$, where $\cA_{\mathrm{Dep}}^q$ in that case corresponds to the identity channel. 

With that, we arrive at
\begin{equation}
\inf_{\cB} \left \| \cI- \cB \circ \cA_{\mathrm{Dep}}^p \right\|_\diamond= \left \| \cI-  \cA_{\mathrm{Dep}}^p \right\|_\diamond.    
\end{equation}
With the property of joint covariance of $\cI$ and $\cA_{\mathrm{Dep}}^p$ under unitaries \cite[Proposition~7.82]{khatri2020principles}, we simplify this to
\begin{equation}
\left \| \cI-  \cA_{\mathrm{Dep}}^p \right \|_\diamond=\frac{1}{d} \left \| \Gamma_{AD}-\Gamma_{AD}^{\cA_{\mathrm{Dep}}^p} \right \|_1.
\end{equation}
Then consider that
\begin{align}
    & \frac{1}{d} \left \| \Gamma_{AD}-\Gamma_{AD}^{\cA_{\mathrm{Dep}}^p} \right \|_1
    \notag \\
    &\stackrel{(a)}
     = \frac{1}{d} \left \| \Gamma_{AD}- \left( (1-p) \Gamma_{AD} + \frac{p}{d} I_{d^2} \right) \right \|_1 \\
    & = \frac{p}{d}\left \| \Gamma_{AD}- \frac{1}{d} I_{d^2} \right \|_1 \\
    &= p \left \| \frac{\Gamma_{AD}}{d} \left(1-\frac{1}{d^2} \right) - \frac{1}{d^2} \left(I_{d^2} -\frac{\Gamma_{AD}}{d} \right) \right \|_1 \\
    & \stackrel{(b)}= p \left( 1- \frac{1}{d^2}\right) \left( \left\|\frac{\Gamma_{AD}}{d}  \right\|_1 + \left\|\frac{I_{d^2}- \frac{\Gamma_{AD}}{d}}{d^2 -1}  \right\|_1 \right)\\
    & \stackrel{(c)}= 2p \left( 1- \frac{1}{d^2}\right) ,
\end{align}
where: (a) from $\Gamma_{AD}^{\cA_{\mathrm{Dep}}^p}=(1-p) \Gamma_{AD} + \frac{p}{d} I_{d^2}$; (b) from $\frac{\Gamma_{AD}}{d}$, and $ I_{d^2} -\frac{\Gamma_{AD}}{d}$ being orthogonal; and (c) from trace norm of quantum states being equal to one. 

Combining the above chain of arguments together completes the proof. 
\end{IEEEproof}

\medskip

Next, we focus on understanding the privacy-utility tradeoff with respect to the parameter $p$ governing a depolarization mechanism. 
From \cref{prop:Utility from Depolarization mechanism}, to achieve $\gamma$-utility, we require that
\begin{equation}
p \leq \frac{ (1-\gamma) d^2}{(d^2-1)}.
\end{equation}
Conversely, to achieve $\varepsilon$-QPP in the chosen privacy framework $(\cS,\cQ,\Theta, \cM)$, from \cref{thm:epsilonQPPDeploarization},it suffices for $p$ to satisfy
\begin{equation}
p \geq \frac{dK}{dK+e^\varepsilon -1}.
\end{equation} 
These two inequalities provide insights into the privacy-utility tradeoff associated with the depolarization mechanism. Consequently, it is essential to carefully tune the parameter $p$ based on the desired utility, characterized by $\gamma$, as well as the privacy parameter $\varepsilon$.

\medskip

\begin{figure}[!t]
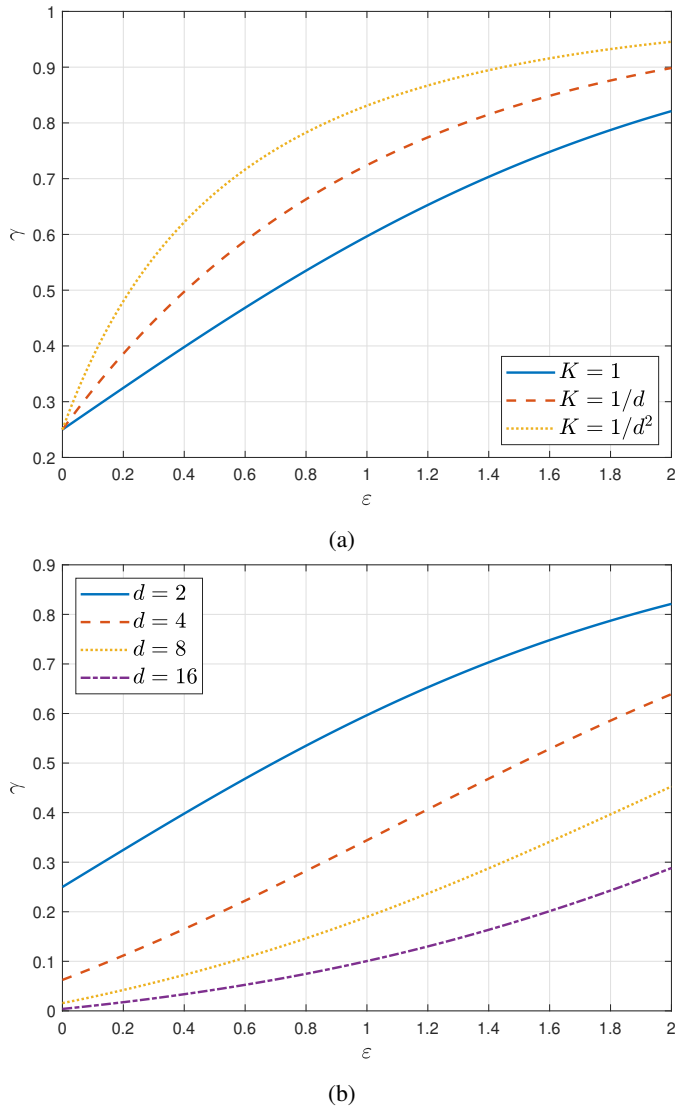

\begin{subfigure}{0.5\textwidth}
 \centering
    \includegraphics[width=\linewidth]
    { epsgammad2.pdf}
    \caption{}
    \label{fig:PVsUtilityPrivacy}
\end{subfigure}
\hspace{4mm}
\begin{subfigure}{0.5\textwidth}
  \centering
    \includegraphics[width=\linewidth]
    { epsgammad2Varying.pdf}
    \caption{}
    \label{fig:PwithDimension} 
    \end{subfigure}
    \caption{(a) For fixed $d=2$, the figure depicts the optimum utility $\gamma$ for $\varepsilon$ achievable with the depolarization mechanism in \cref{thm:epsilonQPPDeploarization}.
    The value of $K$ encodes the domain knowledge available, where $K=1$ corresponds to no such additional information being available. 
    (b) For fixed $K=1$, the figure depicts the optimum utility $\gamma$ for $\varepsilon$ achievable with the depolarization mechanism in \cref{thm:epsilonQPPDeploarization} for $d\in \{2,4,8,16\}$. }
\end{figure}

\noindent\textbf{Effect of domain knowledge:} 
 \cref{fig:PVsUtilityPrivacy} illustrates the optimal utility achievable using the $\varepsilon$-QPP depolarization mechanism presented in \cref{thm:epsilonQPPDeploarization}. Notably, as the value of $K$ reduces, the attainable utility region expands. The parameter $K$ derived from \cref{thm:epsilonQPPDeploarization} represents the domain knowledge accessible and incorporated into the privacy model of the $(\mathcal{S},\mathcal{Q},\Theta,\mathcal{M})$ QPP framework. This observation underscores the significance of incorporating such domain knowledge to enhance utility gains while simultaneously ensuring the necessary privacy assurances.

\medskip 

\noindent\textbf{Effect of dimension:}
In \cref{fig:PwithDimension}, we observe a prominent privacy-utility tradeoff as the dimension increases for the depolarization mechanism presented in \cref{thm:epsilonQPPDeploarization}.
Regarding the utility of the depolarization mechanism (given by $1-\frac{p (d^2-1)}{d^2}$), we can always establish the following lower bound for every $d$:
\begin{equation}
1-\frac{p (d^2-1)}{d^2} \geq 1-p ,
\end{equation}
where this lower bound is attained as $d \to \infty$. However, the achievable privacy level $\varepsilon$ in~\eqref{eq:privacy-dep-2} degrades at most by an order of $\ln(d)$. Hence, it is crucial to identify the optimal privacy parameters achieved by private mechanisms, particularly in high-dimensional scenarios.

\medskip

\begin{remark}[Application specific privacy-utility tradeoffs]
    In the previous analysis concerning the depolarization mechanism in \cref{fig: depolarizedChannel }, we chose $\mathcal{E}=\mathcal{I}$, the identity channel. However, it would be an interesting future work to explore the utility for user-specific $\mathcal{E}$ channels. Specifically, we can choose $1- \frac{1}{2}\inf_{\mathcal{B} \in \mathrm{CPTP}} \| \mathcal{E} - \mathcal{B} \circ \mathcal{A}^p_{\mathrm{Dep}} \circ \mathcal{E} \|_\diamond$ as the utility metric. If $\mathcal{E}$ possesses certain symmetries, one can potentially utilize arguments akin to those presented in the proof of \cref{prop:Utility from Depolarization mechanism}. This investigation could shed light on tailoring privacy mechanisms to specific application needs,  leading to more effective privacy-utility tradeoffs.
\end{remark}

\section{Auditing Privacy Frameworks} \label{Sec:Auditing-Privacy-Frameworks}

Auditing for privacy aims to detect violations in privacy guarantees and reject incorrect algorithms (see~\cite{ding2018detecting,jagielski2020auditing,domingo2022auditing,nuradha2022pufferfishJ} for classical approaches). In this section, our focus is on utilizing quantum information theory tools and quantum algorithms to audit the privacy of quantum systems. Specifically, we concentrate on auditing algorithms for QDP guarantees, and it should be noted that these ideas can be extended to audit algorithms for privacy guarantees demanded by QPP (see \cref{rem:QPP_auditing}).

The main idea behind auditing classical privacy frameworks (DP and PP) involves translating the privacy requirement into a weaker privacy notion that can be efficiently computed. For example, in~\cite{nuradha2022pufferfishJ}, sliced mutual information based DP is used to audit for DP. By doing so, algorithms failing to meet the privacy conditions imposed by the relaxed privacy notion are concluded to violate the original privacy requirement. However, a pitfall of this approach is the inability to quantify the gap between the constraints stemming from the original DP or PP notion and the relaxed privacy notions. In other words, even if we verify that the relaxed privacy notion is satisfied, we cannot determine whether the original privacy requirement is also satisfied. In contrast, in this work, we focus on QDP without translating it into a relaxed privacy notion. 

\subsection{Techniques for Auditing QDP}

\underline{\textbf{Using semi-definite programs:}}

\medskip
By leveraging the equivalent formulation from \cref{prop: Equivalent formulation with DS} and adopting the specific choices of $(\cS,\cQ,\Theta,\cM)$ as provided in \cref{rem: Quantum DP set}, for
$(\varepsilon,\delta)$-QDP,  we have that
\begin{equation}\sup_{\rho \sim \sigma} \ \overline{\sD}^{\delta}\!\left( \cA(\rho) \Vert\cA(\sigma) \right) \leq \varepsilon.
\end{equation}
Then, we can compute the left-hand side above by using the SDP formulation of $\overline{\sD}^{\delta}\!\left( \cdot \Vert \cdot \right)$ presented in \cref{lem: SDP formulation ISD}.  This approach is particularly beneficial for low-dimensional setups as the time complexity of SDP computation is polynomial in the dimension of the quantum states. However, it is essential to note that the complexity of this approach grows exponentially with the number of qubits, making it less feasible for higher-dimensional systems.

Additionally, using the equivalent formulation in \cref{rem:Equivalent formulation with hockey-stick divergence}, consider that
$(\varepsilon,\delta)$-QDP is equivalent to 
\begin{equation} \sup_{\rho \sim \sigma} \ \sE_{e^{\varepsilon}}\!\left( \cA(\rho) \| \cA(\sigma) \right)\leq \delta,\end{equation}
where
\begin{align}
\label{eq:hockey-stick-def-trace}
    \sE_{\gamma}(\rho\| \sigma) & \coloneqq  \Tr\!\left[(\rho-\gamma \sigma)_{+}\right]
    \\
    & = \frac{1}{2} \left \| \rho - \gamma \sigma \right \|_1 + \frac{(1-\gamma)}{2},
    \label{eq:HS-to-TD}
\end{align}
with $\gamma \geq 1$ for quantum states $\rho$ and $\sigma$ \cite[Eq.~(II.2)]{hirche2023quantum}.
As shown in~\eqref{eq:SDP-positive-part} and~\eqref{eq:dual-positive-part}, the quantity on the right-hand side of~\eqref{eq:hockey-stick-def-trace} can be evaluated by means of an SDP.
Then, auditing QDP reduces to computing $\sE_{\gamma}\big(\cA(\rho)\| \cA(\sigma)\big)$ for $\rho \sim \sigma$. 
However, similar to the previous approach, the time complexity of this SDP also grows exponentially with the number of qubits. Thus, computing these SDPs remains challenging for higher-dimensional quantum systems.

\medskip
\underline{\textbf{Using quantum circuits:}}
\medskip

Another approach is to borrow the results of~\cite{W02,W09zkqa} and use the connection of $\sE_{\gamma}(\rho \Vert \sigma)$ to the trace distance given in~\eqref{eq:HS-to-TD}.  Despite this connection, evaluating ${\sE}_{\gamma}(\rho \Vert \sigma)$
remains computationally challenging, even for quantum computers~\cite{W02, W09zkqa}. Nevertheless, there are proposals for evaluating the trace distance using variational quantum algorithms~\cite{ditinguishabilityMeasure2021estimate,chen2021variational} (which however do not give particular runtimes), and for cases in which the quantum states have low rank~\cite{wang2023fastTraceEstimation}.

In the subsequent analysis, we explore an approach from~\cite{ditinguishabilityMeasure2021estimate,chen2021variational} (via variational algorithms with parameterized quantum circuits) to estimate the quantity $ \left\|\cA(\rho) - e^\varepsilon \cA(\sigma) \right\|_1$. Using such an estimate, for a fixed value of the privacy parameter $\varepsilon$, we can validate on which values of $\delta$ the needed guarantees are satisfied. 
\begin{figure}
    \centering
    \includegraphics[width=\linewidth]{ 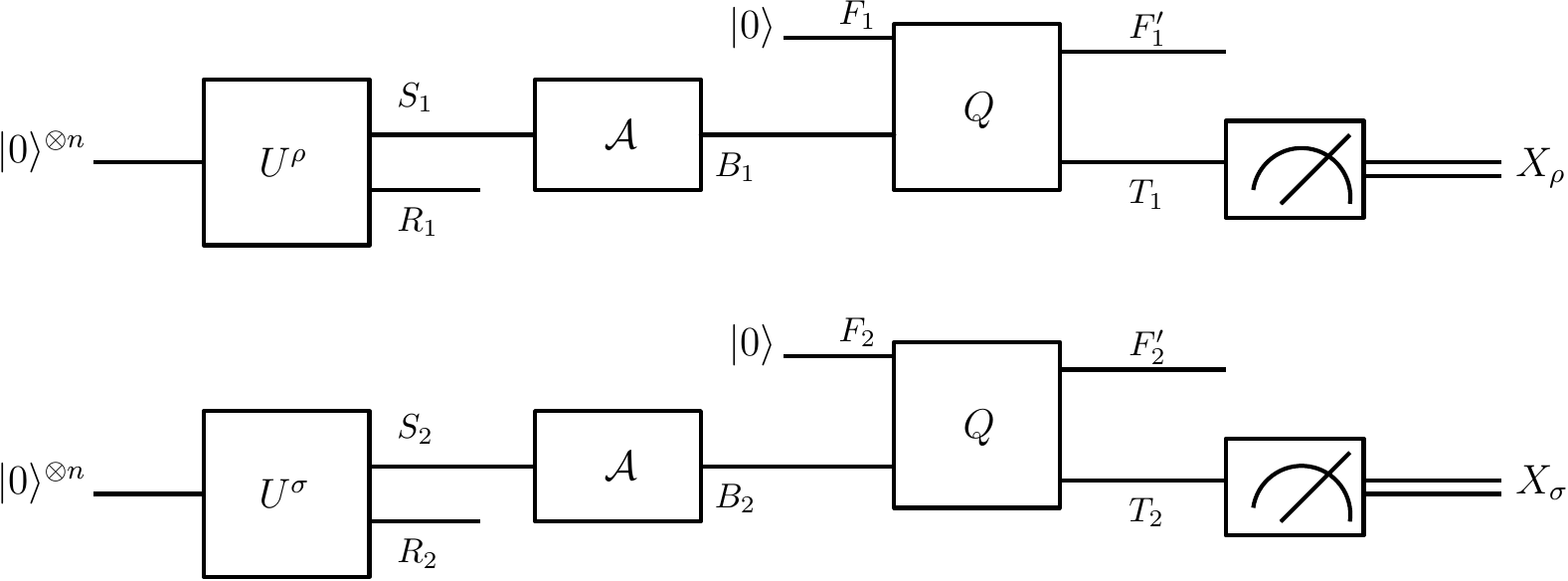}
    \caption{Quantum circuit assisted in estimating QDP: $U^\rho,U^\sigma$ are the unitaries used to prepare $\rho$ and $\sigma$ by tracing out $R_1,R_2$ systems, respectively. Then $\cA$ is applied on the systems $S_i$ for $i\in\{1,2\}$. The unitary $Q$ takes inputs $F_i,B_i$ and outputs $F_i', T_i$, where $F_i$ and $T_i$ are qubit systems. Finally, each of the $T_i$ systems is measured and the (classical) output random variable is denoted as $X_\rho$ for $i=1$ and $X_\sigma$ for $i=2$. Here $X_\rho, X_\sigma \in \{0,1\}$. This procedure is repeated a sufficient number of times, and the outcomes of the trials are used to estimate $\PP(X_\rho=0)$ and $\PP(X_\sigma=1)$.}
    \label{fig:Quantumcircuit_trace_for_QDP}
\end{figure}

Firstly, let us focus on how to estimate $\sE_{e^{\varepsilon}}\!\left( \cA(\rho) \| \cA(\sigma) \right) $ for a fixed $\rho \sim \sigma$ and $\varepsilon$. With the ideas developed in \cite[Algorithm~14]{ditinguishabilityMeasure2021estimate}, we discuss how a process similar to quantum interactive proof can be used for estimating the privacy level. For that, refer to the quantum circuit in \cref{fig:Quantumcircuit_trace_for_QDP}: the unitaries $U^\rho$ and $U^\sigma$ are used to prepare the states $\rho$ and~$\sigma$, by tracing out  systems $R_1$ and $R_2$, respectively. Then the algorithm $\cA$ is applied on the systems $S_i$ for $i\in\{1,2\}$. The unitary $Q$ takes inputs $F_i,B_i$ and outputs $F_i', T_i$, where $F_i$ and $T_i$ are qubit systems. Finally, both of the $T_i$ systems are measured in the standard basis $\{|0\rangle,|1\rangle\}$, and the (classical) output random variable is denoted as $X_\rho$ for $i=1$ and $X_\sigma$ for $i=2$. Here $X_\rho$ and $X_\sigma$ take values in  $ \{0,1\}$. This procedure is repeated a sufficient number of times, and we use the results to estimate $\PP(X_\rho=0)$ and $\PP(X_\sigma=1)$.

Next, consider a scenario in which one could maximize the following utility function over all possible choices of $Q$:
\begin{equation}\label{eq:rhoSigmaUtility}
    g(Q,\rho,\sigma,\cA,\varepsilon) \coloneqq \frac{1}{e^\varepsilon +1} \PP(X_\rho=0) + \frac{e^\varepsilon}{e^\varepsilon +1} \PP(X_\sigma=1) .
\end{equation}
In quantum complexity terminology, this action could be conducted by a quantum prover who has unbounded computational power (we discuss how to relax this assumption in \cref{rem:relaxing-prover-assumption}). From \cite[Eq.~(128)]{ditinguishabilityMeasure2021estimate} and the discussion therein, we conclude that
\begin{align}\label{eq:relationWithTraceQDP}
    f(\rho,\sigma,\cA,\varepsilon) & \coloneqq \sup_{Q} g(Q,\rho,\sigma,\cA,\varepsilon) \\ 
    &= \frac{1}{2} \left( 1+ \frac{1}{e^\varepsilon +1} \left\| \cA(\rho) -e^\varepsilon \cA(\sigma) \right\|_1 \right),
\end{align}
where the optimization is over every unitary $Q$.

If 
\begin{equation}
    f(\rho,\sigma,\cA,\varepsilon) \leq \frac{1}{2}\left( 1+ \frac{2\delta+e^\varepsilon-1}{e^\varepsilon +1}\right),
\end{equation}
then $\sE_{e^{\varepsilon}}\!\left( \cA(\rho) \| \cA(\sigma) \right) \leq \delta $, due to~\eqref{eq:hockey-stick-def-trace}.
Then, 
by changing the role of $\rho$ and $\sigma$ in~\eqref{eq:rhoSigmaUtility}, we obtain $ f(\sigma,\rho,\cA,\varepsilon)$. 
Next, we select the maximum out of these quantities 
\begin{align}
 & {\widehat{f}} (\rho,\sigma,\cA, \varepsilon)  
  \coloneqq  \max\left\{    f(\rho,\sigma,\cA,\varepsilon),  f(\sigma,\rho,\cA,\varepsilon)\right\}.
\end{align}
To this end, if \begin{equation} \label{eq:privacycheck}
    \sup_{\rho \sim \sigma} \widehat{f}(\rho,\sigma,\cA, \varepsilon) 
 \leq \frac{1}{2}\left( 1+ \frac{2\delta+e^\varepsilon-1}{e^\varepsilon +1}\right),
\end{equation}
 it can be verified that $\cA$ is $(\varepsilon, \delta)$-QDP. 

 \begin{remark}[Relaxing the computationally unbounded assumption]\label{rem:relaxing-prover-assumption}
  As the number of qubits increases, classical methods (e.g., using SDPs) often become intractable due to the exponential growth in computational complexity. So in light of this, the above approach is desired with the increasing dimension of the quantum system. 

  However, finding the $Q$ that achieves the optimum utility in~\eqref{eq:rhoSigmaUtility} is practically infeasible. To relax this assumption, we replace the action of the prover (who finds this $Q$) with a parameterized circuit $Q_\theta$. Then we can use~\eqref{eq:rhoSigmaUtility} as a utility function for training a variational quantum algorithm~\cite{CABBEFMMYCC20,bharti2021noisy} to estimate a lower bound for~\eqref{eq:relationWithTraceQDP}. With that we obtain a lower bound on $\sE_{e^{\varepsilon}}\!\left( \cA(\rho) \| \cA(\sigma) \right)$, which we denote as $\sE^{\mathrm{LB}}_{e^{\varepsilon}}\!\left( \cA(\rho) \| \cA(\sigma) \right)$. A lower bound gives a sufficient condition for ruling out algorithms that do not satisfy $(\varepsilon,\delta)$-QDP. This claim follows because the estimated lower bound $\sE^{\mathrm{LB}}_{e^{\varepsilon}}\!\left( \cA(\rho) \| \cA(\sigma) \right) > \delta$ implies that $\sE_{e^{\varepsilon}}\!\left( \cA(\rho) \| \cA(\sigma) \right) > \delta$.
 \end{remark}

 \begin{remark}[Neighboring pairs]
  To verify   $(\varepsilon, \delta)$-QDP, it is required to compute whether $\widehat{f}(\rho,\sigma,\cA, \varepsilon) 
 \leq \frac{1}{2}\left( 1+ \frac{2\delta+e^\varepsilon-1}{e^\varepsilon +1}\right)$ for all neighboring pairs $(\rho,\sigma) \in \cQ$. However, checking this requirement has increasing computational complexity as the cardinality of the set $\cQ$ increases. If the privacy requirements can be relaxed so as to encode domain knowledge as in QPP framework, the effective set $\cQ$ may be a small set in some applications of interest. {For an example, consider an application where hypothesis testing is carried out between the states $\rho$ and $\sigma$ under privacy constraints where $\cQ=\{ (\rho,\sigma), (\sigma,\rho)\}$}. 
\end{remark}

\begin{remark}[Auditing QPP\footnote{Note that the following ideas can also be extended to auditing for variants of QPP in \cref{sec:variants-of-QPP} as well (See also \cref{rem:auditing-variants}).}]\label{rem:QPP_auditing}
    To audit for $(\varepsilon,\delta)$-QPP in the framework $(\cS,\cQ,\Theta,\bar{\cM})$, one can use the ideas described above by choosing $\rho^\cR$ and $\rho^\cT$ for all $P_X \in \Theta, (\cR,\cT) \in \cQ$ instead of $\rho$ and $\sigma$. In that case, the complexity of the approach relies on the set of $\cQ$ as well as distributions contained in $\Theta$. 
\end{remark}

\subsection{Formal Guarantees for Auditing QDP}

The question of quantifying the success of a privacy auditing approach, specifically in correctly accepting and rejecting an algorithm with given privacy requirements, is a crucial consideration in privacy auditing research.
The authors of~\cite{domingo2022auditing,nuradha2022pufferfishJ}  have worked towards answering this question for auditing classical DP, but using a relaxed privacy definition. 

To tackle this for the quantum setting, we propose a hypothesis testing-based auditing pipeline tailored specifically for QDP (also for QPP). In this pipeline, we use the trace-norm estimation quantum algorithm proposed in~\cite{wang2023fastTraceEstimation}. This quantum algorithm provides an estimation with at most $\alpha$ additive error from the exact value, with high probability, which allows us to achieve the desired significance in the hypothesis test.

Let us define
\begin{equation}
 \sT^\varepsilon(\rho,\sigma,\cA)\coloneqq \frac{1}{e^\varepsilon+1} \left\| \cA(\rho) - e^\varepsilon \cA(\sigma) \right \|_1.  
\end{equation}
Fix $(\rho,\sigma) \in \cQ$.
If algorithm $\cA$ satisfies $(\varepsilon,\delta)$-QDP, then 
$\sE_{e^{\varepsilon}}\!\left( \cA(\rho) \| \cA(\sigma) \right) \leq \delta$. By applying~\eqref{eq:hockey-stick-def-trace}, we arrive at 
\begin{equation} \sT^\varepsilon(\rho,\sigma,\cA) \leq \frac{2 \delta +e^\varepsilon -1}{e^\varepsilon +1} \eqqcolon \sg(\varepsilon,\delta).\end{equation}
Next, by estimating the quantity on the left-hand side, and using $\sg(\varepsilon,\delta)$ as a threshold, we design an auditing pipeline for QDP by means of the following null and alternative hypotheses:
\begin{align}
    H_0 &: \max \{ \sT^\varepsilon(\rho,\sigma,\cA), \sT^\varepsilon(\sigma,\rho,\cA) \} \leq \sg(\varepsilon,\delta), \\
     H_1 &:  \max \{ \sT^\varepsilon(\rho,\sigma,\cA), \sT^\varepsilon(\sigma,\rho,\cA) \} > \sg(\varepsilon,\delta) .
\end{align}
Let the estimates of $\sT^\varepsilon(\rho,\sigma,\cA)$ and $ \sT^\varepsilon(\sigma,\rho,\cA) $ from a randomized algorithm (in our analysis we use the algorithm corresponding to \cite[Corollary~3.4]{wang2023fastTraceEstimation}) be $\hat{\sT}^\varepsilon(\rho,\sigma,\cA)$ and $ \hat{\sT}^\varepsilon(\sigma,\rho,\cA) $, respectively. 
We choose
\begin{equation}
\hat{\sT}^\varepsilon_{\max}(\rho,\sigma,\cA)\coloneqq  \max\{ \hat{\sT}^\varepsilon(\rho,\sigma,\cA), \hat{\sT}^\varepsilon(\sigma,\rho,\cA) \}    
\end{equation}
as our test statistic. 

\medskip

\underline{\textbf{Estimation of the test statistic:}}

\begin{lemma}[Estimating trace distance using samples of $\rho$, $\sigma$ --- restatement of {\cite[Corollary~3.4]{wang2023fastTraceEstimation}}] \label{lem: Estimating trace distance improved}
    Given access to identical copies of $d$-dimensional quantum states $\rho$ and $\sigma$, there is a quantum algorithm that estimates the normalized trace distance $\sT(\rho,\sigma)$ (recall~\eqref{eq:normalized-TD}) within additive error $\alpha$  and with probability not less than $1-\beta$, by using 
    \begin{equation}
    O\!\left( \log\!\left( \frac{1}{\beta} \right)\frac{r^2}{\alpha^5} \log^2\!\left( \frac{r}{\alpha}\right) \log^2\!\left( \frac{1}{\alpha}\right)\right)\end{equation} samples of $\rho$ and $\sigma$,
    where $r$ is an upper bound on the rank of $\rho$ and $\sigma$.
\end{lemma}
\cref{lem: Estimating trace distance improved} is obtained by using the existing result in Corollary~3.4 of~\cite{wang2023fastTraceEstimation}, and combining its argument in Theorem~2.6 therein on estimating  $\Tr\!\left[A\rho\right]$ within additive error $\alpha$ with probability $1-\beta$, by using $O\!\left( \frac{1}{\alpha^2}\log\!\left(\frac{1}{\beta}\right)\right)$ identical samples of~$\rho$. The algorithm proposed in~\cite{wang2023fastTraceEstimation} is designed based on the following idea. 
Let $V \coloneqq (\rho-\sigma)/2$. Consider its singular value decomposition as $V=W \Sigma U^\dagger $. Then the trace distance can be expressed by the following identity: 
\begin{equation}
\sT(\rho,\sigma) = \frac{1}{2} \left( \Tr\!\left[\rho \, \mathrm{sgn}(V) \right]-  \Tr\!\left[\sigma \, \mathrm{sgn}(V) \right]\right),
\end{equation}
where $\mathrm{sgn}(V) \coloneqq W \mathrm{sgn}(\Sigma) U^\dagger$, and $\mathrm{sgn}(\cdot)$ is the sign function. Then, $\Tr\!\left[\rho \, \mathrm{sgn}(V) \right]$ and $ \Tr\!\left[\sigma \, \mathrm{sgn}(V) \right]$ can be estimated separately, by combining the techniques of quantum singular value transformation~\cite{GSLW19} and the Hadamard test~\cite{AJL09}. To this end, to implement unitary block-encodings of $\rho$ and $\sigma$ approximately, the technique of density matrix exponentiation~\cite{Lloyd2014QuantumAnalysis} is used.

The same techniques can be employed to compute $\sT^\varepsilon(\rho,\sigma,\cA)$ since 
\begin{equation} \sT^\varepsilon(\rho,\sigma,\cA) = \\
\frac{1}{e^\varepsilon +1} \left( \Tr\!\left[\cA(\rho) \ \mathrm{sgn}(\rV^\varepsilon) \right]-  e^\varepsilon\Tr\!\left[\cA(\sigma) \ \mathrm{sgn}(\rV^\varepsilon) \right]\right),
\end{equation}
where $\rV^\varepsilon \coloneqq (\cA(\rho)- e^\varepsilon \cA(\sigma))/(e^\varepsilon +1) $.

\medskip 
\underline{\textbf{Type-1 error analysis}}

We arrive at the following bound on the Type-1 error of the proposed hypothesis testing pipeline. 

\begin{proposition}[Type-I error] \label{prop:type I error improved} 
Fix arbitrary $\alpha,\delta>0$ and consider the above hypothesis testing pipeline. Then 
\begin{equation} \sup_{\rho \sim \sigma} \PP\!\left(  \hat{\sT}^\varepsilon_{\max}(\rho,\sigma,\cA) > \sg(\varepsilon,\delta) +\alpha\, \middle| \,H_0\right) \leq \beta,
\end{equation}
if the algorithm from \cref{lem: Estimating trace distance improved} has access to 
\begin{equation}
O\!\left( \log\!\left( \frac{1}{\beta} \right)\frac{r^2}{\alpha^5} \log^2\!\left( \frac{r}{\alpha}\right) \log^2\!\left( \frac{1}{\alpha}\right)\right)
\end{equation}
identical copies of the states $\rho$ and $\sigma$, such that $\rho \sim \sigma$ and
where
\begin{equation}
   r \coloneqq \sup_{\rho \sim \sigma} \max\{\mathrm{rank}\!\left(\cA(\rho)\right), \mathrm{rank}\left(\cA(\sigma)\right)\}.
\end{equation}

\end{proposition}

\begin{IEEEproof}
   Fix $\rho$ and $\sigma$ such that $\rho \sim \sigma$. 
Under the null hypothesis and the assumption that $ \hat{\sT}^\varepsilon_{\max}(\rho,\sigma,\cA)={\hat{\sT}}^\varepsilon(\rho,\sigma,\cA)$, we have that
\begin{align}
    & \PP\!\left(  \hat{\sT}^\varepsilon_{\max}(\rho,\sigma,\cA) > \sg(\varepsilon,\delta) +\alpha \right)  \notag \\
    &=\PP\!\left( {\hat{\sT}}^\varepsilon(\rho,\sigma,\cA) > \sg(\varepsilon,\delta) +\alpha \right)  \\
    &\stackrel{(a)}\leq \PP\!\left( {\hat{\sT}}^\varepsilon(\rho,\sigma,\cA) - {{\sT}}^\varepsilon(\rho,\sigma,\cA)  > \alpha \right) \\ 
    & \leq \PP\!\left(  |{\hat{\sT}}^\varepsilon(\rho,\sigma, \cA) - {{\sT}}^\varepsilon(\rho,\sigma, \cA) | > \alpha \right) \\ 
    & \stackrel{(b)}\leq \beta,
\end{align}
where: (a) follows since ${{\sT}}^\varepsilon(\rho,\sigma,\cA) \leq  \sg(\varepsilon,\delta)$ under the null hypothesis; (b) from the high probability statement in \cref{lem: Estimating trace distance improved}. 
Similarly, the above inequality holds when $ \hat{\sT}^\varepsilon_{\max}(\rho,\sigma,\cA)={\hat{\sT}}^\varepsilon(\sigma,\rho,\cA)$ concluding the proof.
\end{IEEEproof}

\medskip

\cref{prop:type I error improved} provides a bound on the number of samples of the states required to achieve type-I error (significance) of $\beta$. In that case, we would use a threshold of $g(\varepsilon,\delta) + \alpha$ for accepting the null hypothesis, such that the null hypothesis is accepted when the test statistic is less than or equal to $g(\varepsilon,\delta) + \alpha$.

\begin{remark}[Computational complexity with rank $r$]
From \cref{prop:type I error improved}, it is evident that the copy complexity of the algorithm grows as $O(r^2 \log^2(r))$. 
Let $\cA: \cL(\cH_A) \to \cL(\cH_B)$ be a quantum channel. Then, $r \leq d_B$, where $d_B$ is the dimension of the Hilbert space~$\cH_B$. To handle computational complexity, one possibility is to compose $\cA$ with another quantum channel $\cN$ that translates the space to a low-dimensional setting. However, due to the data-processing inequality for the trace distance, it will only provide a lower bound. With that, it is possible to reject algorithms if $\sT^\varepsilon(\rho,\sigma,\cN \circ \cA) > g(\varepsilon,\delta) $, which implies that $\sT^\varepsilon(\rho,\sigma,\cA) > g(\varepsilon,\delta) $. Consequently, it may lead to limitations similar to classical auditing approaches that use relaxed privacy notions, since the contraction gap between $\sT^\varepsilon(\rho,\sigma,\cN \circ \cA)$, and $\sT^\varepsilon(\rho,\sigma,\cA)$ is hard to quantify. In the quantification of the gap, finding the contraction coefficient $\eta_\cN$ of the channel $\cN$ would be useful if $\eta_\cN <1$ {(recall that {the} contraction coefficient of a channel $\cN$ under a generalized divergence $\boldsymbol{\sD}$, as given in \cref{eq:generalized-divergence}, is defined as 
$
   \eta_\cN \coloneqq  \sup_{\rho,\sigma \in \cD , \boldsymbol{\sD}(\rho \Vert \sigma) \neq 0}\frac{\boldsymbol{\sD}\left( \cN(\rho) \Vert \cN(\sigma)\right)}{\boldsymbol{\sD}(\rho \Vert \sigma)}
$)}. 
\end{remark}

In summary, we proposed a hypothesis testing pipeline for auditing the privacy of quantum systems, offering formal guarantees on auditing QDP using quantum algorithms designed for estimating trace distance. However, an essential task for further investigation is analyzing the Type-II error of this approach. This analysis would allow us to quantify the power of the test and assess its ability to correctly accept algorithms with the desired privacy requirements, which is left for future work.  

\section{Information-Theoretic Bounds from QPP and Connections to Quantum Fairness}

\label{Sec:Connections-to-the-Tools-of-Interest-to-Quantum-community}

In this section, we begin by investigating several information-theoretic bounds that stem from an algorithm satisfying QPP constraints. We then establish connections between QPP and quantum fairness~\cite{Fairness_quantum21,fairnessQ_verifying22} using these bounds. 
Later on we utilize the derived bounds to assess the relative strength of the QPP variants introduced in \cref{sec:variants-of-QPP}. 
\subsection{Information-Theoretic Bounds from QPP}
In~\cite{hirche2023quantum}, it was highlighted that finding bounds on quantum relative entropy and mutual information resulting from QDP is an interesting open problem. We address this question in a general setting, encompassing QDP as a special case. We offer bounds for relative entropy and Holevo information, along with bounds for R\'enyi relative entropies and trace distance. For the rest of the discussion, we adopt the fixed privacy framework to be $(\cS,\cQ,\Theta,\bar{\cM})$.

\begin{proposition}[Bounds on quantum R\'enyi relative entropy and quantum relative entropy due to QPP]
\label{prop: Bounds on Quantum divergences with QPP}
    Fix $\alpha >1$. If $\cA$ is $\varepsilon$-QPP {(i.e., $(\varepsilon,0)$-QPP)} in the framework $(\cS,\cQ,\Theta,\bar{\cM})$ for all $P_X \in \Theta$ and $(\cR,\cT) \in \cQ$, then 
    \begin{equation}
        \sD_\alpha\!\left( \cA(\rho^\cR) \| \cA(\rho^\cT) \right) \leq \min\! \left\{ \frac{\varepsilon^2 \alpha}{2}, \varepsilon \right\},
    \end{equation}
    where $\sD_\alpha(\cdot \Vert \cdot)$ is an arbitrary quantum R\'enyi relative entropy satisfying data processing.\footnote{Note that Petz--R\'enyi in~\eqref{eq:petz renyi} satisfies data processing for $\alpha \in (0,1) \cup (1,2]$ and sandwiched R\'enyi in~\eqref{eq:sandwiched-renyi-def} satisfies data processing for $\alpha \in [1/2,1) \cup (1,\infty)$.}
    Furthermore, 
    \begin{equation}
       \sD\!\left( \cA(\rho^\cR) \| \cA(\rho^\cT) \right) \leq  \min\! \left\{ \frac{\varepsilon^2}{2}, \varepsilon \right\},
    \end{equation}
    where $\sD$ is the quantum relative entropy in~\eqref{eq: limit alpha 1}.
\end{proposition}

\begin{IEEEproof}
    $\varepsilon$-QPP of the framework $(\cS,\cQ,\Theta,\bar{\cM})$ implies that for all $P_X \in \Theta$ and $(\cR,\cT) \in \cQ$
    \begin{equation}
    \sD_{T}\!\left(\cA(\rho^\cR) \Vert \cA(\rho^\cT) \right) \leq \varepsilon, \label{eq: equivalence-of-epsilon-QPP}
    \end{equation}
   where $\sD_{T}$ is the Thompson metric from~\eqref{eq:Thompson-metric-def}. {This  follows by the definition of the max-relative entropy defined in~\eqref{eq:D-max-def_ALT} and Thompson metric in~\eqref{eq:Thompson-metric-def}, and recalling the definition of QPP framework for $\delta=0$ and $\cM= \bar{\cM}$ (see \cref{def: qpp}).}
By this implication and the fact that every quantum R\'enyi-divergence $\sD_\alpha$ of order $\alpha$ satisfying data processing is bounded from above by $\sD_{\max}$ \cite[Eq.~(4.36)]{tomamichel2015quantum}, $\varepsilon$-QPP implies that for all $P_X \in \Theta$ and $ (\cR,\cT) \in \cQ$,
\begin{equation}
\sD_\alpha\!\left( \cA(\rho^\cR) \| \cA(\rho^\cT) \right) \leq  \varepsilon.
\end{equation}
By a different argument via the maximal extension presented in \cref{lem:Bound-on-Renyi}, we can further arrive at the following, for all $P_X \in \Theta$,  $(\cR,\cT) \in \cQ$ and $\varepsilon \alpha <2$,
\begin{equation}
\sD_\alpha\!\left( \cA(\rho^\cR) \| \cA(\rho^\cT) \right)
\leq  \frac{\varepsilon^2 \alpha}{2}.
\end{equation}
This completes the proof of the first inequality.

Next, noting that the Petz--R\'enyi relative entropy in~\eqref{eq:petz renyi} satisfies data processing for $\alpha \in (0,1) \cup (1,2]$, 
and then taking the limit $\alpha \to 1^{+}$, we arrive at the bound on quantum relative entropy by using the equality in~\eqref{eq: limit alpha 1}.
\end{IEEEproof}

\begin{lemma}\label{lem:Bound-on-Renyi}
    Fix $\alpha >1$, and $\rho,\sigma$ PSD operators. For $\alpha \mathsf{D}_T(\rho \Vert \sigma) \leq 2$, the following inequality holds: 
    \begin{equation}
       \sD_\alpha(\rho \| \sigma) \leq \frac{\alpha}{2} (\sD_T(\rho \Vert \sigma))^2,
    \end{equation}
where $\mathsf{D}_\alpha$ is an arbitrary quantum R\'enyi relative entropy satisfying data processing.
\end{lemma}
\begin{IEEEproof}
See Appendix~\ref{proof:bound-on-quantum-renyi}.    
\end{IEEEproof}

\begin{remark}[Operational interpretation of Thompson metric]
    An operational interpretation of the Thompson metric
    has appeared in symmetric postselected hypothesis testing (a setting allowing for an inconclusive outcome along with two general conclusive outcomes and postselecting on the conclusive outcomes) as the asymptotic error exponent of discriminating two quantum states $\rho$ and $\sigma$~\cite{regula2022postselected}, as well as in the resource theory of symmetric distinguishability~\cite{salzmann2021symmetric}. Here by referring to~\eqref{eq: equivalence-of-epsilon-QPP}, the QPP framework also provides another operational interpretation of the Thompson metric.
   In the framework $(\cS,\cQ,\Theta, \bar{\cM})$, for fixed $P_X \in \Theta$ and $(\cR,\cT)~\in~\cQ$,\footnote{by definition also for $(\cT,\cR) \in \cQ$} the Thompson metric given by $\sD_{T}\!\left(\cA(\rho^\cR) \Vert \cA(\rho^\cT)\right)$
  is equal to the minimal $\varepsilon$ needed to achieve $\varepsilon$-QPP.
\end{remark}

\begin{proposition}[Bounds on trace norm]\label{prop:Bounds on trace norm}
If $\cA$ is $\varepsilon$-QPP, then
    \begin{equation}
    \sup_{\substack{\Theta, 
    (\cR,\cT) \in \cQ}} \left \| \cA(\rho^\cR) -\cA(\rho^\cT) \right \|_1  
    \leq \min\! \left\{ \varepsilon, \sqrt{2\varepsilon} \right\},
    \end{equation}
and if $\cA$ is $(\varepsilon,\delta)$-QPP, 
   we have
    \begin{equation}
    \sup_{\Theta, (\cR,\cT) \in \cQ}  \left\| \cA(\rho^\cR) -\cA(\rho^\cT) \right \|_1 
    \leq 2- \frac{4(1-\delta)}{e^\varepsilon +1}.
    \end{equation} 
\end{proposition}

\begin{IEEEproof}
    The first inequality holds by applying the quantum Pinsker inequality ($\frac{1}{2} \left \| \rho -\sigma \right\|^2_1 \leq \sD(\rho \| \sigma)$) \cite[Theorem~1.15]{ohya1993quantum} and \cref{prop: Bounds on Quantum divergences with QPP}. 

    For the second inequality: $(0,\delta')$-QPP is equivalent to 
    \begin{equation}
       \sup_{\Theta, (\cR,\cT) \in \cQ}  \frac{\left\| \cA(\rho^\cR) -\cA(\rho^\cT) \right \|_1 }{2} \leq \delta'.
    \end{equation}
    Then, adapting  \cref{lem:strength_QPP} and fixing~$\varepsilon'$ therein to zero leads to the desired result.  
\end{IEEEproof}

\begin{lemma}\label{lem:strength_QPP}
    Fix $(\cS,\cQ,\Theta,{\cM})$ privacy framework. Then, we have 
    \begin{equation}
        (\varepsilon,\delta)\textnormal{-QPP} \implies   (\varepsilon',\delta')\textnormal{-QPP},
    \end{equation}
   { where $\varepsilon' < \varepsilon$ with
    \begin{equation}\label{eq:delta_prime}
    \delta' \coloneqq 1- \frac{(e^{\varepsilon'}+1)(1-\delta)}{(e^\varepsilon+1)}. 
    \end{equation}}
\end{lemma}
\begin{IEEEproof}
    The proof  follows similarly to the proof of \cite[Property~3]{CY16} for classical DP and is presented in Appendix~\ref{app:lemma_strength_proof}.
\end{IEEEproof}

\begin{remark}[Bounds on Holevo information]
    In Appendix~\ref{app:Bounds_Holevo}, we provide bounds on the Holevo information in the settings of QDP and QLDP (recall this privacy notion from \cref{rem:QLDP}).
\end{remark}

\subsection{Quantum Fairness and QPP}
We now demonstrate that quantum fairness can be viewed as a special case of QPP, which should encourage the design of customized fairness models via the QPP framework. 

Quantum fairness seeks to treat all input states equally, meaning that all pairs of input states that are close in some metric (e.g., close in trace distance) should  yield similar outcomes 
when processed by an algorithm~\cite{fairnessQ_verifying22}.
For a quantum decision model $\cA=\{\cE, \{\rM_i\}_{i \in \cO} \}$, where a quantum channel $\cE$ is followed by a POVM $\{\rM_i\}_{i \in \cO}$ (i.e., a quantum algorithm described by a quantum to classical channel), quantum fairness is defined in~\cite{fairnessQ_verifying22} as follows.

\begin{definition}[$(\alpha,\beta)$-fairness~\cite{fairnessQ_verifying22}]
Suppose we are given a quantum decision model $\cA=\{\cE, \{\rM_i\}_{i \in \cO} \}$, two distance metrics $D(\cdot \| \cdot)$ and $d(\cdot \| \cdot)$ on $\cD(\cH)$ and $\cD(\cO)$ respectively. Fix $0 < \alpha,\beta \leq 1$. Then the decision model $\cA$ is $(\alpha,\beta)$ fair if for all $\rho,\sigma \in \cD(\cH)$ such  that $D(\rho \| \sigma) \leq \alpha$, then
\begin{equation}
d\!\left( \cA(\rho)\| \cA(\sigma) \right) \leq \beta.
\end{equation}
\end{definition}

\begin{proposition}[Fairness guarantee from QPP]\label{rem:Privacy-implying-fairness}
     Let $D(\rho \| \sigma)= \left \| \rho- \sigma \right\|_1 /2$ and $d\!\left( \cA(\rho)\| \cA(\sigma) \right)= \frac{1}{2} \sum_i \!\left| \Tr\!\left[\rM_i \cE(\rho-\sigma) \right]\right |$. Fix  
     \begin{equation} \label{eq:fairness-QPP-framework}
   \begin{aligned}
    \cS&=\{ \rho: \rho \in \cD(\cH) \}, \\
    \cQ&= \{(\rho,\sigma) : \rho,\sigma \in \cD(\cH), \ D(\rho\| \sigma) \leq \alpha \} , \\
    \Theta&={ \cP_2\big(\cD(\cH)\big)} ,\\
    \cM&=\{\rM: 0 \psd \rM  \psd \rI \}.
\end{aligned} 
\end{equation}
    If $\cE$ satisfies $\varepsilon$-QPP with $(\cS,\cQ,\Theta,{\cM})$ above, then $\cA=\{\cE, \{\rM_i\}_{i \in \cO} \}$ is $(\alpha, \sqrt{\varepsilon'/2})$-fair, where $\varepsilon'= \min\{\varepsilon,\varepsilon^2/2\}$.
\end{proposition}
\begin{IEEEproof}
    From \cref{prop:Bounds on trace norm}, $\cE$ being $\varepsilon$-QPP implies
    \begin{equation}
     \| \cE(\rho) - \cE(\sigma) \|_1 \leq \min\{\varepsilon, \sqrt{2 \varepsilon}\} = \sqrt{2\varepsilon'},   
    \end{equation}
    for $\rho$ and $\sigma$ such that $D(\rho \| \sigma) \leq \alpha$. 
    
    Then, consider the measurement channel 
    that performs the following transformation:
    \begin{equation}
    \cE(\rho) \to \sum_{i \in \cO} \Tr\!\left[\rM_i \cE(\rho)\right] |i \rangle\!\langle i|.
    \end{equation}
    It follows from the data-processing inequality for the trace distance that 
    \begin{equation}
    \left \| \sum_{i \in \cO} \left( \Tr\!\left[\rM_i \cE(\rho)\right]  -  \Tr\!\left[\rM_i \cE(\sigma)\right] \right) |i \rangle\!\langle i| \right \|_1 \leq \sqrt{2 \varepsilon'}.
    \end{equation}
    This leads to 
    \begin{equation}
    d\!\left( \cA(\rho)\| \cA(\sigma) \right)= \frac{1}{2} \sum_i \left| \Tr\!\left[\rM_i \cE(\rho-\sigma) \right] \right | \leq \sqrt{\frac{\varepsilon'}{2}},\end{equation}
    concluding the proof. 
\end{IEEEproof}

\medskip
This shows how $\varepsilon$-QPP also provides fairness to the decision models of interest. However, fairness guarantees may not be sufficient to guarantee privacy in general. Next, we show that when $\cM$ satisfies a relationship related to the POVM of the quantum decision model where fairness needs to be ensured, fair algorithms also act as privacy mechanisms.

\begin{proposition}[Privacy guarantees obtained from fairness]
    Consider the same privacy framework as in~\eqref{eq:fairness-QPP-framework} but with the modification $\cM=\{ \cup_{i \in \cB} \rM_i| \cB \subseteq \cO\}$. If $\cA=\{\cE, \{\rM_i\}_{i \in \cO} \}$ is $(\alpha,\beta)$-fair, then $\cE$ is $(\varepsilon, 2 \beta)$-QPP for every $\varepsilon \geq 0$. 
\end{proposition}

\begin{IEEEproof}
 Since $\cA$  is $(\alpha,\beta)$-fair, it follows that 
 \begin{equation}\frac{1}{2} \sum_{i \in \cO} \left| \Tr\!\left[\rM_i \cE(\rho-\sigma) \right] \right | \leq \beta. \numberthis \label{eq:fairness guarantee}\end{equation}
 Then, for every $\cB \subseteq \cO$
 \begin{align}
 \left| \Tr\!\left[\sum_{i \in \cB} \rM_i \cE(\rho-\sigma) \right] \right |  & \leq   \sum_{i \in \cB} \left | \Tr\!\left[\rM_i \cE(\rho-\sigma) \right] \right |  \\
 & \leq \sum_{i \in \cO} \left | \Tr\!\left[\rM_i \cE(\rho-\sigma) \right] \right |\\
 & \leq 2\beta,
 \end{align} 
 where the first inequality follows from the triangular inequality, the second from $\cB \subseteq \cO$, and the last from~\eqref{eq:fairness guarantee}.
\end{IEEEproof}

\section{Variants of Quantum Pufferfish Privacy Framework} 

\label{sec:variants-of-QPP}

We now propose variants of QPP via generalized divergences (as defined in~\eqref{eq:generalized-divergence}).\footnote{Due to the definition of generalized divergences, the privacy notions defined based on them inherently satisfy post-processing.}
We provide an operational interpretation of generalized divergences as privacy metrics and characterize the relative strength between them.

Here, we focus on QPP  in the framework $(\cS,\cQ,\Theta,\bar{\cM})$
and formulate QPP based on generalized divergences, which we denote by $(\boldsymbol{\mathsf{D}}, \varepsilon)$-QPP, where $\boldsymbol{\mathsf{D}}$ is a placeholder for the generalized divergence being used.

\begin{definition}[$(\boldsymbol{\mathsf{D}}, \varepsilon)$-QPP] \label{def: D-QPP}
Fix a privacy framework $(\cS,\cQ,\Theta,\bar{\cM})$ and $\varepsilon > 0$.
An algorithm $\cA$ is $(\boldsymbol{\mathsf{D}}, \varepsilon)$-QPP if
\begin{equation} \sup_{\Theta,(\cR,\cT) \in \cQ} \boldsymbol{\sD} \!\left( \cA(\rho^\cR) \| \cA(\rho^\cT)\right) \leq \varepsilon,\end{equation}
where $\boldsymbol{\sD}$ is an arbitrary generalized divergence  and $\rho^\cR$  and $\rho^\cT$ are defined as in \cref{def: qpp}. Note that these two density matrices depend on elements of $\Theta$ and $\cQ$.
\end{definition}

Indeed, \cref{def: D-QPP} encompasses variants of classical DP~\cite{mironov2017Renyi, bun2016concentrated} and QDP~\cite{hirche2023quantum}, which rely on R\'enyi divergences as the generalized divergence along with the appropriate choice of  QPP framework $(\cS,\cQ,\Theta,\cM)$ stated in \cref{rem:classical-PP} and \cref{rem: Quantum DP set}, respectively.

\begin{remark}[Properties of $(\boldsymbol{\sD},\varepsilon)$-QPP] \label{rem: properties of D-QPP}
Post-processing holds, by the definition of generalized divergences. 
Convexity follows if $\boldsymbol{\sD}$ satisfies the direct-sum property, as defined in \cite[Eq.~(4.3.7)]{khatri2020principles}, from which it follows that  $\boldsymbol{\sD}$ is jointly convex \cite[Proposition~4.15]{khatri2020principles}.

Parallel composability: If $\cA_i$  satisfies $(\varepsilon_i,\delta_i)$-QPP in $(\cS,\cQ,\Theta,\bar{\cM})$ for $i\in\{1,\ldots,k\}$, then the composed mechanism, as defined in \cref{thm:QPP_properties},
    satisfies $\left(\boldsymbol{\sD}, \sum_{i=1}^k \varepsilon_i\right)$-QPP in the framework $\left(\cS, \cQ^{(k)},\Theta, \bar{\cM}^k \right)$, if $\boldsymbol{\sD}$ satisfies subadditivity (i.e., if $\boldsymbol{\sD}(\rho_1 \otimes \rho_2 \Vert \sigma_1 \otimes \sigma_2) \leq \boldsymbol{\sD}(\rho_1  \Vert \sigma_1 ) + \boldsymbol{\sD}( \rho_2 \Vert  \sigma_2)$ for all states $\rho_1$, $\rho_2$, $\sigma_1$, and $\sigma_2$).
    It is worth noting that by employing this privacy notion based on generalized divergences, we achieve improved composability results, even in scenarios involving joint measurements (recall property~3 of \cref{Cor: Properties of QPP all measurements}). 
\end{remark}

\begin{remark}[Auditing variants of QPP] \label{rem:auditing-variants}
   The methodologies proposed in \cref{Sec:Auditing-Privacy-Frameworks} can be used to audit the variants of QPP (based on generalized divergences) as well. In this regard, quantum algorithms and procedures for estimating respective generalized divergences (e.g., R\'enyi relative entropies) would be useful. 
   This motivates the development of novel techniques for estimating them efficiently and accurately, beyond those already established in~\cite{wang2022newEstimateDistances}. 
\end{remark}

\subsection{Variants Based on R\'enyi Divergences}

First, let us recall definitions of  the following quantities. The measured R\'enyi divergence of order $\alpha \in (0,1) \cup(1,\infty)$ is defined  as \cite[Eqs.~(3.116)--(3.117)]{F96} 
\begin{equation}
\label{eq:measured-divergence}
{ \check{\sD}}_\alpha ( \rho \| \sigma) \coloneqq  \sup_{\cM} \sD^c_\alpha \!\left( \cM(\rho) \| \cM(\sigma)\right),\end{equation}
where
\begin{equation}
    \sD^c_\alpha(\cM(\rho) \| \cM(\sigma)) \coloneqq~\frac{1}{\alpha-1} \ln\!\left(\sum_{x \in \cX} \!\left(p(x)\right)^\alpha \!\left( q(x) \right)^{1-\alpha} \right),
\end{equation}
with $p(x) \coloneqq \Tr[\rM_x \rho]$ and $q(x) \coloneqq \Tr[\rM_x \sigma]$ for $\cM$ corresponding to a POVM $\{ \rM_x\}_{x \in \cX}$. The R\'enyi preparation divergence of order $\alpha \in (0,1) \cup(1,\infty)$ is defined as~\cite{Matsumoto2018}
\begin{equation}
\label{eq:preparation-divergence}
{ \hat{\sD}}_\alpha ( \rho \| \sigma) \coloneqq  \inf_{P,Q,\cP} \sD^c_\alpha ( P \| Q ),\end{equation}
where $\cP$ is a classical--quantum channel, $\cP(P)=\rho$, $\cP(Q)=\sigma$, and the classical R\'enyi divergence is defined as
\begin{equation}
    \sD^c_\alpha(P \| Q) \coloneqq~\frac{1}{\alpha-1} \ln\!\left(\sum_{x \in \cX} \!\left(P(x)\right)^\alpha \!\left( Q(x) \right)^{1-\alpha} \right).
\end{equation}
The quantities in~\eqref{eq:measured-divergence} and~\eqref{eq:preparation-divergence} satisfy the data-processing inequality for all $\alpha \in (0,1)\cup(1,\infty)$ by construction, following from this property holding for the underlying classical divergence.
Moreover, the following bounds hold
\begin{equation} \label{eq:Maximal_minimal_D_relation}
    \check{\sD}_\alpha ( \rho \| \sigma) \leq {\sD}_\alpha ( \rho \| \sigma)  \leq { \hat{\sD}}_\alpha ( \rho \| \sigma).
\end{equation}
where $ {\sD}_\alpha$ is an arbitrary quantum  R\'enyi divergence that satisfies data processing \cite[Eq.~(3.7)]{hiai2017different}. 

\begin{remark}[Relative strength of privacy metrics]
  Choosing the preparation divergence $\hat{\sD}_\alpha =\boldsymbol{\sD}$ in $(\boldsymbol{\sD},\varepsilon)$-QPP gives {a stronger} privacy metric that satisfies the post-processing property for the family of quantum R\'enyi divergences of order~$\alpha$, while $\check{\sD}_\alpha=\boldsymbol{\sD}$ gives {a weaker} privacy metric from that same family of divergences. That is, we have that
\begin{align}
& \sup_{\Theta,(\cR,\cT) \in \cQ} \check{\sD}_\alpha \!\left( \cA(\rho^\cR) \| \cA(\rho^\cT)\right) \notag \\
& \leq \sup_{\Theta,(\cR,\cT) \in \cQ} \sD_\alpha \!\left( \cA(\rho^\cR) \| \cA(\rho^\cT)\right) \\
& \leq \sup_{\Theta,(\cR,\cT) \in \cQ} \hat{\sD}_\alpha\!\left( \cA(\rho^\cR) \| \cA(\rho^\cT)\right), 
\end{align} 
so that
\begin{equation}
(\hat{\sD}_\alpha,\varepsilon)\textnormal{-QPP} \implies (\sD_\alpha,\varepsilon)\textnormal{-QPP} \implies (\check{\sD}_\alpha,\varepsilon)\textnormal{-QPP}.
\end{equation}
{The above relations follow by the direct application of~\eqref{eq:Maximal_minimal_D_relation} and \cref{def: D-QPP}.}
\end{remark}

\begin{remark}[Operational interpretation as privacy metrics]
Choose
\begin{align}
    \cS&=\{ \rho,\sigma\}, \\
    \cQ&= \{(\rho,\sigma)  \},  \\
    \Theta&=\{ \{ P_X(x), \rho^x \}_{x \in \cX}:  P_X \in \cP(\cX), \  \rho^x \in \{ \rho, \sigma \}\},  \\
    \cM&= \bar{\cM}.
\end{align}
The strongest privacy metric that can be generated from the family of quantum R\'enyi divergences of order $\alpha$, which is also a generalized divergence, is devised by setting $\boldsymbol{\sD}=\hat{\sD}_\alpha$. In the chosen QPP framework $(\cS,\cQ,\Theta,\cM)$, the value of $\varepsilon$ that divides the region where $(\sD,\varepsilon)$-QPP is achieved and the region where it is violated, for an identity channel, is $\hat{\sD}_\alpha( \rho \| \sigma)$.  
\end{remark}

The sandwiched R\'enyi relative entropy $\widetilde{\sD}_\alpha$ in~\eqref{eq:sandwiched-renyi-def} satisfies data processing for $\alpha \in[1/2,1) \cup (1,\infty)$~\cite{FL13,W18opt}, and the quantum relative entropy $\sD$ in~\eqref{eq: limit alpha 1} also satisfies data processing~\cite{Lin75}. Thus, both of these are candidates for a generalized divergence.

\begin{proposition}
   Fix $(\cS,\cQ,\Theta,\bar{\cM})$, $\alpha \in[1/2,1) \cup (1,\infty)$, and $\delta\in(0,1)$. Then, for an algorithm $\cA$, we have that
   \begin{equation}
   \label{eq:sandwiched-Renyi-privacy-chain}
       \varepsilon\textnormal{-QPP} \implies (\widetilde{\sD}_\alpha,\varepsilon')\textnormal{-QPP} \implies (\varepsilon^{\star},\delta)\textnormal{-QPP},
   \end{equation}
   where 
   \begin{align}
       \varepsilon' & \coloneqq \min\! \left\{\varepsilon,\frac{\varepsilon^2 \alpha}{2}\right\}, \\ 
       \varepsilon^{\star} & \coloneqq  \varepsilon' + \frac{1}{\alpha-1}\ln\!\left(\frac{1}{\delta^2}\right) + \ln\!\left( \frac{1}{1-\delta^2} \right).
   \end{align}
   We also have 
   \begin{equation}\label{eq:relative-entropy-privacy-chain}
       \varepsilon\textnormal{-QPP} \implies ({\sD},\varepsilon'')\textnormal{-QPP} \implies (\hat{\varepsilon},\delta)\textnormal{-QPP},
   \end{equation}
   where 
   \begin{align}
       \varepsilon'' & \coloneqq \min\! \left\{\varepsilon,\frac{\varepsilon^2}{2}\right\}, \\ 
       K & \coloneqq  \sup_{\Theta, (\cR,\cT) \in \cQ} \sT \!\left( \cA(\rho^\cR), \cA(\rho^\cT)\right)  \\ 
       \hat{\varepsilon} & \coloneqq  \frac{1}{\delta^2} \left(\varepsilon'+ K  \right)+  \ln\!\left( \frac{1}{1-\delta^2} \right).
   \end{align}
\end{proposition}

\begin{IEEEproof}
    The first implication in both~\eqref{eq:sandwiched-Renyi-privacy-chain} and~\eqref{eq:relative-entropy-privacy-chain} follows from \cref{prop: Bounds on Quantum divergences with QPP}. Then for the second implication, recall from item 3 of \cref{prop: Properties of DL divergence} that $\overline{\sD}^{\delta}(\rho\Vert\sigma)\leq \sD_{\max}^{\delta
}(\rho\Vert\sigma)$. Then applying  \cite[Propositions~5 and 6]{wang2019resource}, we arrive at: 
\begin{align}
    \overline{\sD}^{\delta}(\rho\Vert\sigma) &\leq \widetilde{\sD}_\alpha(\rho \Vert \sigma) + \frac{1}{\alpha-1}\ln\!\left(\frac{1}{\delta^2}\right) + \ln\!\left( \frac{1}{1-\delta^2} \right), \\ 
    \overline{\sD}^{\delta}(\rho\Vert\sigma) &\leq \frac{1}{\delta^2}\left( \sD(\rho \Vert \sigma) + \sT(\rho,\sigma)  \right) + \ln\!\left( \frac{1}{1-\delta^2} \right).
\end{align}
Finally, to conclude the proof, invoke \cref{prop: Equivalent formulation with DS} to establish the required relationship to the QPP framework from there. 
\end{IEEEproof}

\medskip
Note that the chain of implications in~\eqref{eq:sandwiched-Renyi-privacy-chain} holds for every R\'enyi divergence $\sD_\alpha$ satisfying data processing, beyond just $\widetilde{\sD}_\alpha$, because data processing is the key property to adapt \cref{prop: Bounds on Quantum divergences with QPP} and \cite[Proposition~6]{wang2019resource} (see Eqs.~(K51) and (K52) therein). 

\begin{remark}[Comparison to existing results for QDP]
In the special case of QDP, the dependence on the $(\varepsilon,\alpha)$ parameters in~\eqref{eq:sandwiched-Renyi-privacy-chain} provides a strict improvement over previous results. Specifically, Lemmas~V.4 and V.5 of~\cite{hirche2023quantum} show that
\begin{equation}
     \varepsilon\textnormal{-QDP} \implies ({\sD}_\alpha,\varepsilon)\textnormal{-QDP} \implies (\bar{\varepsilon},\delta)\textnormal{-QDP},
\end{equation} with
\begin{equation}
    \bar{\varepsilon} \coloneqq \varepsilon + \frac{\ln\left( 1/(1- \sqrt{1-\delta^2})\right)}{\alpha-1} \approx \varepsilon + \frac{\ln(2 / \delta^2)}{\alpha-1},
\end{equation}
where the approximation holds for small $\delta$. 
Our first implication in~\eqref{eq:sandwiched-Renyi-privacy-chain} is {tighter} compared to this since $\varepsilon' \leq \varepsilon$ and the second implication is also {tighter} (i.e, $\varepsilon^\star \leq \bar{\varepsilon}$) if we choose $\delta^2 \leq 1- 2^{\left(- \frac{1}{\alpha-1}\right)}$ in the small $\delta$ regime.
    
\end{remark}

\subsection{Variant Incorporating Entanglement} \label{SubS: Incorporating Entanglement}

In this subsection, we introduce a novel variant of QPP that incorporates reference systems. This extension potentially allows us to explore the impact of entanglement on the privacy of the system of interest.  

\begin{definition} [$(\boldsymbol{\sD}^R,\varepsilon)$-QPP with reference systems]\label{def:qpp-with-R}
    With the same setup $(\cS,\cQ,\Theta,\cM)$ in \cref{def: qpp}, a quantum algorithm~$\cA: \cL(\cH_A) \to \cL(\cH_B)$ is $(\varepsilon, \boldsymbol{\sD})$-QPP with reference systems  if 
    \begin{equation}
    \sup_{\Theta, (\omega_{RA}^\cR, \omega_{RA}^\cT) \in \cG } \boldsymbol{\sD} \!\left( (\cI_R\otimes \cA)(\omega_{RA}^\cR) \| (\cI_R\otimes \cA)(\omega_{RA}^\cT)\right) \leq \varepsilon,
    \end{equation}
   where the set $\cG$ is defined in~\eqref{eq: set L for reference systems}.
\end{definition}
Since there is no restriction on the reference systems, the supremum in \cref{def:qpp-with-R} is also taken over the dimension of the reference system $R$, which is an unbounded set. 
However, following from  isometric invariance of generalized divergences, together with purification and the Schmidt decomposition, 
 it suffices to take the supremum over pure bipartite states with the reference system $R$ isomorphic to the channel input system $A$. 

 \begin{remark}[Properties]\label{rem:properties_referenceSystemVariant}
     Similar to variants of QPP (recall \cref{rem: properties of D-QPP}), this variant incorporating reference systems also satisfies post-processing, convexity, and parallel composability. 

     Adaptive composition: Let $\cA_i$ be a $(\boldsymbol{\sD}^R,\varepsilon_i)$-QPP algorithm for $i\in \{1,2\}$. Consider the adaptive composition of $\cA_1$ and $\cA_2$, as shown in \cref{fig:adaptive-reference-DR}:  First $\cA_1$ is applied on the upper system, then the quantum channel $\cN$ acts on both systems, and lastly $\cA_2$ acts on the upper system.
Adaptive composition of $\cA_1$ and $\cA_2$ also satisfies $(\boldsymbol{\sD}^R,\varepsilon_1)$-QPP. This shows that the adaptive composition illustrated in \cref{fig:adaptive-reference-DR} does not degrade privacy, showcasing the strength of the privacy framework in \cref{def:qpp-with-R}.
   \begin{figure}
       \centering
       \includegraphics[width=\linewidth]{ 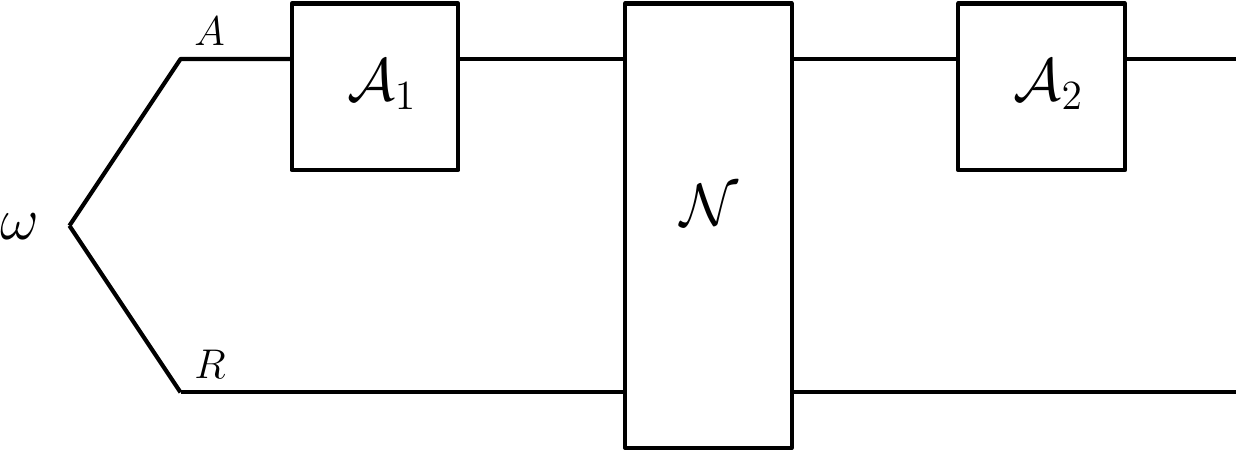}
       \caption{Adaptive composition with reference systems: First $\cA_1$ is applied on the upper system, then the quantum channel $\cN$ on both the systems, and lastly $\cA_2$ on the upper system. Then the adaptive composition is QPP if $\cA_1$ and $\cA_2$ are.}
       \label{fig:adaptive-reference-DR}
   \end{figure} 
 \end{remark}

Next, we observe that \cref{def:qpp-with-R} is a stronger privacy guarantee than \cref{def: D-QPP}, which does not take into account reference systems. 
 \begin{corollary}[Strength compared to $(\boldsymbol{\sD},\varepsilon)$-QPP]
    $(\boldsymbol{\sD}^R,\varepsilon)$-QPP implies $(\boldsymbol{\sD},\varepsilon)$-QPP. 
 \end{corollary}
 This follows from the data-processing inequality for the underlying generalized divergence $\boldsymbol{\sD}$, choosing the channel as the partial trace over the reference system $R$. It highlights that in certain scenarios where the system of interest is entangled with other reference systems, the general QPP guarantees defined in earlier sections may not be sufficient.
 
 The choice of $\boldsymbol{\sD}$ could heavily affect the design of useful privacy frameworks with entanglement (recall the example in \cref{rem: Incorporating entanglement through reference systems} with the choice $\boldsymbol{\sD}= \sD_{\max}$, along with the equivalence of $\sD_{\max}$ to $\varepsilon$-QPP, as given in~\eqref{eq: equivalence-of-epsilon-QPP}). Therefore, careful consideration of the appropriate generalized divergence is essential for developing effective and meaningful privacy frameworks that account for entanglement effects.

\section{Concluding Remarks and Future Directions}

\label{Sec:summary}

This work proposed QPP as a flexible privacy framework for quantum systems. We showed that QPP is captured exactly by the DL divergence, endowing the latter with an operational interpretation of the DL~divergence. The DL divergence representation was used to study properties of QPP mechanisms and characterize privacy-utility tradeoffs. As a concrete case study, we explored the depolarization QPP mechanisms and characterized the parameter values to achieve privacy. A methodology for auditing quantum privacy was also developed.

Future research directions are abundant and include privacy-utility analysis of specific quantum estimation tasks, designing efficient quantum algorithms that achieve QPP, analysing the type-II error of our quantum privacy auditing pipeline, providing  tight characterizations of parallel composability of QPP mechanisms, and devising efficient methods for computing the DL~divergence using quantum algorithms to enable quantum privacy auditing. In the longer term, the proposed framework could lay the foundations for privacy-preserving learning in quantum systems.

\section*{Acknowledgements}
We acknowledge helpful discussions with Rochisha Agarwal, Hansadi Jayamaha, Kaiyuan Ji, Margarite LaBorde, Hemant Mishra, Dhrumil Patel, Aby Philip, Soorya Rethinasamy, and Vishal Singh. We are also indebted to Prof.~\c Ci\c sik Balyk for sharing his valuable insights. 
\bibliographystyle{IEEEtran}
\bibliography{reference}

\appendices

\section{Alternative Proof for Joint-quasi Convexity of DL divergence}\label{APP:Alternative_proof}

{By adapting the strong convexity of Hockey-Stick divergence (Proposition II.5 of~\cite{hirche2023quantum}), with the substitutions $p=q$ and $\gamma_2=1$  and $\gamma_1= \gamma$ therein, we obtain 
\begin{multline}
    \Tr\!\left[\left( \sum_{i=1}^k p_i \rho_i - \gamma \sum_{i=1}^k  p_i \sigma_i \right)_{+} \right] \\ \leq \sum_{i=1}^k  p_i \Tr\!\left[ \left(\rho_i - \gamma \sigma_i \right)_{+} \right] \leq \max_{i} \Tr\!\left[ \left(\rho_i - \gamma \sigma_i \right)_{+} \right] .
\end{multline}

Then by assuming $\max_{i} \Tr\!\left[ \left(\rho_i- \gamma \sigma_i  \right)_{+} \right]  \leq \delta$, $\gamma$ is a candidate for the optimization of \\ $\overline{\sD}^{\delta}\!\left(  \sum_{i=1}^k p_i \rho_i \middle \Vert   \sum_{i=1}^k p_i \sigma_i \right) $.
This leads to 
\begin{equation}
    \overline{\sD}^{\delta}\!\left(  \sum_{i=1}^k p_i \rho_i \middle \Vert   \sum_{i=1}^k p_i \sigma_i \right)  \leq \ln( \gamma).
\end{equation}
The above holds for all $\gamma$ such that $\max_{i} \Tr\!\left[ \left(\rho_i - \gamma \sigma_i \right)_{+} \right] \leq \delta$. Then, optimizing over such $\gamma$, we arrive at joint quasi-convexity:

\begin{equation}
\overline{\sD}^{\delta}\!\left(  \sum_{i=1}^k p_i \rho_i \middle \Vert   \sum_{i=1}^k p_i \sigma_i \right) \leq \max_i \overline{\sD}^{\delta}\!\left(   \rho_i \Vert   \sigma_i \right).
\end{equation}
}

\section{Proof of \cref{lem:info-spec-to-sm-dmax-helper}} \label{Sec: proof lemma supporting connecting DL to smooth max}
The proof given below is closely related to the proof of \cite[Lemma~6.21]{tomamichel2015quantum}, but there are some subtle differences and so we provide it here for completeness.

Let%
\begin{align}
\Sigma &  \coloneqq\left(  \rho-\lambda\sigma\right)  _{+},\\
G  &  \coloneqq\left(  \lambda\sigma\right)  ^{1/2}\left(  \lambda\sigma
+\Sigma\right)  ^{-1/2}.
\end{align}
Note that%
\begin{align}
0  &  \leq G^{\dag}G\\
&  =\left(  \lambda\sigma+\Sigma\right)  ^{-1/2}\left(  \lambda\sigma\right)
\left(  \lambda\sigma+\Sigma\right)  ^{-1/2}\\
&  \leq\left(  \lambda\sigma+\Sigma\right)  ^{-1/2}\left(  \lambda
\sigma+\Sigma\right)  \left(  \lambda\sigma+\Sigma\right)  ^{-1/2}\\
&  \leq I.
\end{align}
From the fact that $\rho-\lambda\sigma\leq\Sigma$, it follows that%
\begin{equation}
\rho\leq\lambda\sigma+\Sigma.
\end{equation}
Define the following state:%
\begin{equation}
\widetilde{\rho}\coloneqq\frac{G\rho G^{\dag}}{\operatorname{Tr}[G^{\dag}G\rho]}.
\end{equation}
Consider that%
\begin{align}
&  1-\operatorname{Tr}[G\rho G^{\dag}]\nonumber\\
&  =\operatorname{Tr}[(I-G^{\dag}G)\rho]\\
&  \leq\operatorname{Tr}[(I-G^{\dag}G)\left(  \lambda\sigma+\Sigma\right)  ]\\
&  =\operatorname{Tr}[(I-G^{\dag}G)\left(  \lambda\sigma+\Sigma\right)  ]\\
&  =\operatorname{Tr}[\lambda\sigma+\Sigma] \notag \\
& \quad 
-\operatorname{Tr}\!\left[  \left(
\lambda\sigma+\Sigma\right)  ^{-1/2}\left(  \lambda\sigma\right)  \left(
\lambda\sigma+\Sigma\right)  ^{-1/2}\left(  \lambda\sigma+\Sigma\right)
\right] \\
&  =\operatorname{Tr}[\lambda\sigma+\Sigma]-\operatorname{Tr}\!\left[
\lambda\sigma\right] \\
&  =\operatorname{Tr}[\Sigma]\\
&=\delta
\end{align}
This implies that%
\begin{equation}
\operatorname{Tr}[G\rho G^{\dag}]\geq1-\delta.
\end{equation}
Then it follows that%
\begin{align}
\widetilde{\rho}  &  =\frac{G\rho G^{\dag}}{\operatorname{Tr}[G^{\dag}G\rho
]}\\
&  \leq\frac{G\left(  \lambda\sigma+\Sigma\right)  G^{\dag}}{\operatorname{Tr}%
[G^{\dag}G\rho]}\\
&  =\frac{\lambda\sigma}{\operatorname{Tr}[G^{\dag}G\rho]}\\
&  \leq\frac{\lambda\sigma}{1-\delta}.
\end{align}
Let $\psi_{RA}=\sqrt{\rho_{A}}\Gamma_{RA}\sqrt{\rho_{A}}$ be the canonical
purification of $\rho_{A}$, with $\Gamma_{RA}\coloneqq\sum_{i,j}|i\rangle\!\langle
j|_{R}\otimes|i\rangle\!\langle j|_{A}$, and let $\widetilde{\psi}_{RA}%
=\frac{G_{A}\psi_{RA}G_{A}^{\dag}}{\operatorname{Tr}[G^{\dag}G\rho]}$ purify
$\widetilde{\rho}$. Then%
\begin{align}
\sqrt{F(\rho,\widetilde{\rho})}  &  \geq\frac{1}{\sqrt{\operatorname{Tr}[G^{\dag
}G\rho]}}\left\vert \langle\psi|_{RA}I_{R}\otimes G_{A}|\psi\rangle
_{RA}\right\vert \\
&  \geq\left\vert \langle\psi|_{RA}I_{R}\otimes G_{A}|\psi\rangle
_{RA}\right\vert \\
&  =\left\vert \langle\Gamma|_{RA}I_{R}\otimes\sqrt{\rho_{A}}G_{A}\sqrt
{\rho_{A}}|\Gamma\rangle_{RA}\right\vert \\
&  =\left\vert \operatorname{Tr}[G\rho]\right\vert \\
&  \geq\operatorname{Re}[\operatorname{Tr}[G\rho]]\\
&  =\operatorname{Tr}[\overline{G}\rho]\\
&  =1-\operatorname{Tr}[\left(  I-\overline{G}\right)  \rho],
\end{align}
where
\begin{equation}
\overline{G}\coloneqq\frac{G+G^{\dag}}{2}.
\end{equation}
The first inequality follows from Uhlmann's theorem for fidelity, and the
second follows because $\operatorname{Tr}[G^{\dag}G\rho]\leq1$. Observe that
$\overline{G}\leq I$ because $\left\Vert G\right\Vert _{\infty}\leq 1$ and by
applying the triangle inequality. So this means that $I-\overline{G}\geq0$.
Now consider that
\begin{align}
&  \operatorname{Tr}[\left(  I-\overline{G}\right)  \rho]\nonumber\\
&  \leq\operatorname{Tr}[\left(  I-\overline{G}\right)  \left(  \lambda
\sigma+\Sigma\right)  ]\\
&  =\operatorname{Tr}[\lambda\sigma+\Sigma]-\operatorname{Tr}[\overline
{G}\left(  \lambda\sigma+\Sigma\right)  ]\\
&  =\operatorname{Tr}[\lambda\sigma+\Sigma]\nonumber\\
&  \mspace{2mu} -\frac{1}{2}\operatorname{Tr}\Big[  \left(  \left(  \lambda
\sigma\right)  ^{1/2}\left(  \lambda\sigma+\Sigma\right)  ^{-1/2}+\left(
\lambda\sigma+\Sigma\right)  ^{-1/2}\left(  \lambda\sigma\right)
^{1/2}\right)  \times  \notag \\
& \qquad 
\left(  \lambda\sigma+\Sigma\right)  \Big] \\
&  =\operatorname{Tr}[\lambda\sigma+\Sigma]-\frac{1}{2}\operatorname{Tr}%
\left[  \left(  \lambda\sigma\right)  ^{1/2}\left(  \lambda\sigma
+\Sigma\right)  ^{-1/2}\left(  \lambda\sigma+\Sigma\right)  \right]
\nonumber\\
&  \qquad
-\frac{1}{2}\operatorname{Tr}\!\left[  \left(  \lambda\sigma
+\Sigma\right)  ^{-1/2}\left(  \lambda\sigma\right)  ^{1/2}\left(
\lambda\sigma+\Sigma\right)  \right] \\
&  =\operatorname{Tr}[\lambda\sigma+\Sigma]-\frac{1}{2}\operatorname{Tr}%
\left[  \left(  \lambda\sigma\right)  ^{1/2}\left(  \lambda\sigma
+\Sigma\right)  ^{1/2}\right] \nonumber\\
&  \qquad
-\frac{1}{2}\operatorname{Tr}\!\left[  \left(  \lambda\sigma\right)
^{1/2}\left(  \lambda\sigma+\Sigma\right)  ^{1/2}\right] \\
&  =\operatorname{Tr}[\lambda\sigma+\Sigma]-\operatorname{Tr}\!\left[  \left(
\lambda\sigma\right)  ^{1/2}\left(  \lambda\sigma+\Sigma\right)  ^{1/2}\right]
\\
&  \leq\operatorname{Tr}[\lambda\sigma+\Sigma]-\operatorname{Tr}\!\left[
\left(  \lambda\sigma\right)  ^{1/2}\left(  \lambda\sigma\right)
^{1/2}\right] \\
&  =\operatorname{Tr}[\lambda\sigma+\Sigma]-\operatorname{Tr}\!\left[
\lambda\sigma\right] \\
&  =\operatorname{Tr}[\Sigma]\\
&  =\delta.
\end{align}
So all of this implies that%
\begin{equation}
\sqrt{\sF(\rho,\widetilde{\rho})}\geq1-\delta,
\end{equation}
and in turn that%
\begin{equation}
\sF(\rho,\widetilde{\rho})\geq\left(  1-\delta \right)  ^{2}.
\end{equation}
By applying the inequality%
\begin{equation}
\frac{1}{2}\left\Vert \rho-\widetilde{\rho}\right\Vert _{1}\leq\sqrt
{1-\sF(\rho,\widetilde{\rho})},
\end{equation}
we conclude that%
\begin{align}
\frac{1}{2}\left\Vert \rho-\widetilde{\rho}\right\Vert _{1}  &  \leq
\sqrt{1-\left(  1-\delta\right)  ^{2}}\\
&  =\sqrt{1-\left(  1-2\delta+\delta^{2}\right)  }\\
&  =\sqrt{2\delta-\delta^{2}}\\
&  =\sqrt{\delta\left(  2-\delta\right)  }.
\end{align}
Putting everything together, we see that $\widetilde{\rho}$ is a quantum state
satisfying%
\begin{align}
\frac{1}{2}\left\Vert \rho-\widetilde{\rho}\right\Vert _{1}  &  \leq
\sqrt{\delta\left(  2-\delta\right)  },\\
\widetilde{\rho}  &  \leq\frac{\lambda\sigma}{1-\delta}.
\end{align}
This means that $\widetilde{\rho}$ and $\frac{\lambda}{1-\delta}$ are
feasible for ${\sD_{\max}^{\sqrt{\delta\left(  2-\delta\right)  }%
}(\rho\Vert\sigma)}$, and so it follows that%
\begin{align}
\sD_{\max}^{\sqrt{\delta\left(  2-\delta\right)  }}(\rho\Vert\sigma)
&  \leq\ln \!\left( \frac{\lambda}{1-\delta}\right)\\
&  =\ln \lambda+\ln \!\left(  \frac{1}{1-\delta}\right)  .
\end{align}
This concludes the proof.

\section{Subadditivity of Smooth Max-Relative Entropy}\label{app:proof-sub-additivity-lemma}

\begin{lemma}\label{lem:subAdditivityofDmax}
Given $\delta_{1},\delta_{2}\in\left[  0,1\right]  $ such that $\delta_1+ \delta_2 \leq 1$, states $\rho_{1}$ and
$\rho_{2}$, and PSD operators $\sigma_{1}$ and $\sigma_{2}
$, the following subadditivity relation holds%
\begin{equation}
\sD_{\max}^{\delta_1 + \delta_2}(\rho_{1}\otimes\rho_{2}\Vert\sigma_{1}\otimes
\sigma_{2})\leq \sD_{\max}^{\delta_{1}}(\rho_{1}\Vert\sigma_{1})+ \sD_{\max
}^{\delta_{2}}(\rho_{2}\Vert\sigma_{2}).
\end{equation}
\end{lemma}
\begin{IEEEproof}
 Let $\overline{\rho}_{i}$ and $\lambda_{i}$ be optimal choices for $D_{\max
}^{\delta_{i}}(\rho_{i}\Vert\sigma_{i})$, for $i\in\left\{  1,2\right\}  $.
Then, consider that%
\begin{equation}
\overline{\rho}_{1}\otimes\overline{\rho}_{2}\leq\lambda_{1}\sigma_{1}%
\otimes\overline{\rho}_{2}\leq\lambda_{1}\sigma_{1}\otimes\lambda_{2}%
\sigma_{2}=\lambda_{1}\lambda_{2}\sigma_{1}\otimes\sigma_{2}.
\end{equation}   
Furthermore, consider that 
\begin{align}
 & \frac{1}{2}\left\Vert \overline{\rho}_{1}\otimes\overline{\rho}_{2}-\rho
_{1}\otimes\rho_{2}\right\Vert _{1}\nonumber\\ 
& =\frac{1}{2}\left\Vert \overline{\rho}_{1}\otimes\overline{\rho}%
_{2}-\overline{\rho}_{1}\otimes\rho_{2}+\overline{\rho}_{1}\otimes\rho
_{2}-\rho_{1}\otimes\rho_{2}\right\Vert _{1}\label{eq:triangle-etc-1}\\
& =\frac{1}{2}\left\Vert \overline{\rho}_{1}\otimes\left(  \overline{\rho}%
_{2}-\rho_{2}\right)  +\left(  \overline{\rho}_{1}-\rho_{1}\right)
\otimes\rho_{2}\right\Vert _{1}\\
& \leq\frac{1}{2}\left\Vert \overline{\rho}_{1}\otimes\left(  \overline{\rho
}_{2}-\rho_{2}\right)  \right\Vert _{1}+\frac{1}{2}\left\Vert \left(
\overline{\rho}_{1}-\rho_{1}\right)  \otimes\rho_{2}\right\Vert _{1}\\
& =\frac{1}{2}\left\Vert \overline{\rho}_{1}\right\Vert _{1}\left\Vert
\overline{\rho}_{2}-\rho_{2}\right\Vert _{1}+\frac{1}{2}\left\Vert
\overline{\rho}_{1}-\rho_{1}\right\Vert _{1}\left\Vert \rho_{2}\right\Vert
_{1}\\
& =\frac{1}{2}\left\Vert
\overline{\rho}_{2}-\rho_{2}\right\Vert _{1}+\frac{1}{2}\left\Vert
\overline{\rho}_{1}-\rho_{1}\right\Vert _{1} \\
&\leq \delta_1 + \delta_2, 
\end{align}
where~\eqref{eq:triangle-etc-1} follows from the triangular inequality for the trace norm and the final inequality from the assumption that $\overline{\rho}_{i}$ are the optimizers for $D_{\max
}^{\delta_{i}}(\rho_{i}\Vert\sigma_{i})$, for $i\in\left\{  1,2\right\}  $.

Finally we have shown that $\overline{\rho}_{1}\otimes\overline{\rho}_{2}$ and
$\lambda_{1}\lambda_{2}$ are candidates for the optimization for $D_{\max
}^{\delta_1+ \delta_2}(\rho_{1}\otimes\rho_{2}\Vert\sigma_{1}\otimes\sigma_{2})$,
 thus concluding the proof. 
\end{IEEEproof}

\section{Proof of \cref{thm:QPP_properties} } \label{App:Proof-of-thm-of-QPP-Properties}

\textit{For (1):}  Fix $(\rho^\cR, \rho^\cT)$ in~\eqref{eq:qpp_def}, $\rM \in \cM$. Consider that
\begin{align}
    \Tr\!\left[ \rM \cA(\rho^\cR) \right] &= \Tr\!\left[\rM \sum_{i=1}^k p_i \cA_i(\rho^\cR)\right] \\
    &=\sum_{i=1}^k p_i  \Tr\!\left[\rM \cA_i(\rho^\cR) \right] \\
    & \stackrel{(a)} \leq \sum_{i=1}^k p_i  \left(e^\varepsilon \Tr\!\left[\rM \cA_i(\rho^\cT) \right] + \delta \right)\\
    &=\sum_{i=1}^k p_i  e^\varepsilon \Tr\!\left[ \rM \cA_i(\rho^\cT) \right] + \delta \\
    &= e^\varepsilon \Tr\!\left[\rM \cA(\rho^\cT) \right] + \delta,
\end{align}
where (a) follows due to each $\cA_i$ being $(\varepsilon,\delta)$-QPP. This inequality holds for every $(\rho^\cR, \rho^\cT)$, and so it holds for all such pairs generated from $(\cS,\cQ,\Theta,\cM)$.

\medskip 
\textit{For (2):} Fix $0 \psd \rM' \psd \rI$ such that $\rM'\in \cM'$ as stated in the property. With that assumption, there exists $\rM \in \cM$ such that $\rM=\cN^\dagger(\rM')$. Consider that
\begin{align}
\Tr\!\left[\rM' \cN\!\left(\cA(\rho^\cR) \right) \right] & =  \Tr\!\left[\cN^\dagger(\rM')  \cA(\rho^\cR) \right] \\
& = \Tr\!\left[\rM \cA(\rho^\cR) \right],
\end{align}
where $\cN^\dagger$ is the adjoint  of $\cN$, implying that 
\begin{equation}
  0 \psd \cN^\dagger(\rM') =\rM \psd \rI  
\end{equation}
because $\cN^\dagger$ is positive and unital by the assumption that $\cN$ is a quantum channel.
Similarly, we have that $\Tr\!\left[\rM' \cN \left(\cA(\rho^{\cT})\right) \right]=\Tr\!\left[\rM \cA(\rho^\cT) \right]$, and we conclude that the processed mechanism satisfies $(\varepsilon,\delta)$-QPP with the choice of $\cM' \subseteq \{ \rM': \cN^\dagger(\rM') \in \cM\}$. 

\medskip 
\textit{For (3):} 
Fix $(\cR_i,\cT_i) \in \cQ$ for $i \in \{1, \ldots,k\}$, and $\ \bigotimes_{i=1}^k \rM_i \in \bigotimes_{i=1}^k \cM_i$. 
Denote
\begin{equation}
\cA^{(k)}(\rho^{\cR^{(k)}})\coloneqq  \cA_1(\rho^{\cR_1} )\otimes \cA_2(\rho^{\cR_2}) \otimes \ldots \cA_k(\rho^{\cR_k})    
\end{equation}
 and $\cA^{(k)}(\rho^{\cT^{(k)}})$ similarly by replacing $\cR$ with $\cT$.

Fix $i \in \{1, \ldots, k\}$. Consider that $\Tr\!\left[\rM_i\cA_i(\rho^{\cR_i})\right] \leq 1 \leq 1+\delta_i$ because $\Tr\!\left[\rM_i\cA_i(\rho^{\cR_i})\right]$ is a probability. Combining with the inequality $\Tr\!\left[\rM_i\cA_i(\rho^{\cR_i})\right] \leq  e^{\varepsilon_i} \Tr\!\left[\rM_i\cA_i(\rho^{\cT_i})\right] + \delta_i$, which holds from the assumption that QPP holds, we conclude that
\begin{equation}
 \Tr\!\left[\rM_i\cA_i(\rho^{\cR_i})\right] \leq \min \!\left\{ e^{\varepsilon _i}\Tr\!\left[\rM_i\cA_i(\rho^{\cT_i})\right],1\right\} + \delta_i.
\end{equation}
 Consider that
\begin{align}
    & \prod_{i=1}^k  \Tr\!\left[\rM_i \cA_i(\rho^{\cR_i}) \right] \notag \\
    &\leq \left(\min\!\left\{1,e^{\varepsilon_1}  \Tr\!\left[ \rM_1 \cA_1(\rho^{\cR_1}) \right] \right\} + \delta_1 \right)      \prod_{i=2}^k  \Tr\!\left[\rM_i \cA_i(\rho^{\cR_i}) \right] \\ 
    & \leq \min\!\left\{1,e^{\varepsilon_1}  \Tr\!\left[\rM_1 \cA_1(\rho^{\cR_1}) \right] \right\}  \prod_{i=2}^k  \Tr\!\left[ \rM_i \cA_i(\rho^{\cR_i}) \right] + \delta_1 \\ 
    & \leq \min\!\left\{1,e^{\varepsilon_1}  \Tr\!\left[\rM_1 \cA_1(\rho^{\cR_1}) \right]\right\} \times \notag \\
    & \quad 
    \left(\min\!\left\{1,e^{\varepsilon_2}  \Tr\!\left[\rM_2 \cA_2(\rho^{\cT_2}) \right] \right\} + \delta_2 \right)      \prod_{i=3}^k  \Tr\!\left[\rM_i \cA_i(\rho^{\cR_i}) \right] \notag \\
    & \qquad 
    + \delta_1 \\
    & \leq \prod_{j=1}^2  \min\!\left\{e^{\varepsilon_j}\Tr\!\left[\rM_j \cA_j(\rho^{\cT_j}) \right], 1 \right\}   \prod_{i=3}^k  \Tr\!\left[\rM_i \cA_i(\rho^{\cR_i}) \right] \notag \\
    & \qquad
    + \delta_1 + \delta_2 \\ 
    & \leq e^{\sum_{i=1}^k \varepsilon_i}  \prod_{i=1}^k  \Tr\!\left[\rM_i \cA_i(\rho^{\cT_i}) \right] + \sum_{i=1}^k \delta_i,
\end{align}
where the last inequality follows by proceeding with similar expansions for each remaining term of the product as carried out in the first three steps. 

\section{Proof of \cref{Prop:Adaptive-composition-of-QPP}}
\label{App:Proof-of-adaptive-com}

   Fix $\rM',\rM \in \bar{\cM}$, $P_X \in \Theta$, and $(\cR_1,\cT_1), (\cR_2,\cT_2)~\in \cQ$. Let $\rM_y\coloneqq  (\langle y| \otimes \rI) \rM'(|y\rangle \otimes \rI) $ and note that $\rM_y$ is a measurement operator in $\bar{\cM}$. Recall the definition of the channel
\begin{equation}
\overline{\mathcal{E}}\coloneqq \sum_{y\in\mathcal{Y}}\mathcal{E}^{y}.%
\end{equation}
Consider that
\begin{align}
&  \operatorname{Tr}\!\left[  (\rM\otimes \rM^{\prime})\left(  \sum_{y\in
\mathcal{Y}}\mathcal{E}^{y}(\mathcal{A}_{1}(\rho^{\mathcal{R}_{1}}%
))\otimes|y\rangle\!\langle y|\otimes\mathcal{A}_{2}^{y}(\rho^{\mathcal{R}_{2}%
})\right)  \right]  \notag \\
&  =\sum_{y\in\mathcal{Y}}\operatorname{Tr}\!\left[  \rM\mathcal{E}^{y}%
(\mathcal{A}_{1}(\rho^{\mathcal{R}_{1}}))\otimes \rM^{\prime}\!\left(
|y\rangle\!\langle y|\otimes\mathcal{A}_{2}^{y}(\rho^{\mathcal{R}_{2}})\right)
\right]  \\
&  =\sum_{y\in\mathcal{Y}}\operatorname{Tr}[\rM\mathcal{E}^{y}(\mathcal{A}%
_{1}(\rho^{\mathcal{R}_{1}}))]\operatorname{Tr}[\rM^{\prime}\!\left(
|y\rangle\!\langle y|\otimes\mathcal{A}_{2}^{y}(\rho^{\mathcal{R}_{2}})\right)
] \\
&  =\sum_{y\in\mathcal{Y}}\operatorname{Tr}[\rM\mathcal{E}^{y}(\mathcal{A}%
_{1}(\rho^{\mathcal{R}_{1}}))]\operatorname{Tr}[\rM_{y}^{\prime}\mathcal{A}%
_{2}^{y}(\rho^{\mathcal{R}_{2}})]\\
&  \stackrel{(a)}\leq\sum_{y\in\mathcal{Y}}\operatorname{Tr}[\rM\mathcal{E}^{y}(\mathcal{A}%
_{1}(\rho^{\mathcal{R}_{1}}))]  \notag \\ 
& \qquad \qquad  \times 
\left(  \min\! \left\{  1,e^{\varepsilon_{2}%
}\operatorname{Tr}[\rM_{y}^{\prime}\mathcal{A}_{2}^{y}(\rho^{\mathcal{T}_{2}%
})]\right\}  +\delta_{2}\right)  \\
&  =\sum_{y\in\mathcal{Y}}\operatorname{Tr}[\mathcal{E}^{y\dag}(\rM)\mathcal{A}%
_{1}(\rho^{\mathcal{R}_{1}})] 
\notag \\ 
 & \qquad \qquad \times 
 \left(  \min\! \left\{  1,e^{\varepsilon_{2}%
 }\operatorname{Tr}[\rM_{y}^{\prime}\mathcal{A}_{2}^{y}(\rho^{\mathcal{T}_{2}%
})]\right\}  +\delta_{2}\right) \\
 &  \stackrel{(b)}=\operatorname{Tr}[\rM\overline{\mathcal{E}}(\mathcal{A}_{1}(\rho^{\mathcal{R}_{1}}))]\delta_{2} \notag \\
 &
+\sum_{y\in\mathcal{Y}}\left(\operatorname{Tr}[\mathcal{E}^{y\dag
 }(\rM)\mathcal{A}_{1}(\rho^{\mathcal{R}_{1}})]\min\! \left\{  1,e^{\varepsilon_{2}%
 }\operatorname{Tr}[\rM_{y}^{\prime}\mathcal{A}_{2}^{y}(\rho^{\mathcal{T}_{2}%
 })]\right\}  \right)  \\
 & \stackrel{(c)} \leq \delta_{2} \notag \\
&  \ 
+ \sum_{y\in\mathcal{Y}}\operatorname{Tr}[\mathcal{E}^{y\dag
 }(\rM)\mathcal{A}_{1}(\rho^{\mathcal{R}_{1}})]\min\! \left\{  1,e^{\varepsilon_{2}%
 }\operatorname{Tr}[\rM_{y}^{\prime}\mathcal{A}_{2}^{y}(\rho^{\mathcal{T}_{2}%
 })]\right\} \\
 &  \stackrel{(d)} \leq \delta_2 +   \sum_{y\in\mathcal{Y}}\left(  e^{\varepsilon_{1}}%
 \operatorname{Tr}[\mathcal{E}^{y\dag}(\rM)\mathcal{A}_{1}(\rho^{\mathcal{T}_{1}%
 })]+\delta_{1}\right)  \notag \\ 
 & \qquad \qquad \qquad \qquad \times
 \min\! \left\{  1,e^{\varepsilon_{2}}\operatorname{Tr}%
[\rM_{y}^{\prime}\mathcal{A}_{2}^{y}(\rho^{\mathcal{T}_{2}})]\right\}  
 \\
 &  \leq\sum_{y\in\mathcal{Y}}\left(  e^{\varepsilon_{1}}\operatorname{Tr}%
 [\mathcal{E}^{y\dag}(\rM)\mathcal{A}_{1}(\rho^{\mathcal{T}_{1}})]e^{\varepsilon
 _{2}}\operatorname{Tr}[\rM_{y}^{\prime}\mathcal{A}_{2}^{y}(\rho^{\mathcal{T}%
 _{2}})]+\delta_{1}\right)\notag \\
 & \qquad \qquad
 +\delta_{2}\\
&  =\sum_{y\in\mathcal{Y}}\left(  e^{\varepsilon_{2}}e^{\varepsilon_{1}%
 }\operatorname{Tr}[\mathcal{E}^{y\dag}(\rM)\mathcal{A}_{1}(\rho^{\mathcal{T}%
 _{1}})]\operatorname{Tr}[\rM_{y}^{\prime}\mathcal{A}_{2}^{y}(\rho^{\mathcal{T}%
 _{2}})]+\delta_{1}\right)  \notag \\  & \qquad \qquad 
 +\delta_{2}\\
 &  =e^{\varepsilon'} \times \notag\\
 & 
 \operatorname{Tr}\!\left[  (\rM\otimes
 \rM^{\prime})\!\left(  \sum_{y\in\mathcal{Y}}\mathcal{E}^{y}(\mathcal{A}_{1}%
 (\rho^{\mathcal{T}_{1}}))\otimes|y\rangle\!\langle y|\otimes\mathcal{A}_{2}%
 ^{y}(\rho^{\mathcal{T}_{2}})\right)  \right] \notag  \\
 & \qquad \qquad  
 +\delta_{2}+\delta_{1}\left\vert \mathcal{Y}\right\vert,
\end{align}
where: (a) from $\operatorname{Tr}[\rM_{y}^{\prime}\mathcal{A}
_{2}^{y}(\rho^{\mathcal{R}_{2}})] \leq 1 \leq 1+\delta_2$ and $\cA_2^y$ being $(\varepsilon_2,\delta_2)$-QPP; (b) and (c) from
\begin{align}
\operatorname{Tr}\!\left[\sum_{y\in\mathcal{Y}}\mathcal{E}^{y\dag}(\rM)\mathcal{A}%
_{1}(\rho^{\mathcal{T}_{1}})\right]  & =\operatorname{Tr}\!\left[\rM\sum_{y\in\mathcal{Y}%
}\mathcal{E}^{y}\mathcal{A}_{1}(\rho^{\mathcal{T}_{1}})\right]\\
& =\operatorname{Tr}[\rM\overline{\mathcal{E}}(\mathcal{A}_{1}(\rho
^{\mathcal{T}_{1}}))]\\
& \leq1;
\end{align} and (d) from $\cA_1$ being $(\varepsilon_1,\delta_1)$-QPP and the fact that $\cE^{y\dag}(\rM)$ is a measurement operator in $\bar{\cM}$.

\section{Composability with {Classically} Correlated States} \label{App:Composability_with_correlated_states}
 In Property~3 of \cref{thm:QPP_properties} and \cref{Prop:Adaptive-composition-of-QPP}, we considered the case in which two mechanisms, composed either in parallel or adaptively, receive independent inputs (i.e., the input being $\rho^{X_1} \otimes \rho^{X_2}$ where $X_i \sim P_X \in \Theta$ for $i=\{1,2\}$, which are chosen independently). 
 We now focus on the setting in which the inputs are {classically} correlated. The input is chosen as a separable state of the form 
\begin{equation}\label{eq:input-correlated-compose}
    \sigma_I \coloneqq \sum_{z \in \cZ} q(z) \, \omega^z \otimes \tau^z,
\end{equation}
 where $q$ represents a probability distribution with $q(z) \geq 0$ and $\sum_{z \in \cZ} q(z) =1$, and $\omega^z$ and $ \tau^z$ are quantum states for all $z \in \cZ$\footnote{Note that~\eqref{eq:input-correlated-compose} covers the case of having input states of the form $\sigma_I \coloneqq \sum_{(x,y)\in \cX\times \cY} q(x,y) \ \omega^x \otimes \tau^y$ where for all $x \in \cX$ and $y \in \cY$ $\omega^x$ and $\tau^y$ are states, by considering $z$ to be an index for multiple variables, i.e., setting $z=(x,y)$.}.
One special case of interest is as follows:
\begin{equation}
   \sigma_I \coloneqq 
   \sum_{x \in \cX} P_X(x) \, \rho^x \otimes \rho^x.
\end{equation}

In this setting, QDP ensures indistinguishability of the input states 
\begin{equation}
    \sigma^1_I \coloneqq  \sum_{z \in \cZ} q(z) \, \omega_1^z \otimes \tau_1^z \quad \textnormal{and} \quad
     \sigma^2_I  \coloneqq \sum_{z \in \cZ} q(z) \, \omega_2^z \otimes \tau_2^z ,
\end{equation}
where $\omega_1^z \sim \omega_2^z$ and $\tau_1^z \sim \tau_2^z$ are neighbors for all $z \in \cZ$. 

We consider an instance of the QPP framework, called flexible QDP, where $(\cS,\cQ,\Theta,\cM)$ is such that $\Theta$ and $\cM$ are chosen based on user needs, while the other parameters are as given in \cref{rem: Quantum DP set}. Flexible QDP then satisfies the following composability properties. 

\begin{corollary}[Composability of flexible QDP] \label{cor:composability-flexible-QDP}
Let the initial input to the two mechanisms $\cA_1$ and $\cA_2$ be of the form $\sum_{z \in \cZ} q(z) \, \omega^z \otimes \tau^z$. The following composability properties hold for the QDP framework.

   \underline{Parallel composability}: Consider the parallel composed mechanism $\sum_{z \in \cZ} q(z) \, \cA_1(\omega^z) \otimes \cA_2(\tau^z)$.
   \begin{enumerate}
       \item If $\cA_i$ is $(\varepsilon_i, \delta_i)$-QDP in the framework $\left(\cS, \cQ,\Theta,  \cM_i\right)$, for $i\in \{1,2\}$, then the composed mechanism satisfies $\left( \varepsilon_1 + \varepsilon_2,  \delta_1 + \delta_2 \right)$-QDP in $\left(\cS, \cQ^{(2)},\Theta, \bigotimes_{i=1}^2 \cM_i\right)$ 
       \item If $\cA_i$ is $(\varepsilon_i,\delta_i)$-QDP in the framework $(\cS,\cQ,\Theta,\bar{\cM})$, for $i\in \{1,2\}$, then the composed mechanism satisfies 
       $(\varepsilon',\delta')$-QDP in $\left(\cS, \cQ^{(2)},\Theta, \bar{\cM}^2 \right)$  with
       \vspace{-1mm}
    \begin{align}
         \varepsilon'& \coloneqq \varepsilon_1 + \varepsilon_2 + \ln\!\left( \frac{1}{(1-\delta_1)(1-\delta_2)}\right), \\
         \delta'& \coloneqq \sqrt{\delta_1(2-\delta_1)} + \sqrt{\delta_2(2-\delta_2)}.    
    \end{align}
and  also satisfies $(\varepsilon_1 + \varepsilon_2, \delta)$ in the same framework with $\delta \coloneqq \min\{ \delta_1 +e^{\varepsilon_1} \delta_2, \delta_2 + e^{\varepsilon_2} \delta_1 \}$.
\end{enumerate}

\medskip
\underline{Adaptive composability}: Suppose that $\cA_1$ satisfies $(\varepsilon_1,\delta_1)$-QDP and 
   $\cA_2$ chosen adaptively satisfies $(\varepsilon_2,\delta_2)$-QDP, as in~\eqref{eq:adaptivelyComposable-distinguishability}. Then, the composed mechanism in \cref{fig:adaptive composition} with $\sigma_I$ in~\eqref{eq:input-correlated-compose} satisfies $(\varepsilon_1+ \varepsilon_2, \delta_2 + \delta_1 | \cY|)$ in the framework $(\cS,\cQ \times \cQ,\Theta,\bar{\cM} \otimes \bar{\cM} )$.
\end{corollary}

\begin{IEEEproof}
    Item 1 in the parallel composability part follows by a similar argument as given  in the proof of Property~3 from \cref{thm:QPP_properties}. For the proof of Item 2,
    first, we use quasi-convexity of the DL~divergence (property 2 in \cref{prop: Properties of DL divergence}) and then adapt Item 3 of \cref{Cor: Properties of QPP all measurements}. The adaptive composition result follows along the same lines as the proof of \cref{Prop:Adaptive-composition-of-QPP} for fixed $z$, and then averaging over all $z \in \cZ$ gives the desired result.
    \end{IEEEproof}

\begin{remark}[Extensions beyond flexible QDP]
    \cref{cor:composability-flexible-QDP} does not hold for the general QPP framework. Indeed, it fails to hold, for instance, for the classical PP framework \cite[Theorem~9.1]{KM14}. Nevertheless, \cref{cor:composability-flexible-QDP} can be extended
    to account for input states 
    $ \sum_{x \in \cX} P_X(x) \, \rho^x \otimes \rho^x$ subjected to additional structural assumptions on the class of admissible distributions:
    \begin{equation}
        \Theta \subseteq \left\{ P_X \in \cP(\cX): \begin{array}{cc}
           \forall \ (\cR,\cT) \in \cQ, \ \exists \ x,x' \in \cX   \quad &  \\
            \textnormal{s.t.} \ q_\cR(x)= q_\cT(x')=1   
        \end{array}  \hspace{-8mm}\right\} 
    \end{equation}
    where $q_\cR$ and $q_\cT$ are defined as in \cref{def: qpp}. The classical version of this condition for PP is known as \textit{``universally composable scenarios''} \cite[Corollary~9.4]{KM14}.
\end{remark}

\section{Characterizing Optimal Privacy-Utility Tradeoff 
} \label{app:privacyUtilityTradeoff}

In this Appendix, we focus on identifying the optimal utility that can be obtained by applying an $(\varepsilon,\delta)$-QPP mechanism. Here, we first focus on the setting in which 
$\cQ=\{ (\cR_1,\cR_2), (\cR_2,\cR_1) \}$, $\bar{\cM}=\{\rM: 0 \psd  \rM \psd \rI\}$, and $\Theta=\{ P_X\}$, but the following ideas can be extended to the case when $\cQ$ is an arbitrary finite set and $\Theta$ includes a finite number of probability distributions. However, the computational complexity involved in identifying the optimal utility increases with the cardinality of the set $\cQ$ and $\Theta$, due to the addition of more constraints to the optimization problem.

To incorporate privacy requirements, we use the equivalent formulation of QPP via the DL~divergence presented in \cref{prop: Equivalent formulation with DS}. To this end, first, we employ the SDP  formulated in \cref{lem: SDP formulation ISD} to compute the relevant DL~divergence and then use that in the optimization of utility. 
We showcase the use of this SDP in characterizing optimal utility next. 

\begin{proposition}[Optimal utility for fixed privacy constraints]
    The optimal utility, as quantified by the $\gamma$-utility metric, for every privacy mechanism that is $(\varepsilon,\delta)$-QPP in the $(\cS,\cQ,\Theta,\bar{\cM})$ framework, where $\cQ=\{ (\cR_1,\cR_2), (\cR_2,\cR_1)\}$, is given by the following:
    \begin{multline}
\sU\!\left(\varepsilon,\delta,\cR_1,\cR_2 \right) \coloneqq 1 -\\
\inf_{\substack{\mu \geq 0 \\ Z_{AD} \geq 0 \\ \Gamma_{CD}^\cB \geq 0 \\ \Gamma_{AC}^\cA \geq 0 \\ \lambda_1 \geq 0, Y_1 \geq 0 \\ \lambda_2 \geq 0, Y_2 \geq 0  }}
\left\{ \begin{array}[c]{c}
\mu :\\
     Z_{AD} \geq \Gamma_{AD} - \Tr_C \!\left[ \Gamma_{CD}^\cB \T_C(\Gamma_{AC}^\cA) \right], 
    \\  \mu I_A \geq \Tr_D\!\left[Z_{AD}\right],\\ \Tr_D\!\left[\Gamma_{CD}^\cB\right] = I_C, \\   \Tr_D\!\left[\Gamma_{AC}^\cA\right] = I_A, \\  \ln(\lambda_1)\leq \varepsilon, \\
    \Tr\!\left[Y_1\right] \leq \delta, \\ 
     Y_1 \geq \Tr_A \!\left[ \big(\T(\rho^{\cR_1}) \otimes I_C \big) \Gamma_{AC}^\cA \right] \\
     \quad - \lambda_1 \Tr_A \! \left[(\T(\rho^{\cR_2}) \otimes I_C) \Gamma_{AC}^\cA \right], \\
     \ln(\lambda_2)\leq \varepsilon, \\
     \Tr\!\left[Y_2\right] \leq \delta, \\ 
     Y_2 \geq \Tr_A \!\left[ (\T\big(\rho^{\cR_2}\big)\otimes I_C) \Gamma_{AC}^\cA \right] \\
     \quad - \lambda_2 \Tr_A \!\left[\big(\T(\rho^{\cR_1}) \otimes I_C\big) \Gamma_{AC}^\cA \right]
    \end{array}
    \right\}.
\end{multline}
\end{proposition}

\begin{IEEEproof}
    The proof follows from the SDP formulation of the $\gamma$-utility given in \cref{prop:SDP formulation of gamma-utility}, and the privacy constraints (i.e., $\max\{\overline{\sD}^{\delta}\!\left( \cA(\rho^{\cR_1}) \Vert\cA( \rho^{\cR_2}) \right), \overline{\sD}^{\delta}\!\left( \cA(\rho^{\cR_2}) \Vert \cA(\rho^{\cR_1}) \right)\} \leq \varepsilon$) imposed through the SDP formulation of DL~divergence presented in \cref{lem: SDP formulation ISD}. We also used the fact that for a superoperator $\cA$ from system $A$ to $C$, the following equality holds
    \cite[Eq.~(3.2.14)]{khatri2020principles}
    \begin{equation}
       \cA(\rho^{\cR_1})=\Tr_A\!\left[\big(\T(\rho^{\cR_1})\otimes I_C \big) \Gamma_{AC}^\cA \right]. 
    \end{equation}
\end{IEEEproof}

\begin{remark}[Privacy constraints via equivalent formulation through hockey-stick divergence]\label{rem: Privacy constraints via the equivalent formulation through the hockey-stick divergence}
  Instead of using the DL~divergence, we can also encode the privacy constraints through the equivalent formulation in \cref{rem:Equivalent formulation with hockey-stick divergence}. To this end, the dual formulation of the hockey-stick divergence (can be obtained by~\eqref{eq:dual-positive-part}), as
  \begin{equation}
  \sE_{\lambda}(\rho \Vert \sigma)= \inf_{Z\geq
0}\left\{  \Tr\!\left[Z\right]:Z\geq\rho-\lambda\sigma\right\} ,     
  \end{equation}
  can also be incorporated to compute the optimum utility.
\end{remark}

\begin{remark}[Optimal privacy parameters for fixed utility]
    To find out the optimal (minimal) privacy parameter $\varepsilon^\star$ for a given mechanism $\cA$ with the utility constraint $\gamma$, and fixed tolerance $\delta$,
    first we compute the following quantity:
     \begin{multline}
\lambda^\star_1(\cA,\gamma,\delta) \coloneqq \\
\inf_{\substack{\lambda \geq 0 \\ Z_{AD} \geq 0 \\ \Gamma_{CD}^\cB \geq 0 \\   Y_1 \geq 0   }}
\left\{ \begin{array}[c]{c}
\lambda :\\
    Z_{AD} \geq \Gamma_{AD} - \Tr_C \! \left[\Gamma_{CD}^\cB \T_C(\Gamma_{AC}^\cA) \right], \\
    ( 1-\gamma) I_A \geq \Tr_D\!\left[Z_{AD}\right], \\ \Tr_D\!\left[\Gamma_{CD}^\cB\right]= I_C,\\  
     \Tr\!\left[Y_1\right] \leq \delta, \\ 
      Y_1 \geq \Tr_A \! \left[ \big(\T(\rho^{\cR_1}) \otimes I_C \big) \Gamma_{AC}^\cA \right] \\
     \quad - \lambda \Tr_A \! \left[(\T(\rho^{\cR_2}) \otimes I_C) \Gamma_{AC}^\cA \right]
    \end{array}
    \right\}.
\end{multline} 
    Similarly $\lambda^\star_2$ can be obtained by exchanging  $\rho^{\cR_1}$ and $\rho^{\cR_2}$. 
Then the optimal value is given by
\begin{equation}
\varepsilon^\star(\cA,\gamma,\delta)\coloneqq  \ln\!\left(\max\{ \lambda^\star_1(\cA,\gamma,\delta), \lambda^\star_2(\cA,\gamma,\delta) \}\right). 
\end{equation}
The optimal (minimal) $\delta$ for a fixed $\varepsilon$ with a utility constraint can be obtained by encoding the privacy constraint through the dual form of hockey-stick divergence, as given in \cref{rem: Privacy constraints via the equivalent formulation through the hockey-stick divergence}.
\end{remark}

\section{Proof of \cref{lem:Bound-on-Renyi}}
\label{proof:bound-on-quantum-renyi}

The proof follows analogously to the classical version of this bound in \cite[Proposition~3.3]{bun2016concentrated}, along with the upper bound for an arbitrary $\sD_\alpha(\cdot \Vert \cdot)$ satisfying data processing. Set $\alpha > 1$. Then, for such $\sD_\alpha(\cdot \Vert \cdot)$, 
from \cite[Equation~4.34]{tomamichel2015quantum}, which is 
obtained by choosing a specific preparation channel, we have that
\begin{equation} \sD_\alpha(\rho \| \sigma) \leq \frac{1}{\alpha-1} \log \Tr \! \left[\sigma^{1/2} \!\left( \sigma^{-1/2} \rho \sigma^{-1/2} \right)^\alpha \sigma^{1/2} \right].
\label{eq: upperbound on D alpha}
\end{equation} 
Let us use the following substitution:
\begin{equation}\label{eq:substitution}
    \sD_T(\rho \Vert \sigma) = \varepsilon.
\end{equation}
Then we have $\sD_{\max}(\rho \| \sigma) \leq \varepsilon$ and $\sD_{\max}(\sigma \| \rho) \leq \varepsilon$.  Moreover, with the definition of $\sD_{\max}(\cdot \Vert \cdot)$ in~\eqref{eq:D-max-def}, we have $\rho \psd e^\varepsilon \sigma $ and $\sigma \psd e^\varepsilon \rho $. Then we find that
\begin{equation} e^{-\varepsilon} \rI \psd  \sigma^{-1/2} \rho \sigma^{-1/2} \psd  e^{\varepsilon} \rI.\end{equation}
Suppose that $\sigma^{-1/2} \rho \sigma^{-1/2}$ has the following spectral decomposition $\sum_{i} t_i |\phi_i\rangle\!\langle \phi_i|$. Then $e^{-\varepsilon} \leq t_i \leq e^\varepsilon$, and so for all $i$, $\exists \lambda_i\in[0,1]$ such that 
\begin{equation}\label{eq:t_i-def}
    t_i=\lambda_i e^\varepsilon + (1-\lambda_i) e^{-\varepsilon}.
\end{equation}
Consider that
\begin{align}
    & e^{(\alpha -1)\sD_\alpha(\rho \| \sigma)} \notag \\
    &\leq \Tr\!\left[\sigma^{1/2} \!\left( \sigma^{-1/2} \rho \sigma^{-1/2} \right)^\alpha \sigma^{1/2} \right] \\ 
    &=\Tr\! \left[\sigma^{1/2} \sum_{i} \!\left(\lambda_i e^\varepsilon + (1-\lambda_i) e^{-\varepsilon}\right )^\alpha |\phi_i\rangle\!\langle \phi_i| \ \sigma^{1/2} \right] \\
    &= \sum_{i} \!\left(\lambda_i e^\varepsilon + (1-\lambda_i) e^{-\varepsilon}\right )^\alpha \Tr \!\left[\sigma |\phi_i\rangle\!\langle \phi_i| \right]\\
    & \leq \sum_{i} \!\left(\lambda_i e^{\varepsilon \alpha} + (1-\lambda_i) e^{-\varepsilon \alpha}\right ) \Tr \!\left[\sigma |\phi_i\rangle\!\langle \phi_i| \right]  \label{eq: convexity x alpha} \\
    &= e^{\varepsilon \alpha} c_1 + e^{-\varepsilon \alpha} c_2 \numberthis \label{eq: alpha relation}
\end{align}
where the first inequality follows from the inequality in~\eqref{eq: upperbound on D alpha}, the second from the convexity of the function $x \mapsto x^\alpha$ for $\alpha >1$, and the definitions
\begin{align}
    c_1 & \coloneqq \sum_{i} \lambda_i \Tr \! \left[\sigma |\phi_i\rangle\!\langle \phi_i| \right],\\
    c_2 & \coloneqq  \sum_{i} (1-\lambda_i) \Tr \!\left[\sigma |\phi_i\rangle\!\langle \phi_i| \right]. 
\end{align}

In~\eqref{eq: alpha relation}, we arrive at a function of $\alpha$ (i.e., $e^{\varepsilon \alpha} c_1 + e^{-\varepsilon \alpha} c_2$). Observing that $c_1 + c_2=1$, we can find $c_1$ and $c_2$ by evaluating this function of $\alpha$ at $\alpha=1$, which turns out to be equal to one because 
\begin{equation}
    \sum_i t_i \Tr [\sigma |\phi_i\rangle\!\langle \phi_i| ]= \Tr[\rho] =1,
\end{equation}
where $t_i$ is given in~\eqref{eq:t_i-def}.
Proceeding with this we get 
$c_1= \frac{1-e^{-\varepsilon}}{e^\varepsilon-e^{-\varepsilon}} $.
Then collecting all these relations and simplifying we obtain, 
\begin{equation}
e^{(\alpha -1)\sD_\alpha(\rho \| \sigma)} \leq \frac{\sinh(\alpha \varepsilon)- \sinh\!\left((\alpha-1) \varepsilon \right)}{\sinh(\varepsilon)}.
\label{eq:sinh-bnd}
\end{equation}
Together with \cite[Lemma~B.1]{bun2016concentrated} and the assumption that $\alpha \varepsilon \leq 2$, we can further bound~\eqref{eq:sinh-bnd} from above by $e^{\alpha(\alpha-1)\varepsilon^2/2}$. With the substitution in~\eqref{eq:substitution} we conclude the proof. 

\section{Proof of \cref{lem:strength_QPP}}\label{app:lemma_strength_proof}

Fix $P_X \in \Theta$ and $(\cR,\cT) \in \cQ $. 
By assumption, we have that
\begin{equation}
    \Tr\!\left[\rM \cA(\rho^\cR)\right] \leq e^\varepsilon \Tr\!\left[\rM \cA(\rho^\cT)\right]+ \delta.
\end{equation}
{With the choice for $\delta'$ as in the Lemma statement (i.e.,~\eqref{eq:delta_prime})}, we have { $\delta- \delta'=(1-\delta)(e^{\varepsilon'}-e^\varepsilon)/ (e^\varepsilon +1)$}. 
Plugging this in, we find that 
\begin{align}
    & \Tr\!\left[\rM \cA(\rho^\cR)\right] \notag \\
    &\leq e^{\varepsilon'} \Tr\!\left[\rM \cA(\rho^\cT)\right] + \delta' +(\delta- \delta') \notag \\
    & \qquad
    +(e^\varepsilon- e^{\varepsilon'}) \Tr\!\left[\rM \cA(\rho^\cT)\right] \\
    &=e^{\varepsilon'} \Tr\!\left[\rM \cA(\rho^\cT)\right] + \delta' \notag \\
   & \qquad 
   + (e^\varepsilon- e^{\varepsilon'}) \left( \Tr\!\left[\rM \cA(\rho^\cT)\right] -\frac{1-\delta}{e^\varepsilon +1}\right).
\end{align}
Since $\varepsilon' \leq \varepsilon$, we get the desired inequality if $\Tr\! \left[\rM \cA(\rho^\cT)\right]  \leq \frac{1-\delta}{e^\varepsilon +1}$.

By choosing the measurement operator $\rI-\rM$, we also have by assumption that
\begin{equation} \label{eq: refer equation 2}
     \Tr\! \left[(\rI-\rM) \cA(\rho^\cT) \right ] \leq e^\varepsilon \Tr \! \left[ (\rI- \rM) \cA(\rho^\cR) \right ] + \delta. 
\end{equation}
Rewriting~\eqref{eq: refer equation 2}, we arrive at 
\begin{equation}  \Tr\! \left[\rM \cA(\rho^\cR)\right]\leq 1-e^{-\varepsilon}(1-\delta) + e^{-\varepsilon} \ \Tr \! \left[\rM \cA(\rho^\cT)\right]. \end{equation}
Similar to the previous manipulations, we get 
\begin{multline}
\Tr\!\left[\rM \cA(\rho^\cR) \right] \leq e^{\varepsilon'} \Tr \!\left[\rM \cA(\rho^\cT)\right] + \delta' \\
\quad \quad + (e^{\varepsilon'}- e^{-\varepsilon}) \left(- \Tr\!\left[\rM \cA(\rho^\cT) \right]+\frac{1-\delta}{e^\varepsilon +1}\right).
\end{multline}
Since $\varepsilon' \leq \varepsilon$, we arrive at the desired inequality when $\Tr\!\left[\rM \cA(\rho^\cT)\right]  \geq \frac{1-\delta}{e^\varepsilon +1}$.
By these two arguments, the desired inequality holds for either of the cases, proving its validity. 

A similar inequality holds for every $(\cR,\cT) \in \cQ$ and $P_X \in \Theta$. Thus, the desired implication has been proved.

\section{Bounds on Holevo Information from QDP and QLDP} \label{app:Bounds_Holevo}

\begin{figure}
    \centering
    \includegraphics[width=\linewidth]{ 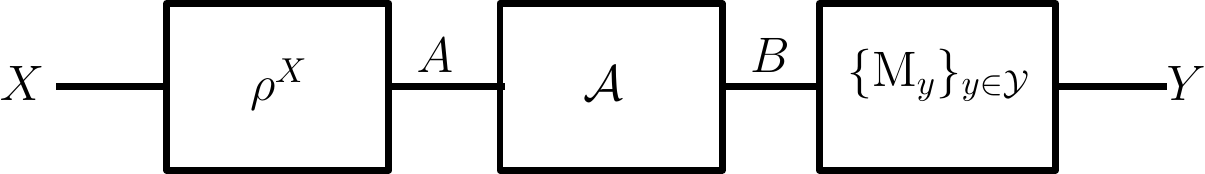}
     \caption{Setup relevant to \cref{prop:Bounds on mutual information due to QLDP}: $X$ is a classical random variable that determines $\rho^X$, which is the input into the channel $\cA: \cL(\cH_A) \to \cL(\cH_B)$. Then $Y$ is a random variable describing the outcome after applying the POVM $\{\rM_y\}_{y \in \cY}$. Note that here the classical systems related to $X,Y$ are also given by the same system labels $X,Y$.}
    \label{fig:classicalXY}
\end{figure}

Let $X \sim P_X$ be a random variable, which can take values in an alphabet $\cX$.
Depending on $X$, the state $\rho^X$ is chosen from the set $\{ \rho^1, \ldots, \rho^{|\cX|} \}$. Then the state $\rho^X$ is sent through a quantum channel $\cA: \cL(\cH_A) \to \cL(\cH_B)$ satisfying QLDP (recall this notion from \cref{rem:QLDP}). Afterwards the goal is to identify $X$ by performing a measurement described by the POVM $\{\rM_y\}_{y\in \cY}$, which realizes the output $Y$. The flow diagram relevant to this setup is shown in \cref{fig:classicalXY}.

Here, we focus on how much information about $X$ can be learned from the output of the quantum privacy mechanism $\cA(\rho^X)$ and the classical output $Y$, with an emphasis on the quantities $\sI\!\left(X;B\right)_{\sigma}$ and $\sI(X;Y)$, respectively, where 
\begin{equation}
    \sigma_{XB} =\sum_{x \in \cX} P_X(x) \, |x \rangle \! \langle x| \otimes \cA(\rho^x).
\end{equation}
We define the Holevo information of the classical--quantum state $\sigma_{XB}$ as
\begin{multline}
    \sI\!\left(X;B\right)_{\sigma} 
    \coloneqq 
    \sS \!\left( \sum_{x \in \cX} P_X(x) \, \cA(\rho^x) \right) \\
    - \sum_{x \in \cX} P_X(x) \, \sS\!\left(\cA(\rho^x)\right).
\end{multline}
By data processing of the Holevo information, we have that $\sI(X;Y) \leq \sI\!\left(X;B\right)_{\sigma}$.
Next, we provide bounds for $\sI\!\left(X;B\right)_{\sigma}$ when $\cA$ satisfies $\varepsilon$-quantum local DP (QLDP). Recall that, to achieve QLDP, the algorithm $\cA$ is designed so that the output of $\cA$ may not release much information about the input states (recall \cref{rem:QLDP}).

\begin{proposition}[Bounds on Holevo information due to QLDP] \label{prop:Bounds on mutual information due to QLDP}
Let $\cA: \cL(\cH_A) \to \cL(\cH_B)$ be a quantum channel. If $\cA$ satisfies $\varepsilon$-QLDP, then
\begin{equation}
  \sI\!\left(X;B\right)_{\sigma} \leq \min\! \left\{\varepsilon, \frac{\varepsilon^2}{2} \right \}.  
\end{equation}
Furthermore, if $\cA$ is $(\varepsilon,\delta)$-QLDP, then 
\begin{equation}
\sI\!\left(X;B\right)_{\sigma} \leq \delta' \ln(d-1) + \sh(\delta'),
\end{equation}
if $\delta' \in [0,1-1/d]$,
where $\delta'= 1- {2(1-\delta)}/{(e^\varepsilon +1)}$, 
$d$ is the dimension of the Hilbert space $\cH_B$, and 
\begin{equation}
 \sh(\delta')\coloneqq  -\delta' \ln(\delta')-(1-\delta') \ln(1-\delta')   
\end{equation}
is the binary Shannon entropy in nats.
\end{proposition}

\begin{IEEEproof}
The following proof ideas are inspired by the works on mutual information based classical DP and PP presented in \cite[Theorem~1]{CY16} and \cite[Theorem~1]{nuradha2022pufferfishJ}, respectively. 
For the proof of the first part, consider that
\begin{align}
  & \sI\!\left(X;B\right)_{\sigma} \notag \\
  &= \sum_{x \in \cX} P_X(x) \,  \sD\!\left( \cA(\rho^x) \middle \Vert \sum_{x' \in \cX} P_X(x') \cA(\rho^{x'}) \right)   \\ 
  & \leq \sum_{x \in \cX} \sum_{x' \in \cX} P_X(x) P_X(x') \  \sD \!\left( \cA(\rho^x) \middle \Vert  \cA(\rho^{x'}) \right) \\
  & \leq \min\! \left\{\varepsilon, \frac{\varepsilon^2}{2} \right \},
\end{align}
where the first inequality follows from the joint convexity of quantum relative entropy~\cite{LR73}. The final inequality follows because $\cA$ satisfies $\varepsilon$-QLDP, as well as from \cref{prop: Bounds on Quantum divergences with QPP} and the setup for QLDP in \cref{rem:QLDP}.

For the second part, consider that
\begin{align}
    \sI\!\left(X;B\right)_{\sigma} 
    &= \sS \!\left( \cA(\bar{\rho}) \right) - \sum_{x \in \cX} P_X(x) \sS\!\left(\cA(\rho^x)\right)\\ 
    & \leq \delta' \ln(d-1) + \sh(\delta'),
\end{align}
where 
\begin{equation}
    \bar{\rho}= \sum_{x \in \cX} P_X(x) \rho^x, 
\end{equation}
and the final inequality follows from the fact that  $\left\| \cA(\bar{\rho})- \cA(\rho^x) \right\|_1 \leq 2\delta'$ due to $\cA$ satisfying $(\varepsilon,\delta)$-QPP along with \cref{prop:Bounds on trace norm}. Then, to arrive at the final expression, we use the entropy continuity result known as the Fannes–Audenaert Inequality~\cite{A07}:
\begin{equation}
\left|\sS\!\left(\cA(\bar{\rho})\right)- \sS\!\left(\cA(\rho^x)\right)   \right| \leq \delta' \ln(d-1) + \sh(\delta'),
\end{equation} 
concluding the proof.
\end{IEEEproof}

\medskip
Next, we extend the earlier setup to the case in which we generate $n$ i.i.d.~random variables from the distribution of~$P_X$. Then, $\{X_1, \ldots, X_n\}$ forms a database. Depending on the value of the database, we choose $\rho^{X^n}\coloneqq \rho^{X_1} \otimes \cdots \otimes \rho^{X_n} $. Then $\rho^{X^n}$ is passed through a quantum algorithm $\cA$ that satisfies QDP, followed by a POVM $\{\rM_y\}_{y\in \cY}$ that generates the random outcome $Y$ taking values in  $\cY$.
In this formulation, we take the convention that $\rho \sim \sigma$ (i.e., $\rho$ and $\sigma$ are neighbors) in the QDP framework if $\Tr_i\!\left[\rho\right]=\Tr_i\! \left[\sigma \right]$ for all $i\in\{1, \ldots, n\}$.
Let 
\begin{equation}
    \sigma_{X^n B} =\sum_{x \in \cX} P_{X^n}(x^n) \, |x^n \rangle \! \langle x^n| \otimes \cA(\rho^{x^n}).
\end{equation}

\begin{proposition}[Bounds on mutual information due to QDP]
\label{prop:Bounds on mutual information due to QDP}
Let $\cA: \cL(\cH_A) \to \cL(\cH_B)$ be a quantum channel. For all $i\in\{1,\ldots,n\}$, suppose that $X_i$ is drawn i.i.d.~from the distribution $P_X$.
If $\cA$ satisfies $\varepsilon$-QDP, then
\begin{equation} \sup_{i\in\{1,\ldots,n\}, P_X }\sI(X_i;B|X^{n \backslash i})_{\sigma} \leq \min\! \left\{\varepsilon, \frac{\varepsilon^2}{2} \right \}.\end{equation}

Furthermore, if $\cA$ satisfies $(\varepsilon,\delta)$-QDP, then 
\begin{equation}
\sup_{i\in\{1,\ldots,n\}, P_X }\sI(X_i;B|X^{n \backslash i})_{\sigma} \leq \delta' \ln(d-1) + \sh(\delta'),
\end{equation}
if $\delta' \in [0,1-1/d]$, 
where $\delta'= 1- {2(1-\delta)}/{(e^\varepsilon +1)}$.
\end{proposition}

\begin{IEEEproof}
    First, let us consider $\sI(X_i;B|X^{n \backslash i}= z^{n \backslash i})_{\sigma}$.
    Define the shorthands
    \begin{align}
    \rho^{n\backslash i} & \coloneqq  \rho_1^{z_1} \otimes \ldots \rho_{i-1}^{z_{i-1}} \otimes \rho_{i+1}^{z_{i+1}} \otimes \rho_n^{z_n}, \\
    \omega &  \coloneqq \sum_{x \in \cX} P_{X_i}(x) \rho_i^x \otimes \rho^{n\backslash i}, \\
    \omega^x & \coloneqq \rho_i^x \otimes \rho^{n\backslash i} .
    \end{align}
    Then $\omega \sim \omega^x $ because $\Tr_i\!\left[\sigma\right]= \Tr_i\!\left[\sigma^x\right]$. 
    
    For the first part, it thus follows that
    \begin{multline}
     \sI(X_i;B|X^{n \backslash i}= z^{n \backslash i})_{\sigma}   \\ 
    = \sum_{x \in \cX} P_{X_i}(x) \  \sD\!\left( \cA(\omega^x) \middle \Vert \cA(\omega)\right)
    \leq  \min\! \left\{\varepsilon, \frac{\varepsilon^2}{2} \right \}.
    \end{multline}
   The last inequality holds because $\cA$ satisfies $\varepsilon$-QDP, along with \cref{prop: Bounds on Quantum divergences with QPP}. Since this inequality holds for all possible $z^{n \backslash i}$ such that $\{X^{n \backslash i}= z^{n \backslash i}\}$, the desired relation holds.

   For the second part, consider that
   \begin{align}
   & \nonumber \sI(X_i;B|X^{n \backslash i}= z^{n \backslash i})_{\sigma} \\
    &= \sS \!\left( \cA({\omega}) \right) - \sum_{x \in \cX} P_{X_i}(x) \sS\!\left(\cA(\omega^x)\right)\\ 
    & \leq \delta' \ln(d-1) + \sh(\delta'),
\end{align}
where the last inequality holds because 
     $\left\| \cA(\omega) - \cA(\omega^x) \right \|_1 \leq 2\delta'$ with $\cA$ being $(\varepsilon,\delta)$-QDP for the pair $\sigma \sim \sigma^x$, and again applying  the continuity result for quantum entropy, as in the proof of \cref{prop:Bounds on mutual information due to QLDP}.
\end{IEEEproof}

\medskip

By the data processing inequality for mutual information, we have $\sI(X_i;Y| X^{n \backslash i}) \leq \varepsilon' \coloneqq \min\{ \varepsilon, \varepsilon^2/2\}$.
This showcases that the setup of \cref{prop:Bounds on mutual information due to QDP} satisfies $\varepsilon'$-mutual information differential privacy, as proposed in~\cite{CY16}, where 
 a randomized mechanism $A: \cX^{n\times k} \to \cY$ is defined to be $\varepsilon$-mutual information differentially private if 
 \begin{equation}
    \sup_{\substack{P_X \in \cP(\cX^{n \times k}),\\i\in \{1,\ldots,n\}}} \sI(X_i;A(X) | X^{n \backslash i}) \leq \varepsilon.\label{EQ:MI_DP}
 \end{equation}

\end{document}